\documentclass[12pt, a4paper]{article}

\usepackage[english]{babel}
\usepackage[toc,page]{appendix}
\usepackage[utf8]{inputenc}
\usepackage{amsmath}
\usepackage{amssymb}
\usepackage{array}
\usepackage{graphicx}
\usepackage{placeins}
\usepackage{caption}
\usepackage{subcaption}
\usepackage[colorinlistoftodos]{todonotes}
\usepackage{tensor}
\usepackage{fullpage}
\usepackage{braket}
\usepackage{framed}
\usepackage{feynmf}
\usepackage{cancel}
\usepackage{empheq}
\usepackage{dsfont}
\usepackage{bbm}
\usepackage{sectsty}
\usepackage{sectsty}
\usepackage{lipsum}
\usepackage{hyperref}
\usepackage{framed}
\usepackage[affil-it]{authblk}
\usepackage{pgfplots}
\usepackage{tikz}
\usetikzlibrary{positioning}
\usetikzlibrary{decorations.markings}
\usetikzlibrary{decorations.pathmorphing}
\hypersetup{
 colorlinks,
 citecolor=magenta,
 filecolor=blue,
 linkcolor=blue,
 urlcolor=blue
}
\definecolor{shadecolor}{rgb}{0.9,0.9,1}
\newcommand{\La}{{\cal L}}

\newcommand{\Ham}{{\cal H}}
\newcommand{\cW}{{\cal W}}

\newcommand{\Op}{{\cal O}}

\newcommand{\N}{{\cal N}}
\newcommand{\E}{{\cal E}}
\newcommand{\I}{{\cal I}}

\newcommand{\bF}{\mathbb F}

\newcommand{\cR}{{\cal R}}

\newcommand{\half}{\frac{1}{2}}

\newcommand{\Real}{{\mathbb{R}}}
\newcommand{\D}{{\cal D}}

\newcommand{\pvec}[2]{\begin{pmatrix} #1 \\ #2 \end{pmatrix}}

\newcommand{\diff}[2]{\frac{d #1}{d #2}}
\newcommand{\eq}[1]{(\ref{eq:#1})}
\numberwithin{equation}{section} 

\usepackage[font=small, margin=2.5cm]{caption} 

\tikzset{->-/.style={decoration={
 markings,
 mark=at position .5 with {\arrow{>}}},postaction={decorate}}}
 
\usepackage{fancyhdr}
\fancypagestyle{plain}{%
	\fancyhead[R]{DMUS-MP-19-08 \\ LTH 1204}
	
}
\begin{document}

\title{\vspace{0.75cm} From Static to Cosmological Solutions of $\N=2$ Supergravity \vspace{0.5cm} }

\author{J. Gutowski\footnote{J.Gutowski@surrey.ac.uk}${\ }^1$,  T. Mohaupt\footnote{Thomas.Mohaupt@liv.ac.uk}${\ }^2$ and G. Pope\footnote{Giacomo@liv.ac.uk}${\ }^2$ \vspace{0.5cm}}

\affil{\normalsize $^1$Department of Mathematics \\ University of Surrey \\Guildford, GU2 7XH, UK}
\affil{$^2$Department of Mathematical Sciences \\ University of Liverpool \\ Liverpool, L69 7ZL, UK}
\date{August 28, 2019}
\maketitle
\abstract{
We obtain cosmological solutions with Kasner-like asymptotics in $\N=2$ gauged and ungauged
supergravity by maximal analytic continuation of planar versions of non-extremal black hole solutions. Initially, we construct static solutions with planar symmetry by solving the time-reduced field equations.
Upon lifting back to four dimensions, the resulting static regions are incomplete and bounded
by a curvature singularity on one side and a Killing horizon on the other. Analytic continuation reveals the existence of dynamic patches in the past and future, with Kasner-like asymptotics. For the
ungauged STU-model, our solutions contain previously known solutions with the same conformal diagram as a subset. We find explicit lifts to five, six, ten and eleven dimensions which show that in the extremal limit, the underlying brane configuration is the same as for STU black holes.
The extremal limit of the six-dimensional lift is shown to be BPS for special choices of the integration constants. We argue that there is a universal correspondence between spherically symmetric black hole solutions and
planar cosmological solutions which can be illustrated using the Reissner-Nordstr\"om solution of Einstein-Maxwell theory.
}
\pagenumbering{gobble}
\sectionfont{\large}
\subsectionfont{\normalsize}

\newpage
\tableofcontents

%
%
\newpage
\pagenumbering{arabic}
\section{Introduction}

Cosmological solutions of string theory are far less understood than stationary, let alone BPS solutions. It thus came as a surprise to the authors to obtain cosmological solutions by tweaking the horizon geometry of the well-studied class of STU black holes. The original aim of this paper was to continue previous work \cite{Mohaupt:2011aa,Errington:2014bta,Dempster:2015,Dempster:2016} on the construction of non-extremal stationary solutions in theories of four-dimensional ${\cal N}=2$ vector multiplets coupled to gauged and ungauged supergravity, and to make contact in the extremal limit with the classification of supersymmetric near-horizon geometries. At a technical level, this paper extends the work of \cite{Dempster:2015,Dempster:2016} on solutions with planar horizons
to solutions with more than one charge. The single charge solutions are `Nernst branes',
which are solutions with zero entropy in the extremal limit, where they reduce to the solutions of \cite{Barisch:2011ui,Cardoso:2015wcf}.

In this paper, we construct non-extremal planar solutions with more than one charge, and we observe
that the static region for solutions with three or four charges interpolates between a curvature
singularity and a Killing horizon. By analytical continuation, we obtain time-dependent regions which are asymptotic to Kasner-like solutions in the infinite past and infinite
future. These solutions are `inside-out' compared to the causal structures of non-extremal black hole and black brane solutions, and should be interpreted as cosmological. Our
family of solutions overlaps with previously found solutions of Einstein-Maxwell-Dilaton theories
and truncations of supergravity theories, which in particular display the same conformal diagram \cite{Burgess:2002vu, Burgess:2003mk, Cornalba:2003kd}. The three-charge ansatz leads to solutions for a class of gauged vector multiplet theories, including
the gauged STU-model, while the four-charge system is a solution to the ungauged STU model
and thus admits lifts to ten and eleven-dimensional supergravity.
In fact, from the ten- or eleven-dimensional point of view these solutions correspond to the
same brane configurations as the STU-black hole; the only difference is that the planar rather than spherical horizon geometry leads to additional de-localisation along
two non-compact directions. The cosmological character of our solutions crucially depends
on them being non-extremal, and thus on our ability to construct non-extremal solutions. From the ten- and eleven-dimensional point of view, these solutions correspond to well-known BPS brane configurations. Besides non-extremality, the only ingredient
for tweaking a black hole into a cosmological solution is to change the horizon geometry
from spherical to planar. This is a robust feature, which can already be understood and demonstrated using the spherical and planar Reissner-Nordstr\"om solutions of
Einstein-Maxwell theory, which are contained in our family of solutions as special cases.

On taking the extremal limit, defined by the Killing horizon having
zero surface gravity, the distance between singularity and horizon
becomes infinite and the cosmological patches disappear. With some
additional fine-tuning of parameters, we identify our uplifted solution
as a supersymmetric solution of six-dimensional
supergravity. The most general classification of supersymmetric solutions in six-dimensional supergravity was constructed
in \cite{Akyol:2010iz, Akyol:2012cq, Akyol:2013ana} making use of spinorial geometry techniques
\cite{Gillard:2004xq},
see also \cite{Cano:2018wnq, Lam:2018jln}  for classifications constructed via the Fierz identity/spinor bilinear method.
 The 6-dimensional
supergravity which we consider is a simpler truncation of this, as it is coupled only to a single 3-form field strength and dilaton; the supersymmetric solutions
of this theory were analysed in detail using the Fierz identity/spinor bilinears method in \cite{Cariglia:2004kk}.
We determine how the uplifted solution
is described by this classification,
and show how the geometry is determined in terms
of harmonic functions on a specific non-flat Gibbons-Hawking manifold.
Such a construction is valuable in providing a solution generating
technique which could be used to produce a potentially very large
number of generalizations of our solutions.  For example, it
is known that many new black hole solutions, with
regular horizons and nontrivial topology
exterior to the horizon in the form of 2-cycles
supported by magnetic flux, can be found
by taking the standard BMPV black hole, which is written
in terms of harmonic functions on the Gibbons-Hawking base space ${\mathbb{R}}^4$, and making appropriately
chosen modifications to the harmonic functions
appearing in the solution
\cite{Horowitz:2017fyg, Breunholder:2018roc}.
Related methods were also previously employed to find
examples of Black Saturn geometries \cite{Gauntlett:2004wh, Gauntlett:2004qy}, as well as numerous
examples of possible smooth horizonless black hole microstate geometries \cite{Bena:2007kg, Bena:2004de}.
By following similar reasoning, we expect to be able to construct large families of new solutions
which are deformations of the planar brane geometries
considered in this work and may exhibit novel
topological structures.

The outline of this paper is as follows: in Section 2 we review the necessary background on four-dimensional vector multiplets and their dimensional reduction over time. We also explain which restrictions we impose on solutions in order to be able to integrate
the reduced three-dimensional field equations explicitly.  In Section 3 we solve the reduced field equations and obtain solutions with two, three and four charges.
In Section 4 we lift these solutions back to four dimensions and reduce the number
of independent integration constants by imposing that the solutions exhibit a regular Killing horizon.
In Section 5 we show that the lifted, static solutions interpolate between a curvature singularity
and the Killing horizon. By analytic continuation, we extend the solutions beyond the horizon and discover a dynamical patch with Kasner-like asymptotic behaviour at timelike infinity.
In Section 6 we use Kruskal-like coordinates to obtain the
full conformal diagram, which is `Schwarzschild rotated by 90 degrees.'
We observe that timelike geodesics are infinitely extendible. Using the Komar
construction we define a `position-dependent mass' in the
static region which turns out to be negative, which is consistent with the behaviour of the timelike geodesics.
We also compute the Brown-York mass, which while
different from the Komar mass, is also negative. Finally, we study the proper acceleration of static test bodies, again finding that the singularities are repulsive. In Section 7 we provide
explicit lifts of the four-charge solution, which is a solution to the ungauged STU model,
to five and six, and as well to ten and eleven dimensions. By taking the extremal limit, we recover in ten and eleven dimensions well known BPS brane configurations which yield BPS black holes of the STU model. In Section 8 we study the extremal
limit of the six-dimensional lift in detail and show that with additional fine-tuning, the
extremal limit is BPS. Here we make contact with the classification of six-dimensional supersymmetric near horizon geometries. In Section 9 we discuss the physical interpretation of our solution and explain that their qualitative features already arise in the planar Reissner-Nordstr\"om solution.

Some technical material has been relegated to appendices. Appendix A discusses Kruskal-like coordinates in some more detail. Appendix B shows how the planar Reissner-Nordstr\"om solution arises as a special limit. In Appendix C we review non-extremal and extremal black hole solutions of the STU model and give their explicit lifts to five, six, ten and eleven dimensions for comparison with the planar case.

%
%
\section{Planar Solutions with Multiple Gauge Fields}
%
%

We follow the approach developed
in \cite{Mohaupt:2011aa,Errington:2014bta,Dempster:2015,Dempster:2016}
to construct stationary solutions of $\N=2$, $D=4$ supergravity coupled to $n_V$
vector multiplets, using
\begin{enumerate}
\item[$\bullet$]
the dimensional reduction over time to obtain an effective three-dimensional Euclidean
theory.
\item[$\bullet$]
The real, rather than the more commonly used complex formulation of the special geometry of $\N=2$ vector multiplets, which is based on a Hesse potential rather
than a prepotential.
\item[$\bullet$]
A set of conditions which decouples the field equations and allows us to integrate them elementarily; this requires us to impose restrictions on the admissible prepotential/Hesse
potential, and to consistently truncate the field content to a subset of `purely imaginary' (PI) configurations.
\end{enumerate}
Since the procedure follows with only small modifications as in the previous
papers, in particular \cite{Dempster:2015} where planar symmetry and a single charge were considered, we only summarise the essential steps to avoid unnecessary duplication
of material.

\subsection{Background}

As in \cite{Dempster:2015}
our starting point is the general two-derivative Lagrangian for $n_V$
${\cal N}=2$ vector multiplets coupled to Poincar\'e supergravity,
including the most general (dyonic) FI-gauging. The bosonic
Lagrangian is given by
\begin{equation}
\label{eq:4dlag}
\begin{aligned}
     \mathbf{e}_4^{-1} \La &= - \frac{1}{2}R_4 - g_{I\bar{J}} \partial_\mu X^I \partial^\mu \bar{X}^J + \frac{1}{4} \I_{IJ} F^I_{\mu \nu} F^{J|\mu \nu} + \frac{1}{4} \cR_{IJ} F^I_{\mu \nu} \tilde{F}^{J|\mu \nu} \\
 &+ V^{\text{dyonic}}_4(X,\bar{X}),
\end{aligned}
\end{equation}
where $\mu=0,1,2,3$ are the spacetime indices, $\mathbf{e}_4$ is the vierbein, $R_4$ is the Ricci scalar, $F_{\mu \nu}^I $ are the Abelian field strengths, $I,J = 0,1,\ldots, n_V$. In our conventions the tilde represents the Hodge-dual
\begin{equation}
\tilde{F}_{\mu \nu} = \frac{1}{2} \tensor{\varepsilon}{_{\mu \nu \rho \sigma}} F^{\rho \sigma} \; ,
\end{equation}
and the Riemann tensor is
\begin{equation*}
    \tensor{R}{^\mu_{\nu \rho \sigma}} = -(\partial_\rho \tensor{\Gamma}{^\mu_{\nu \sigma}} - \partial_\sigma \tensor{\Gamma}{^\mu_{\nu \rho}} + \tensor{\Gamma}{^\tau_{\nu \sigma}} \tensor{\Gamma}{^\mu_{\tau \rho}} - \tensor{\Gamma}{^\tau_{\nu \rho}} \tensor{\Gamma}{^\mu_{\tau \sigma}} ) ,
\end{equation*}
which introduces the minus sign in the first term of \eq{4dlag}.

We are using the standard formulation of special K\"ahler
geometry in terms of complex scalar fields $X^I$, which are
subject to complex scale transformations $X^I \rightarrow
\lambda X^I$, $\lambda \in \mathbb{C}^*$. Explicit expressions
for the couplings $g_{I\bar{J}}$, ${\cal N}_{IJ} := \cR_{IJ} + i \I_{IJ}$, can be found
in \cite{Mohaupt:2011aa}, but are not needed in the following. All data except the
scalar potential $V_4^{\rm dyonic}$ are
encoded in a {\em prepotential} $F=F(X^I)$, which is a holomorphic
function, homogeneous of degree two in the complex scalar fields
$X^I$. As already mentioned, the scalars $X^I$ are not independent degrees of
freedom. The Poincar\'e supergravity multiplet contains an
Abelian vector field, called the graviphoton, so that
a theory of $n_V$ vector multiplets has $n_V +1$ vector fields, but
only $n_V$ independent complex scalar fields $z^A$. Using the redundant
parametrisation in terms of $X^I$ is akin to using homogeneous coordinates
on a projective complex space and
has the advantage of formally
balancing the number of scalar and vector fields.

The most important feature of ${\cal N}=2$ vector multiplets is that the field equations (though not
the Lagrangian) are invariant under the action of the symplectic group
$Sp(2n_V+2, \mathbb{R})$, which acts linearly on the field strength
$(F^{I}_{\mu \nu}, G_{I|\mu \nu})^T$, where the dual
field strengths are defined by $G_{I|\mu \nu} := {\cal R}_{IJ} F^{J}_{\mu \nu} - {\cal I}_{IJ}
\tilde{F}^J_{\mu \nu}$.
Symplectic transformations generalise the electric-magnetic duality of the source-free
Maxwell equations, and contain stringy symmetries, such as T-duality and
S-duality, if the theory under consideration arises as a low-energy effective
theory from string theory. By supersymmetry, the full set of field equations is symplectically
invariant, with the scalars described by the
symplectic vectors
$(X^I, F_I)^T$, where $F_I = \partial F/\partial X^I$.
One important feature of our method is to preserve manifest symplectic covariance.
The $n_V$ physical scalars $z^A$
can be parametrized as
\begin{equation}
 z^A = \frac{X^A}{X^0} \; ,
\end{equation}
and we can extract them after solving the equations of motion.

One drawback of using the complex scalar fields $X^I$ is that they
do not form a symplectic vector by themselves. Similarly, the
couplings $g_{I\bar{J}}$, $\I_{IJ}$ and $\cR_{IJ}$ do not transform as
tensors under the symplectic group, but have a more complicated
behaviour. We will therefore switch to a formulation in terms of real scalar
fields $q^a$, $a=1, \ldots, 2n_V+2$, which are related to the complex scalars $X^I$
by
\[
 (q^a) = \left( \begin{array}{c}
          x^I \\ y_J \\
          \end{array} \right) = \mbox{Re}
         \left( \begin{array}{c}
             X^I \\ F_I \\
            \end{array} \right) ,
\]
           and transform as a vector under
symplectic transformations. In this formulation, all couplings are
encoded in a real function $H=H(q^a)$, called the Hesse potential,
which is homogeneous of degree two, and which up to a factor is
the Legendre transform of the imaginary part of the prepotential $F$, see 
\cite{Mohaupt:2011aa} for further details.  It is helpful to think of the
complex and real formulations of special geometry 
as being related to one another in the same
way as the Lagrangian and Hamiltonian formulations of mechanics, see for
example \cite{Cardoso:2012nh} for a detailed discussion.

To include a scalar potential we consider the most general (dyonic) FI gauging
of the vector multiplet theory, which depends on
$2n_V+2$ parameters $g^a=(g^I,g_I)$, transforming as a vector
under symplectic transformations. The expression for the potential in terms of real scalar fields was worked out in
appendix A of \cite{Dempster:2015}.

%
%

\subsection{Dimensional Reduction}
\label{sec:dimred}

We will initially construct static solutions, and therefore decompose
the four-dimensional spacetime metric as
\begin{equation}
\label{4d3d}
 ds_4^2 = - e^{\phi} dt^2 + e^{-\phi} ds_3^2 \; ,
\end{equation}
where $\phi$ and all matter fields are independent of time $t$.
The field $\phi$ is the Kaluza-Klein scalar. There is no Kaluza-Klein
vector since we assume the field configuration to be static, that is
stationary (time-independent) and hypersurface-orthogonal (no
time-space cross-terms). We rearrange the fields using the same method as in \cite{Mohaupt:2011aa} in the following way. The Kaluza-Klein scalar $\phi$ is absorbed
into the scalar fields by introducing
\begin{equation}
\label{YX}
Y^I := e^{\phi/2} X^I \;.
\end{equation}
We do not
introduce a new symbol for the corresponding real scalars $q^a$,
which, are subject to the same rescaling, so that from now on
\[
 (q^a) = \left( \begin{array}{c}
          x^I \\ y_J \\
          \end{array} \right) = \mbox{Re}
         \left( \begin{array}{c}
             Y^I \\ F_I(Y) \\
            \end{array} \right) \;.
\]
The advantage of this
field redefinition is that the Kaluza-Klein scalar is now on the same
footing as the four-dimensional scalar fields. When needed, the Kaluza-Klein
scalar can be extracted as $e^\phi = - 2 H$, where $H$ is the Hesse
potential considered as a function of the rescaled real scalars
$q^a$. Upon dimensional reduction, the $n_V+1$ four-dimensional
vector fields split into scalars $\zeta^I$ and three-dimensional
vector fields which can be dualised into a second set of scalars
$\tilde{\zeta}_I$. These $2n_V+2$ scalars can be combined into
the symplectic vector $\hat{q}^a = \frac{1}{2} (\zeta^I,
\tilde{\zeta}_I)^T$. We refrain from giving the explicit relations
between the various fields at this point, and refer the interested
reader to \cite{Mohaupt:2011aa,Errington:2014bta,Dempster:2015}.
What matters is that all dynamical degrees of freedom
are now encoded in the $4n_V+4$ real scalars $q^a, \hat{q}^a$, which
form two symplectic vectors.

\subsection{Restricted Field Configurations}

In order to obtain solutions of the field equations by decoupling and
elementary integration, we make two further assumptions.
\begin{enumerate}
\item
We will need to know the Hesse potential
explicitly, but models are naturally defined in terms of their
prepotential, \textit{e.g.} in the context of Calabi-Yau compactifications of
string theory. Since the Hesse potential is obtained by a Legendre
transformation of the (imaginary part of the) prepotential, it cannot
be computed in closed form for a generic prepotential.
We will therefore restrict the form of the prepotential in such a way
that we can obtain the Hesse potential explicitly. 
\item
We want the field equations to decouple. This is achieved by
imposing a block structure, where the scalar fields within each block
are proportional to each other, and where equations within a block 
do not couple to equations in other blocks. Such block structures 
appear if we consistently truncate out part of the scalar fields.
\end{enumerate}
We remark that the two types of conditions we impose are not independent.
We only need
to know a Hesse potential for the subset of fields which are not truncated out
consistently. The more fields we truncate out, the larger the class of 
prepotentials admissible. Conversely, when switching on more and more
charges, more and more fields need to be kept in the effective three-dimensional
theory, and the prepotentials we can admit become more and more
restricted. 

For the single charged Nernst brane solutions \cite{Dempster:2015}
of gauged supergravity
it is sufficient to restrict the prepotential to be of the so-called very special
form
\begin{equation}
F(Y) = \frac{f(X^1,\ldots X^n)}{X^0} ,
\end{equation}
where $f$ is a polynomial homogeneous of degree three.
This condition is equivalent to imposing that the vector multiplet theory can
be lifted to five dimensions. For solutions with 2, 3 and 4 charges we
will show in the following sections that non-trivial solutions can be found when further
restricting the prepotential to the forms
\begin{equation}
\label{prepotentials}
\begin{aligned}
    F_2(X) = \frac{X^1 f_2(X^i,\ldots,X^n)}{X^0}\;,
    & \quad F_3(X) = \frac{X^1 X^2 f_3(X^i,\ldots,X^n)}{X^0}\;,
     \\ F_4(X) &= \frac{X^1 X^2 X^3}{X^0} \;,
\end{aligned}
\end{equation}
where $f_2$ and $f_3$ are polynomials homogenous of degree 2 and 1
respectively. While $F_2(X)$ is still the generic form for a
compactification of the heterotic string on $K3 \times T^2$ at string
tree level, $F_4(X)$ is the well known STU model, which is also the
minimal example for a prepotential of the form $F_3(X)$. While more
general models can be defined and solved for by relaxing the condition
that $f_3$ is a polynomial, we do not know how such models could be
embedded into string theory and so restrict ourselves to the polynomial
case.

Next we specify the consistent truncation of the scalar fields
$q^a,\hat{q}^a$ that we impose to achieve decoupling. In
\cite{Errington:2014bta} the truncated field configurations were
called `purely imaginary' (PI) because the corresponding four-dimensional
physical scalars $z^A$ are purely imaginary. This type of condition is
also known as `axion-free', as in our parametrization the real parts
of $z^A$ have an axion-like shift symmetry for prepotentials of the
very special form. In terms of three-dimensional scalars the `PI
condition' takes the form
\begin{equation}
\label{PI1}
 (q^a) \big|_{\text{PI}} = (x^0,\ldots,0;0,y_1,y_2,\ldots,y_n),
\end{equation}
This is extended to the scalars $\hat{q}^a$ which correspond to
four-dimensional gauge fields by
\begin{equation*}
 (\partial_\mu \hat{q}^a) \big|_{\text{PI}} = \half (\partial_\mu \zeta^0,\ldots,0;0,\partial_\mu \zeta_1,\partial_\mu \zeta_2,\ldots,\partial_\mu \zeta_n).
\end{equation*}
In \cite{Dempster:2015} it was shown that in the presence of FI
gauging the analogous condition
\begin{equation}
\label{PI3}
 (g^a) \big|_{\text{PI}} = (g^0,\ldots,0;0,g_1,g_2,\ldots,g_n)
\end{equation}
on the gauging parameters extends the factorization to the terms
introduced by the scalar potential. While we will not
directly use this here, we remark that the PI condition reflects the
existence of a distinguished totally goedesic $(2n_V +2)$-dimensional
submanifold of the $(4n_V + 4)$-dimensional scalar manifold of the
three-dimensional effective theory obtained by dimensional reduction.

For notational convenience, we adjust the assignment of indices
$a,b,\dots$ to the scalar fields
$q^a, \hat{q}^a$ such that the non-constant scalars correspond to
indices $a=1, \ldots, n_V+1$. We can further simplify the equations of motion
through some simple manipulations. All terms in the
three-dimensional Lagrangian and field equations which do not involve
the scalar potential either involve the constant anti-symmetric
matrix
\[
 \Omega_{ab} = \left( \begin{array}{cc}
            0 & \mathbbm{1} \\
            - \mathbbm{1} & 0 \\
           \end{array} \right) ,
 \]
 or the Hesse potential $H$ and its derivatives. It is convenient to
 introduce the auxiliary Hesse potential
 \[
  \tilde{H} = -\frac{1}{2} \log (-2H) ,
 \]
 and its derivatives $\tilde{H}_a, \tilde{H}_{ab}$. Moreover we
 replace the scalar fields $q^a$ by their duals $q_a = \tilde{H}_a = -
 \tilde{H}_{ab} q^b$, where we used that $\tilde{H}_{ab}$ are
 homogeneous functions of degree $-2$. While in general we cannot
 lower indices on $\hat{q}^a$ in the same way, we can lower the indices
 after differentiation: $\partial_\mu \hat{q}_a := \tilde{H}_{ab}
 \partial_\mu \hat{q}^b$ \cite{Errington:2014bta}. As the fields $\hat{q}^a$
 are essentially
 the four-dimensional gauge potentials, which only enter into the
 field equations through their derivatives, this is sufficient for
 rewriting all field equations with indices $a$ in the lower
 position.

%
%
 \section{Euclidean Instanton Solutions}

 We are now in the position to formulate the problem that we will solve.
 Starting from the Lagrangian \eq{4dlag}, we impose that the four-dimensional
 metric (\ref{4d3d}) is static,
 the restrictions (\ref{prepotentials}) on the prepotential, the PI truncation
 (\ref{PI1}) -- (\ref{PI3}) and finally that
 the three-dimensional metric has planar symmetry
\begin{equation}
ds_3^2 = e^{4\psi} d\tau^2 + e^{2\psi} (dx^2 + dy^2).
\end{equation}
All fields, including the unknown function $\psi$ depend only on the
overall transverse coordinate $\tau$ in this brane-like ansatz.

The resulting equations of motion, follow from the general three-dimensional
equations for timelike dimensional reduction derived in \cite{Mohaupt:2011aa} 
by imposing the
above conditions:
\begin{equation}
\label{eq:eom1}
\nabla^2 \hat{q}_a = 0,
\end{equation}
\begin{equation}
\label{eq:eom2}
\nabla^2 q_a + \frac{1}{2}\partial_a \tilde{H}^{bc}(\partial_\mu q_b \partial^\mu q_c - \partial_\mu \hat{q}_b \partial^\mu \hat{q}_c) -\frac{1}{2} \partial_a \tilde{H}_{bc} g^b g^c + 4\tilde{H}_{ab}g^b(g^c q_c) = 0,
\end{equation}
\begin{equation}
\label{eq:eom3}
-\frac{1}{2} R_{(3)\mu\nu} - \tilde{H}^{ab}(\partial_\mu q_a \partial_\nu q_b - \partial_\mu \hat{q}_a \partial_\nu \hat{q}_b) + g_{\mu \nu} \big(-\tilde{H}_{ab}g^ag^b + 4(g^aq_a)^2 \big) =0.
\end{equation}
The first line are the equations for the scalars $\hat{q}^a$, which
correspond to the four-dimensional vector field equations. The second line
is the equations for the scalars $q^a$, which encode the four-dimensional
scalars $z^A$ and the Kaluza-Klein scalar $\phi$. The last line are the
three-dimensional Einstein's equations which determines the
three-dimensional warp factor $\psi$.

To solve Einstein's equations we use that the non-zero part of the Ricci tensor is found to be
\begin{equation*}
 R_{\tau \tau} = 2 \ddot{\psi} - 2\dot{\psi}^2, \quad R_{xx} = R_{yy} = e^{-2\psi} \ddot{\psi},
\end{equation*}
where we use a dot to denote differentiation by $\tau$. The equations \eq{eom3} then reduce to the following form for $\mu, \nu \neq \tau$
\begin{equation}
\label{eq:eom4}
-\tilde{H}_{ab} g^a g^b + 4(q_ag^a)^2 - \tfrac{1}{2}e^{-4\psi}\ddot{\psi} = 0 ,
\end{equation}
and for $\mu, \nu = \tau$
\begin{equation}
\label{eq:eom5}
\tilde{H}^{ab} \big(\dot{q}_a \dot{q}_b - \dot{\hat{q}}_a \dot{\hat{q}}_b \big) = \dot{\psi}^2 - \tfrac{1}{2} \ddot{\psi},
\end{equation}
where we have substituted in \eq{eom4} to
reduce this condition. We see that \eq{eom5} is the Hamiltonian
constraint \cite{Errington:2014bta,Dempster:2015}.

%
%
\subsection{Two-Charge Solution}

We now turn our attention to generalising the single charge Nernst solution by starting with a solution
carrying charge under two gauge fields.
The prepotentials we admit take the factorised form
\begin{equation*}
    F_2(X) = \frac{X^1 f_2(X^2,\ldots,X^n)}{X^0} ,
\end{equation*}
where $f_2$ is homogenous of degree two. The corresponding
Hesse potential \cite{Mohaupt:2011aa} is
\begin{equation}
\label{eq:Hesse}
  H = -\frac{1}{4} (-q_0q_1f_2(q_2,\ldots,q_n))^{-\frac{1}{2}} \; .
\end{equation}
The generality of the function $f_2(q)$ prevents us from obtaining each element of the metric $\tilde{H}^{ab}$, however as the fields $q_0$ and $q_1$ decouple we are able to calculate the components we actually need explicitly:
\begin{equation*}
    \tilde{H}^{00} = \frac{1}{4q_0^2} \;, \qquad \tilde{H}^{11} = \frac{1}{4q_1^2} .
\end{equation*}
We start by solving for $\hat{q}_a$. Since all fields are assumed to only depend on $\tau$, equation
 \eq{eom1} reduces to
\begin{equation}
\ddot{\hat{q}}_a = 0.
\end{equation}
Integrating up we obtain
\begin{equation}
\dot{\hat{q}}_a = K_a \; .
\end{equation}
The non-vanishing constants $K_a$ are proportional to the electric charge $Q_0$
and magnetic charge $P^1$ of the two gauge fields in this
solution\footnote{The minus sign in front of $Q_0$ reflects that $K_a$
transforms as a co-vector, and not as a vector, under symplectic transformations.}
\begin{equation}
\label{eq:eom6}
\dot{\hat{q}}_0 = -Q_0, \quad \quad \dot{\hat{q}}_1 = P^1 \;.
\end{equation}

We now turn our attention to \eq{eom2} to solve for the scalar fields $q_a$.
Due to the conditions we have imposed,
the equations for $q_0$ and $q_1$ decouple from the rest and \eq{eom2} reduces to:
\begin{equation}
\ddot{q}_0 - \frac{\dot{q}^2_0 - Q_0^2}{q_0} = 0, \quad \quad \ddot{q}_1 - \frac{\dot{q}^2_1 - (P^1)^2}{q_1} = 0,
\end{equation}
where we used \eq{eom6}. These equations can be integrated up to obtain:
\begin{equation}
\begin{aligned}
\label{eq:sol2}
q_0(\tau) = &\mp \frac{Q_0}{B_0} \sinh \bigg(B_0 \tau + B_0 \frac{h_0}{Q_0} \bigg) ,\\
q_1(\tau) = &\pm \frac{P^1}{B_1} \sinh \bigg(B_1 \tau + B_1 \frac{h^1}{P^1} \bigg),
\end{aligned}
\end{equation}
where we have introduced the integration constants $h_0, B_0, h^1, B_1$, where without loss of generality we set $B_0, B_1 \geq 0$. To avoid curvature singularities associated with zeros of the fields $q_0, q_1$ we require that sign($h_0$) = sign($Q_0$) and sign($h^1$) = sign($P^1$). This ensures that there are no zeros for the domain $0 \leq \tau < \infty$.
The remaining equations of motion corresponding to $q_A$ for $A=2,\ldots,n$ are
\begin{equation}
\label{eq:sc1}
e^{-4\psi} \ddot{q}_A + \tfrac{1}{2}e^{-4\psi} \partial_A \tilde{H}^{BC}\dot{q}_B \dot{q}_C -\tfrac{1}{2} \partial_A \tilde{H}_{BC} g_B g_C + 4 \tilde{H}_{AB}g_B\left(g_C q_C\right)^2 = 0.
\end{equation}
The corresponding components of the Einstein equations \eq{eom4} are
\begin{equation}
\label{eq:ee1}
- \tilde{H}_{AB} g_A g_B +4 \left( g_Aq_A\right)^2 - \frac{1}{2} e^{-4\psi} \ddot{\psi} = 0.
\end{equation}
Contracting \eq{sc1} with $q_A$ and substituting in \eq{ee1} we obtain
\begin{equation*}
   q^A\ddot{q}_A + \tilde{H}^{AB} \dot{q}_A \dot{q}_B =\frac{1}{2}\ddot{\psi} = \diff{}{\tau} \left( q^A \dot{q}_A \right).
\end{equation*}
We have used here that the LHS can be written as a total derivative which upon integration yields
\begin{equation}
\label{eq:sc3}
q^A \dot{q}_A = \frac{1}{2} \dot{\psi} - \frac{1}{4}a_0,
\end{equation}
where $a_0$ is an integration constant and the pre-factor has been chosen for later convenience. This equation can be further rearranged using properties of the Hesse potential \cite{Mohaupt:2011aa} and integrated a second time to obtain an expression for the function $\psi$
\begin{equation*}
      -2\psi + a_0 \tau + b_0 = -2 \log(-4H(-q_0q_1)^{\half} ) \;.
\end{equation*}
Substituting in the explicit form of the Hesse potential for this solution allows the realisation of the condition
\begin{equation}
\label{eq:con1}
\log (f_2(q_2,\ldots,q_n)) = -2\psi + a_0 \tau + b_0.
\end{equation}
Returning to the Hamiltonian constraint \eq{eom5}, substituting in the result from \eq{sol2}, we find
\begin{equation}
\label{eq:con2}
 \tilde{H}^{AB} \dot{q}_A \dot{q}_B = \dot{\psi}^2 - \frac{1}{2} \ddot{\psi} - \frac{B_0^2 + B_1^2}{4} \; .
\end{equation}
To obtain an explicit expression for the remaining scalars $q_A$ we need to make use of our final assumption that the scalars in the `block', $q_2, \ldots, q_n$ are proportional to each other:
\begin{equation*}
    q_A(\tau) = \lambda_A Q(\tau) .
\end{equation*}
This, together with the constraint \eq{eom5} allows
 us to solve the field equations for general homogeneous $f_2$ by manipulating \eq{sc1}
 into the form
\begin{equation}
\label{eq:dife}
\ddot{\psi} - \dot{\psi}a_0 - \dot{\psi}^2 + \frac{a_0}{4} + \frac{B_0^2 + B_1^2}{2} = 0.
\end{equation}
This is solved using the substitution
\begin{equation}
y \equiv \exp \bigg(-\psi - \frac{a_0 \tau}{2} \bigg),
\end{equation}
and thus the general solution is of the form
\begin{equation}
\label{eq:sol1}
y = \frac{\alpha}{\omega} \sinh (\omega \tau + \omega \beta) = \exp \bigg(-\psi - \frac{a_0 \tau}{2} \bigg) ,
\end{equation}
with two new integration constants $\alpha$ and $\beta$, and $\omega^2 := \frac{1}{2} (a_0 + B_0^2 + B_1^2)$ where without loss of generality we set $\omega \geq 0$. This solution of $y$ can then be back-substituted to obtain the form of the scalars
\begin{equation}
q_A(\tau) = \lambda_A e^{a_0} \sinh (\omega \tau + \omega \beta),
\end{equation}
and the warp factor
\begin{equation}
\label{eq:epsi}
e^{-4\psi} = e^{2a_0\tau} \bigg(\frac{\alpha}{\omega} \bigg)^4 \sinh^4 (\omega \tau + \omega \beta).
\end{equation}
The constants $\lambda_A$ are determined by the gauging parameters through
requiring consistency of \eq{ee1}
\begin{equation}
  \lambda_A = \pm \frac{\alpha^2}{2\omega m g_A} \; ,
\end{equation}
where $m=n-1$ is the number of scalar fields belonging to the block $q_2, \ldots, q_n$. 
From the homogeneity of $f_2$ we obtain an expression for the constant $b_0$ as a function of the
gauging parameters
\begin{equation}
e^{b_0} = \frac{\alpha^2}{4m^2} f_2 \left(\frac{1}{g_2}, \ldots, \frac{1}{g_n} \right) \;.
\end{equation}

In summary, we have obtained the following instanton solution of the reduced, three-dimensional
Euclidean field equations, which depend on a single coordinate $\tau$:
\begin{equation*}
  \begin{aligned}
  q_0(\tau) &= \mp \frac{Q_0}{B_0} \sinh \bigg(B_0 \tau + B_0 \frac{h_0}{Q_0} \bigg), \\
q_1(\tau) &= \pm \frac{P^1}{B_1} \sinh \bigg(B_1 \tau + B_1 \frac{h^1}{P^1} \bigg), \\
  q_A(\tau) &= \pm \frac{\alpha^2}{2\omega m g_A} e^{a_0\tau} \sinh (\omega \tau + \omega \beta),
  \end{aligned}
\end{equation*}
\begin{equation*}
  \begin{aligned}
  e^{-4\psi} = e^{2a_0\tau} \bigg(\frac{\alpha}{\omega} \bigg)^4 \sinh^4 (\omega \tau + \omega \beta).
  \end{aligned}
\end{equation*}
We will later find that imposing regularity on the lifted four-dimensional solutions reduces
the number of independent integration constants to five.

%
%
\subsection{Three-Charge Solution}
We can generate further solutions by following the previous method for a system supported by an additional gauge field. To do this we further restrict the form of the prepotential to
\begin{equation}
F(Y) = \frac{X^1 X^2 f_3(X)}{X^0},
\end{equation}
where $f_3(X)$ is a homogeneous polynomial of degree one. The corresponding
Hesse potential is
\begin{equation*}
H = -\frac{1}{4} (-q_0 q_1 q_2 f_3(q_3,\ldots,q_n))^{-\half}\;.
\end{equation*}
The charges are arranged as
\begin{equation*}
K_a = (-Q_0,0, \ldots, 0; 0,P^1,P^2,0\ldots,0),
\end{equation*}
and we additionally switch off the gauging parameter for the $U(1)$ supporting the charge $P^2$,
\begin{equation*}
  (g^a)\big|_{\text{PI}} = (0,\ldots,0;0,0,0,g_3,\ldots,g_n).
\end{equation*}
The solution of the equations of motion (\ref{eq:eom1}-\ref{eq:eom3}) follow the same steps as in the two-charge case with only minor changes. We begin by summarising the steps for this model which are identical to the two-charge solution.

There are now three non-trivial `hatted' scalar fields
\begin{equation}
\dot{\hat{q}}_a = K_a.
\end{equation}
The scalar fields $q_0,q_1,q_2$ have decoupled equations of motion which can be
integrated as before:
\begin{equation}
\begin{aligned}
q_0(\tau) &= \mp \frac{Q_0}{B_0} \sinh \bigg(B_0 \tau + B_0 \frac{h_0}{Q_0} \bigg) ,\\
q_1(\tau) &= \pm \frac{P^1}{B_1} \sinh \bigg(B_1 \tau + B_1 \frac{h^1}{P^1} \bigg), \\
q_2(\tau) &= \pm \frac{P^2}{B_2} \sinh \bigg(B_2 \tau + B_2 \frac{h^2}{P^2} \bigg).
\end{aligned}
\end{equation}
To keep $f_3$ general we assume that the scalar fields $q_A$ for $A=3,\ldots,n$
are proportional to each other, $q_A(\tau) = \lambda_A Q(\tau)$, for a set of constants $\lambda_A$.

The central difference between this and the two-charge case is caused by $f_3$ being of degree one rather than two.
This changes the balance between the scalar and Hamiltonian conditions, and thus the form of the differential equation for the function $\psi$
\begin{equation*}
\ddot{\psi} -2\psi a_0 + \frac{a_0^2 + B_0^2 + B_1^2 + B_2^2}{2} = 0.
\end{equation*}
This difference results in the disappearance of the $\dot{\psi}^2$ term. This missing term allows the differential equation to be solved using standard methods
\begin{equation}
\psi = \alpha + \beta e^{2a_0 \tau} + X \tau ,
\end{equation}
where for simplicity we have collected integration constants together
\begin{equation*}
  X := \bigg(\frac{a_0^2 + B_0^2 + B_1^2 + B_2^2}{4a_0}\bigg) .
\end{equation*}
Using this form of the warp factor $\psi$, the solution for the remaining scalars is found to be
\begin{equation}
q_a(\tau) = \lambda_A e^{b_0-2\alpha} e^{-2\beta e^{2a_0 \tau}} e^{(a_0 -2X)\tau} \;.
\end{equation}
At this point we have fixed the functional form of all non-trivial fields:
\begin{equation*}
  \begin{aligned}
   q_a(\tau) &= \pm \frac{K_a}{B_a} \sinh \left(B_a \tau + B_a \frac{h^a}{K_a} \right) ,\\
   q_A(\tau) &= \lambda_A e^{b_0-2\alpha} e^{-2\beta e^{2a_0 \tau}} e^{(a_0 -2X)\tau} \;,\\
   e^{-4\psi} &= \exp \big(-4\big(\alpha + \beta e^{2a_0 \tau} + X\tau \big)\big) ,\\
  \end{aligned}
\end{equation*}
for $a = 0,1,2$ and $A=3,\ldots,n$.

We can reinsert these expressions into the scalar equation of motion as we did for the two-charge case, but unlike before where we were able to set $\lambda_A$ in terms of the gauge parameters, we instead find that either all $B_i$ must be zero or $\beta=0$. As we will see later, the limit of $B_i \rightarrow 0$ is associated to the extremal limit of the $4D$ solution. To maintain a non-extremal solution we choose $\beta =0$. From the previous condition $C = \lambda_A g_A$, the constant $\Lambda_A$ is written explicitly as inversely proportional to the gauging parameters $g_A$
\begin{equation*}
    \lambda_A = \frac{C}{g_A}.
\end{equation*}
Finally using the same method as in the two-charge system we find that
\begin{equation}
    e^{b_0} = C f_3\left(\frac{1}{g_A}\right),
\end{equation}
where $\alpha$ has been set to zero through a simple redefinition of the coordinate $\tau$.

In summary, the three-dimensional solution is fully described by:
\begin{equation*}
  \begin{aligned}
   q_a(\tau) &= \pm \frac{K_a}{B_a} \sinh \left(B_a \tau + B_a \frac{h^a}{K_a} \right), \\
   q_A(\tau) &= \frac{C^2}{g_A}f_3\left(\frac{1}{g_A}\right) e^{-2\beta e^{2a_0 \tau}} e^{(a_0 -2X)\tau} \; , \\
   e^{-4\psi} &= e^{-4X\tau} .
  \end{aligned}
\end{equation*}
for $a = 0,1,2$ and $A=3,\ldots,n$.

%
%
\subsection{Four-Charge Solution}
We now generate a final solution employing the same method by studying a brane supported by four gauge fields. This requires that the prepotential takes the form
\begin{equation}
        F(X) = \frac{X^1 X^2 X^3}{X^0},
\end{equation}
which is the well known STU model. The charges are chosen such that
\begin{equation*}
  K_a = (-Q_0, 0, 0, 0; 0, P^1, P^2, P^3).
\end{equation*}
We necessarily turn off all gauging parameters, thus obtaining the ungauged STU model. From this, we will obtain a solution with planar symmetry.
The STU prepotential gives us a simple Hesse potential
\begin{equation}
H = -\frac{1}{4} (-q_0 q_1 q_2 q_3)^{-\half} \;,
\end{equation}
and we can now completely solve for the metric which is diagonal, with elements given by
\begin{equation*}
  \tilde{H}^{aa} = \frac{1}{4q_a^2} \; .
\end{equation*}
The scalar equations of motion are now simple to solve. For the hatted scalars we obtain the now familiar solution $\dot{\hat{q}}_a = K_a$. The scalar fields $q_a$ completely decouple from each other, and we obtain the form of the scalars
\begin{equation}
q_a = \pm \frac{K_a}{B_a} \sinh \bigg(B_a \tau + B_a \frac{h_a}{|K_a|} \bigg),
\end{equation}
where
\begin{equation*}
  B_a = (B_0,B_1,B_2,B_3), \quad h_a = (h_0, h^1, h^2, h^3).
\end{equation*}
Looking at the Hamiltonian constraint
\begin{equation}
\frac{1}{4q_a^2} \left(\dot{q}_a^2 - \dot{\hat{q}}_a \right) = \dot{\psi}^2 - \half \ddot{\psi} = \frac{B_0^2 + B_1^2 + B_2^2 + B_3^2}{4} \; ,
\end{equation}
we find
\begin{equation*}
-\half e^{-4\psi} \ddot{\psi} = 0 \quad \Rightarrow \quad \ddot{\psi} = 0.
\end{equation*}
Returning to the Hamiltonian constraint we find
\begin{equation*}
\dot{\psi} = \pm \frac{\sqrt{\sum_iB_i^2}}{2} \quad \Rightarrow \quad \psi = \pm \frac{\sqrt{\sum_iB_i^2}}{2}\tau + a_0.
\end{equation*}
This allows us to calculate
\begin{equation}
e^{-4\psi} = e^{-4a_0} e^{\pm 2\sqrt{\sum_iB_i^2} \tau} = e^{-4a_0} e^{\pm 2\sqrt{\sum_iB_i^2} \tau}\; .
\end{equation}
In summary we have found the following planar solution to the
time-reduced ungauged STU model:
\begin{equation}
    \begin{aligned}
      \dot{\hat{q}}_a &= K_a, \\
      q_a &= \pm \frac{K_a}{B_a} \sinh \bigg(B_a \tau + B_a \frac{h_a}{K_a} \bigg), \\
         e^{-4\psi} &= e^{-4a_0} e^{\pm 2\sqrt{\sum_i B_i^2} \tau} \; .\label{STU1}
    \end{aligned}
\end{equation}

%
%

\section{Four-Dimensional Planar Solutions \label{Sect:4dSolutions}}

The three-dimensional Euclidean solutions found in the previous section can now be
lifted to four dimensions using the dimensional reduction formulae found originally in \cite{Mohaupt:2011aa,Errington:2014bta,Dempster:2015}.
In particular the four-dimensional metric is
\begin{equation}
\label{eq:ansatz}
ds_4^2 = -e^\phi dt^2 + e^{-\phi + 4\psi}d\tau^2 + e^{-\phi + 2\psi} (dx^2 + dy^2).
\end{equation}
The four-dimensional physical scalars are determined by the three-dimensional
scalars through \cite{Errington:2014bta}
\begin{equation*}
    z^A = -i \bigg(-\frac{q_0q_A^2}{f(q)} \bigg)^{\frac{1}{2}}.
\end{equation*}
Finally the four-dimensional gauge fields are calculated using $\hat{q}_a$ through the relation
\begin{equation}
\label{eq:gauge1}
    \hat{q}^a = \half \pvec{\zeta^I}{\tilde{\zeta}_I},
\end{equation}
where as displayed in (\ref{sec:dimred}), $\zeta^I$ are the components of the gauge fields along the reduction dimension and $\tilde{\zeta}_I$ are the Hodge duals of the three-dimensional vectors. Their relation to the four-dimensional gauge fields can be calculated from \cite{Errington:2014bta}
\begin{equation}
\label{eq:gauge2}
    \partial_\mu \zeta^I := F_{\mu \eta}^I, \qquad \partial_\mu \tilde{\zeta}_I := G_{I | \mu \eta},
\end{equation}
where
\begin{equation}
\label{eq:gauge3}
    G_{I|\mu \nu} := \mathcal{R}_{IJ} F^J_{\mu \nu} - \I_{IJ} \tilde{F}^J_{\mu \nu},
\end{equation}
is the dual field strength.

%
%
\subsection{Two-Charge Solution}
We first probe for the existence of a Killing horizon by looking for zeros of the
norm of the Killing vector ($k^t = \partial_t$).
 As $k^\mu$ has only one non-zero component, its norm is given as $k^2=g_{tt}$. 
 In the limit $\tau\rightarrow \infty$, it takes the form
\begin{equation}
e^{\phi}\big|_{\tau\rightarrow \infty} \sim \exp\bigg( -\frac{B_0\tau}{2} - \frac{B_1 \tau}{2} - a_0\tau - \omega\tau \bigg).
\end{equation}
We see this always vanishes in the limit when $a_0 \geq 0$. If this restriction is lifted, the horizons position will change depending on the relative magnitudes of $B_0$, $B_1$ and $a_0$, so for now we choose to keep this restriction in place.

The area of the horizon is given by
\begin{equation}
A = \int dx dy e^{-\phi+2\psi}\bigg|_{\tau \rightarrow \infty}.
\end{equation}
As our $x$ and $y$ coordinates are not compact, this diverges, reflecting the planar symmetry of our ansatz. To obtain finite quantities we could identify $x,y$ periodically, but we prefer to work
with densities instead and take ratios relative to the coordinate volume $\int dx dy$. The
area density of the horizon is
\begin{equation}
a = e^{-\phi+2\psi}\bigg|_{\tau \rightarrow \infty} \sim \exp\bigg(\frac{B_0\tau}{2} + \frac{B_1 \tau}{2} + a_0\tau + \omega\tau - a_0\tau -2\omega\tau \bigg).
\end{equation}
Imposing that the area density is finite and non-zero requires that
\begin{equation}
\omega = \frac{B_0 + B_1}{2} := B_a.
\end{equation}
Recalling: $2\omega^2 = a_0 + B_0^2 + B_1^2$, we can write down $a_0$ in terms of $B_{0}$ and $B_1$, $a_0 = B_0B_1 - \frac{1}{2} (B_0^2 + B_1^2)$. We can condense this expression by including $B_a$
\begin{equation*}
a_0 = 2(B_0B_1 - B_a^2),
\end{equation*}
and notice here that $a_0 =0$ for $B_0=B_1$. We also notice that in \eq{epsi}, the value for the constant $\beta$ can be changed by a shift in the value of $\tau$; we use this to shift $\tau$ such that $\beta$ vanishes.

To constrain $\alpha$ we look at the limit where $B_0,B_1 \rightarrow 0$.\footnote{We will see
later that this is the extremal limit, that is, the limit where the surface gravity of the Killing horizon goes to zero.} Then
\eq{epsi} reduces to
\begin{equation*}
e^{-4\psi} = \bigg(\frac{\alpha}{\omega} \bigg)^4 (\omega \tau)^4,
\end{equation*}
which implies that $\alpha\tau = e^{-\psi}$. This allows us to scale $\tau$ such that $\alpha = 1$.

The integration constants can be further constrained by imposing regularity of the physical scalar fields $z^1, z^A$. According to \cite{Errington:2014bta}
\begin{equation}
Y^1 = -\frac{i}{2}e^\phi q_1, \quad Y^A = -\frac{i}{2}e^\phi q_A, \quad Y^0 = -\frac{1}{4q_0} \; ,
\end{equation}
where $Y^I$ are rescaled scalar fields defined in (\ref{YX}). This yields:
\begin{equation}
z^1 = -i \bigg(\frac{-q_0q_1^2}{q_1 f_2(q)} \bigg)^{\frac{1}{2}}, \quad z^A = -i \bigg(\frac{-q_0q_A^2}{q_1 f_2(q)} \bigg)^{\frac{1}{2}}\;.
\end{equation}
Next we impose that the physical fields take finite values on the horizon. Since
\begin{equation}
z^1 \bigg|_{\tau \rightarrow \infty} \sim e^{-a_0\tau}, \quad z^A \bigg|_{\tau \rightarrow \infty} \sim (e^{B_0\tau}e^{-B_1\tau})^{\frac{1}{2}} \;,
\end{equation}
this can be satisfied by setting $B_0 = B_1 = B \quad \Rightarrow \quad B_a = B$ $\Rightarrow$ $a_0 = 0$.

Having imposed regularity we end up with the following solution:
\begin{equation}
\label{eq:2csummary}
  \begin{aligned}
  q_0(\tau) &= \mp \frac{Q_0}{B} \sinh \bigg(B \tau + B \frac{h_0}{Q_0} \bigg), \\
q_1(\tau) &= \pm \frac{P^1}{B} \sinh \bigg(B \tau + B \frac{h^1}{P^1} \bigg), \\
  q_A(\tau) &= \pm \frac{1}{2B m g_A} \sinh (B\tau), \\
  e^\phi &= \frac{1}{2} (-q_0q_1f_2(q_2,\ldots,q_n))^{-\frac{1}{2}} \; ,\\
  e^{-4\psi} &= \bigg(\frac{1}{B} \bigg)^4 \sinh^4(B\tau),
  \end{aligned}
\end{equation}
\noindent where only 5 out of the original 9 integration constants remain: $h_0,h^1,Q_0,P^1,B$.

%
%
\subsection{Three-Charge Solution}

We proceed as in the two-charge case. Since the computations are similar, we don't need
to give many details. The condition for the existence of a Killing horizon at $\tau\rightarrow
\infty$ is:
\begin{equation*}
 k^2 = e^\phi = \exp \left(-\frac{B_0 \tau}{2} -\frac{B_1 \tau}{2} -\frac{B_2 \tau}{2} + \beta e^{2a_0\tau} - \frac{(a_0 -2X)}{2} \tau \right) \xrightarrow[\tau \rightarrow \infty]{}0 \;,
\end{equation*}
implying
\begin{equation*}
B_0, B_1, B_2, a_0-2X > 0 \quad \text{ and } \quad
\beta \leq 0 \text{ or } a_0<0.
\end{equation*}
Regularity of the physical scalar fields demands that the physical scalars
\begin{equation*}
  z^a = -i\bigg( \frac{- q_0 q_a^2}{q_1q_2 f_3 (q_A)} \bigg)^{\half}, \quad \text{ where } a=1,\ldots,n,
\end{equation*}
take finite values on the horizon. For $\tau \rightarrow \infty$ we find:
\begin{equation*}
  \begin{aligned}
  z^1 \big|_{\tau \rightarrow \infty} &\sim \exp \left[ \left(B_0 + B_1 - B_2 +\beta e^{2a_0\tau} -(a_0 -2X) \right)\tau \right] ,\\
   z^2 \big|_{\tau \rightarrow \infty} &\sim \exp \left[ \left(B_0 + B_2 - B_1 +\beta e^{2a_0\tau} -(a_0 -2X) \right)\tau \right] ,\\
  z^A \big|_{\tau \rightarrow \infty} &\sim \exp \left[ \left(B_0 -\beta e^{2a_0\tau} +(a_0 -2X) -B_1 - B_2 \right)\tau \right] .\\
  \end{aligned}
\end{equation*}
The exponential term is absent or decreasing if $\beta\leq0$ or $a_0<0$. Imposing that
the remaining terms cancel implies the following three simultaneous equations:
\begin{equation*}
  \begin{aligned}
  B_0 + B_1 -B_2 -(a_0 -2X) &= 0 ,\\
  B_0 + B_2 -B_1 -(a_0 -2X) &= 0 ,\\
  B_0 + (a_0 -2X) - B_1 -B_2&= 0 .\\
  \end{aligned}
\end{equation*}
This implies $B_0=B_1=B_2=a_0-2X$. We set $B:= B_1=B_2=B_3$ and
impose that the horizon area density is finite for $\tau \rightarrow \infty$, this is given by
\begin{equation*}
  e^{-\phi + 2\psi} \bigg|_{\tau \rightarrow \infty} \sim \exp\left[\left( \frac{a_0^2 + 3B^2}{2a_0} + \frac{3B}{2} + \frac{a_0}{2} - \frac{a_0^2 + 3B^2}{4a_0} \right) \tau\right] ,
\end{equation*}
giving us one last constraint
\begin{equation*}
\frac{a_0^2 + 3B^2}{2a_0} + \frac{3B}{2} + \frac{a_0}{2} - \frac{a_0^2 + 3B^2}{4a_0} = 0,
\end{equation*}
and we find that
\begin{equation}
B= B_0 = B_1 = B_2 =-a_0=-X .
\end{equation}
Hence our solution takes the form
\begin{equation}
\begin{aligned}
q_0(\tau) &= \mp \frac{Q_0}{B} \sinh \bigg(B \tau + B \frac{h_0}{Q_0} \bigg) ,\\
q_1(\tau) &= \pm \frac{P^1}{B} \sinh \bigg(B \tau + B \frac{h^1}{P^1} \bigg), \\
q_2(\tau) &= \pm \frac{P^2}{B} \sinh \bigg(B \tau + B \frac{h^2}{P^2} \bigg), \\
q_A(\tau) &= \pm \frac{C}{g_A} \exp \big(B\tau \big) , \\
\end{aligned}
\end{equation}
\begin{equation}
\begin{aligned}
e^\phi &= \frac{1}{2} (-q_0q_1q_2 f_3(q_3,\ldots,q_n))^{-\frac{1}{2}} \; , \\
e^{-4\psi} &= e^{4B \tau} ,
\end{aligned}
\end{equation}
we see that we have reduced the total number of integration constants to just 7.

%
%
\subsection{Four-Charge Solution}
The four-dimensional physical scalar fields take the form
\begin{equation}
\label{eq:gensca}
z^A = -i \bigg( -\frac{q_0 q_A^2}{q_1 q_2 q_3} \bigg)^{\half} \; .
\end{equation}
Imposing that these fields are finite in the limit $\tau \rightarrow \infty$ implies:
\begin{equation*}
  \begin{aligned}
      B_0 + B_1 - B_2 - B_3 = 0 ,\\
        B_0 + B_2 - B_1 - B_3 = 0 ,\\
         B_0 + B_3 - B_2 - B_1 = 0 ,
  \end{aligned}
\end{equation*}
which is satisfied when the integration constants obey $B_0 = B_1 = B_2 = B_3 = B$. The integration constant $a_0$ can be removed by a suitable shift in the $\tau$ coordinate which simplifies the warp factor to
\begin{equation*}
    e^{-4\psi} = e^{\pm 2\sqrt{\sum_i B_i^2} \tau} \; .
\end{equation*}
Then, regularity of the horizon area density further dictates that
\begin{equation}
e^{-\phi + 2\psi} \bigg|_{\tau \rightarrow \infty} \sim \exp \bigg(\frac{4B\tau}{2} \mp 2B\tau \bigg)
\end{equation}
is finite, where we used $\sqrt{\sum_i B_i^2} = 2B>0$. To cancel the terms inside
the exponential we should therefore pick the $(+)$-sign in the solution (\ref{STU1})
for $\psi$.
The explicit solution takes the following form:
\begin{equation}
\begin{aligned}
\dot{\hat{q}}_a &= K_a, \\
q_a &= \pm \frac{K_a}{B} \sinh \bigg(B \tau + B \frac{h_a}{|K_a|} \bigg), \\
e^{\phi} &= \frac{1}{2} (-q_0q_1q_2q_3)^{-\half} \;, \\
e^{-4\psi} &= e^{ 4B \tau} .
\end{aligned}
\end{equation}
Later, when oxidising the four-charge solution to ten and eleven dimensions we will need the explicit form of the gauge fields. As we assume all three-dimensional components depend only on the coordinate $\tau$ the non-zero components are found
from (\ref{eq:gauge1}-\ref{eq:gauge3})
\begin{equation}
\label{eq:4dgauge}
    (\dot{A}^0)_{\eta} = 2 \dot{\hat{q}}^0 = 2 \tilde{H}^{00} \dot{\hat{q}}_0 = -\frac{Q_0}{2 q_0^2(\tau)}, \qquad \qquad (\dot{\tilde{A}}_A)_{\eta} = \frac{P^A}{2 q_A^2(\tau)},
\end{equation}
where the dot references differentiation by the parameter $\tau$.

%
%
\section{Properties of Black Planar Solutions}

In this section, we investigate the properties of the four-dimensional
solutions obtained in the previous section. For the two-charge solution
we find that we are able to follow the methodology of \cite{Dempster:2015}
to produce a meaningful discussion of the solution's geometry. For the three- and four-charged
solutions we find that this is not the case. Their resulting four-dimensional spacetime is
geodesically incomplete for both ends of the domain of the transverse coordinate.
Through coordinate changes and analytic continuation, we show that these higher charged solutions
have time-dependent asymptotics and timelike singularities.

%
%
\subsection{Two-Charge Solution}

Following previous experience \cite{Dempster:2015} we introduce a new
transverse coordinate $\rho$ by setting
\begin{equation}
\label{eq:cchange}
e^{-2B\tau} = 1 - \frac{2B}{\rho} =: W(\rho).
\end{equation}
Applying the coordinate change to the scalars $q_a$ in \eq{2csummary} we obtain
\begin{equation}
q_0 = \mp \frac{\Ham_0}{W^{1/2}}, \quad q_1 = \pm \frac{\Ham_1}{W^{1/2}}, \quad q_A = \frac{1}{2mg_A} \frac{1}{\rho W^{1/2}},
\end{equation}
where we have introduced the harmonic functions:
\begin{equation*}
\begin{aligned}
\Ham_0(\rho) &:= Q_0 \left[ \frac{1}{B} \sinh \left(\frac{B h_0}{Q_0}\right) + \frac{e^{-\frac{Bh_0}{Q_0}}}{\rho}\right], \\
\Ham_1(\rho) &:= P^1 \left[\frac{1}{B} \sinh \bigg(\frac{B h^1}{P^1}\bigg) + \frac{e^{-\frac{Bh^1}{P^1}}}{\rho}\right].
\end{aligned}
\end{equation*}
The physical scalars $z^a$ as functions of $\rho$ are given by
\begin{equation*}
z^1 = - 2im\rho \sqrt{\Ham_0 \Ham_1} f_2^{-1/2}\bigg(\frac{1}{g} \bigg) \;, \quad z^A = i\sqrt{\frac{\Ham_0}{\Ham_1}} \frac{1}{g_A} f_2^{-1/2}\bigg(\frac{1}{g} \bigg).
\end{equation*}
Their construction ensures regularity at the horizon. Their asymptotic behaviour in
the limit $\rho \rightarrow \infty$ depends on $h_0, h_1$ and is
summarised in the table below.
\renewcommand{\arraystretch}{1.2}
\begin{table}[h!]
\centering
\begin{tabular}{|c|c|c|}
\hline
  $h_I$    & $z^1$    & $z^A$    \\ \hline \hline
$h_0, \; h_1\not=0$ & $\rho$    & Const.    \\ \hline
$h_0 = 0$ & $\rho^{1/2}$ & $\rho^{-1/2}$      \\ \hline
$h_1 = 0$ & $\rho^{1/2}$ & $\rho^{1/2}$   \\ \hline
$h_0=h_1=0$ & Const.    & Const.     \\ \hline
\end{tabular}
\end{table}

Next, we re-write the four-dimensional metric \eq{ansatz} in terms of our new transverse coordinate $\rho$. Applying \eq{cchange} to the functions
$\psi$ and $\phi$ gives
\begin{equation}
\label{eq:dof}
e^{-4\psi} = \frac{1}{B^4} \sinh^4 (B\tau), \qquad e^\phi = \frac{1}{2} \left[\frac{\Ham_0}{W^{1/2}}\frac{\Ham_1}{W^{1/2}}\frac{1}{4m^2\rho^2W}f_2\bigg(\frac{1}{g} \bigg) \right]^{1/2}.
\end{equation}
Combining this with \eq{ansatz} we obtain
\begin{equation}
\label{eq:2cmet}
ds^2 \; = \; - \frac{W \rho}{ \Ham} dt^2 + \frac{ \Ham}{W\rho} d\rho^2+ \Ham \rho (dx^2 + dy^2),
\end{equation}
where it has been convenient to define a new function
\begin{equation*}
   \Ham (\rho) := \frac{\sqrt{\Ham_0\Ham_1}}{m} f_2^{1/2} \bigg(\frac{1}{g_2}, \ldots , \frac{1}{g_n}\bigg).
\end{equation*}
The function $\Ham$ encodes the contributions of charges and gauging parameters, whereas
$W(\rho)$ is a blackening factor which controls the deviation from extremality.

To study the near-horizon behaviour of the metric \eq{2cmet} we introduce a new transverse coordinate
\begin{equation*}
  r^2 \equiv \rho - 2B.
\end{equation*}
Then for $r \ll 1$
\begin{equation*}
  d\rho^2 = 4r^2 dr^2, \quad W \simeq \frac{r^2}{2B}.
\end{equation*}
The near horizon value for the harmonic functions are calculated
\begin{equation*}
  \Ham_0(2B) = \frac{Q_0}{2B} \exp\left(\frac{Bh_0}{Q_0} \right),
\; \;
  \Ham_1(2B) = \frac{P^1}{2B} \exp\left(\frac{Bh^1}{P^1} \right) \; \Rightarrow \; \Ham \big(2B) = \frac{Z \E}{B} ,
\end{equation*}
where we have defined
\begin{equation*}
Z := \frac{\sqrt{Q_0P^1}}{2m} f_2^{-1/2} \left(\frac{1}{g_2} ,\ldots, \frac{1}{g_n} \right), \qquad \E := \exp\left(\frac{B}{2} \left( \frac{h_0}{Q_0} +\frac{h^1}{P^1} \right)\right).
\end{equation*}
Using this compact form, the near horizon metric can be found by substitution into \eq{2cmet}
\begin{equation}
\begin{aligned}
ds^2 \; = \; &-\frac{Br^2}{Z\E} dt^2
+ \frac{4Z\E}{B} dr^2 + 2Z\E(dx^2 + dy^2).
\end{aligned}
\end{equation}

By Wick rotating the time component $t\rightarrow t_E = -it$ we can find the Hawking temperature of
the Killing horizon\footnote{Computing the temperature via the surface gravity yields the same result.}
\begin{equation}
\label{eq:temp}
4\pi T_H = \frac{B}{Z \E}.
\end{equation}
This confirms the expected interpretation of $B$ as a non-extremality parameter and of
$B\rightarrow 0$ as an extremal limit. The solution can be interpreted as a black brane,
with entropy density
\begin{equation}
\label{eq:edens1}
s = 2Z \E.
\end{equation}
We can then use \eq{temp} and \eq{edens1} to obtain an expression for the integration constant
\begin{equation*}
  B = 2\pi T_H s,
\end{equation*}
which is interestingly the same as in \cite{Dempster:2015}. We note that unlike the Nernst solution, this two-charge solution will have finite entropy density in the extremal limit.

Finally, we consider the behaviour in the limit $\rho \rightarrow \infty$. While $W \rightarrow 1$, the
behaviour of the harmonic functions $\Ham_0, \Ham_1$ depends on whether the integration constants $h_0, \; h^1$ are finite or zero:
\begin{equation}
 \lim_{\rho \rightarrow \infty} \Ham_a \sim
  \begin{cases}
  \text{Constant} &\text{ for } h_a \neq 0, \\
  \rho^{-1} &\text{ for } h_a = 0.
  \end{cases}
\end{equation}
where $a$ runs over $0,1$. This in turn gives the asymptotic form of $ \Ham$. As the asymptotic behaviour of $\Ham$ is only sensitive to the number of finite integration constants, rather than the particular constant itself, there are three cases, namely both constants finite, one finite, or both zero
\begin{equation*}
 \Ham \big|^2_{\rho \rightarrow \infty} = C_2, \quad \Ham \big|^1_{\rho \rightarrow \infty} = \frac{C_1}{\rho^{1/2}}, \quad \Ham \big|^0_{\rho \rightarrow \infty} = \frac{C_0}{\rho}\;,
\end{equation*}
where the superscript index labels the number of non-zero elements. The corresponding line
elements are:
\begin{equation}
\begin{aligned}
h_0, h^1 &&\neq 0 \quad \quad \quad ds^2 &= -\frac{\rho}{C_2} dt^2 + C_2\frac{d\rho^2}{\rho} + C_2\rho (dx^2 + dy^2), \\
h_0 \text { or } h^1 &&\neq 0 \quad \quad \quad ds^2 &= - \frac{\rho^{3/2}}{C_1} dt^2 + C_1  \frac{d\rho^2}{\rho^{3/2}} + C_1 \rho^{1/2} (dx^2 + dy^2)\;, \\
h_0, h^1 &&= 0 \quad \quad \quad ds^2 &= -\frac{\rho^2}{C_0} dt^2 + C_0 \frac{d\rho^2}{\rho^2} + C_0 (dx^2 + dy^2).
\end{aligned}
\end{equation}

These geometries can be brought to standard forms using further coordinate transformations.
For $h_0,h^1 \neq 0$, we define $R$ by
 \begin{equation*}
  \pm R = \log(\rho),
\end{equation*}
and the line element reduces to
\begin{equation*}
  ds^2 \; = \; e^{\pm R} \big(-dt^2 + dR^2 + dx^2 + dy^2 \big),
\end{equation*}
showing that the metric is conformally flat.

When both $h_0$ and $h^1$ are zero we use the coordinate transformation $\rho = R^{-1}$ and obtain
\begin{equation*}
 ds^2 \; = \; \frac{1}{R^2} \bigg(-dt^2 + dR^2 \bigg) + dx^2 + dy^2,
\end{equation*}
which decomposes as AdS$_2$ $\times$ $\Real^2$.

Finally, when looking at the case when only one of the integration constants are finite we use the transformation $\rho = R^{-2}$ to obtain
\begin{equation}
ds^2 \; = \; \frac{\lambda}{R^3} \bigg(- \frac{dt^2}{\lambda^2} + 4 dR^2 \bigg) + \frac{\lambda}{R} \bigg(dx^2 + dy^2 \bigg),
\end{equation}
for some constant $\lambda$. The simplification achieved by the coordinate transformation is that the
terms proportional to $dt^2$ and $dR^2$ have the same dependence on $R$.
%
%
\subsection{Three-Charge Solution}

As mentioned at the beginning of the section, this solution is fundamentally different
from \cite{Dempster:2015} and the two-charge solution. Previously, null geodesics were able to
be extended to an infinite affine parameter in the limit of $\rho \rightarrow \infty$;
this justified the definition of this as the asymptotic limit of the solution.\footnote{
Commonly, an asymptotic limit is defined by the maximally symmetric geometry associated
with the vacuum solution. For a four-dimensional solution with planar symmetry, this type of fall-off is not
expected and so our best definition for an asymptotic region comes from the behaviour of the null geodesics of the solution.}

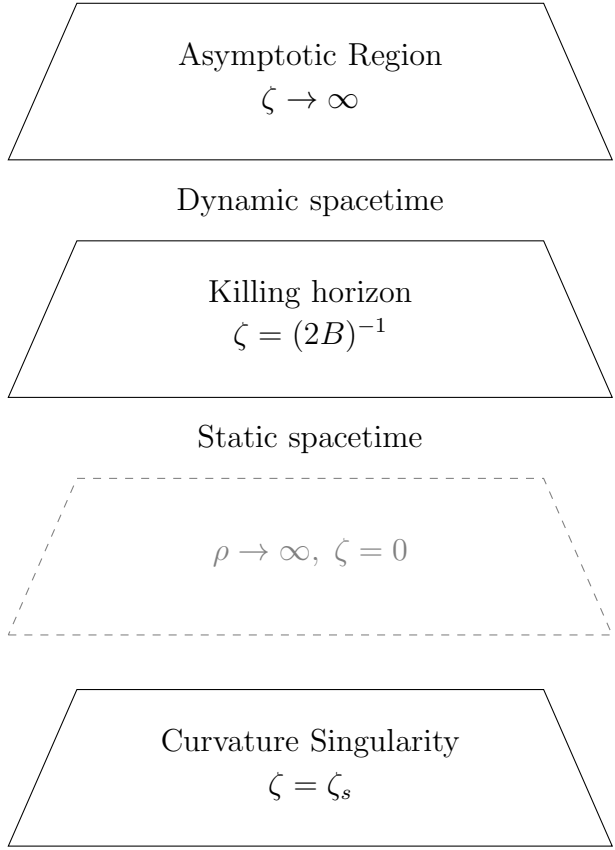
\begin{figure}[h!]
\centering
\begin{tikzpicture}[scale=0.7]
\node (I)  at ( 0,-8.5)  {Curvature Singularity};
\node (O)  at ( 0,-4.5)  {};
\node (II)  at (0,0)  {Killing horizon};
\node (III) at (0, 4.5) {Asymptotic Region};
\node (IV) at (0,1.75) {Dynamic spacetime};
\node (V) at (0,-2.75) {Static spacetime};

\path 
 (I) +(19:-6) coordinate (sbleft)
    +(-19:6) coordinate (sbright)
    +(13:4.5)  coordinate (stright)
    +(-13:-4.5) coordinate (stleft)
    ;
\draw (sbleft) -- node[midway, above, inner sep=6mm] {$\zeta=\zeta_s$} (sbright) -- (stright) -- (stleft) -- (sbleft) -- cycle;

\path 
 (O) +(19:-6) coordinate (sbleft)
    +(-19:6) coordinate (sbright)
    +(13:4.5)  coordinate (stright)
    +(-13:-4.5) coordinate (stleft)
    ;
\draw[gray,dashed] (sbleft) -- node[midway, above, inner sep=9mm] {$\rho \rightarrow\infty, \; \zeta=0$} (sbright) -- (stright) -- (stleft) -- (sbleft) -- cycle;

\path 
 (II) +(19:-6) coordinate (ebleft)
    +(-19:6) coordinate (ebright)
    +(13:4.5)  coordinate (etright)
    +(-13:-4.5) coordinate (etleft)
    ;

\draw (ebleft) -- node[midway, above, inner sep=6mm] {$\zeta= (2B)^{-1}$} (ebright) -- (etright) -- (etleft) -- (ebleft) -- cycle;

\path 
 (III) +(19:-6) coordinate (ableft)
    +(-19:6) coordinate (abright)
    +(13:4.5)  coordinate (atright)
    +(-13:-4.5) coordinate (atleft)
    ;

\draw (ableft) -- node[midway, above, inner sep=6mm] {$\zeta \rightarrow \infty$} (abright) -- (atright) -- (atleft) -- (ableft) -- cycle;

\end{tikzpicture}
\caption{Diagram of the brane solution. When starting from the 3D solution, a static patch of the spacetime is found, parametrised by $\tau$ for $\zeta \in [0,(2B)^{-1}]$. We can extend this spacetime to a singularity where the Kretschmann invariant becomes infinite. Analytically continuing our parameter through the horizon to $\zeta > (2B)^{-1}$ we obtain a time-dependent geometry.}
\label{fig:branesum}
\end{figure}

As we will show below, for the three- and four-charge solutions, $\rho=\infty$ is reached
by transverse null geodesics at finite affine parameter and therefore cannot be interpreted
as the asymptotic region. Introducing a new transverse coordinate $\zeta$, 
which is defined by $\rho = \zeta^{-1}$, we will arrive at the picture summarized in 
Figure \ref{fig:branesum}: the locus $\rho \rightarrow \infty \Leftrightarrow \zeta \rightarrow
0$ is of no particular significance. The Killing horizon is located at $\zeta=(2B)^{-1}$ and
the static region $\zeta < (2B)^{-1}$ terminates at some value $\zeta = \zeta_s$
with a timelike curvature singularity, which is reached by transverse null geodesics 
at finite affine parameter. By analytic continuation the solution can be extended
to the region $(2B)^{-1} < \zeta < \infty$, where the coordinate $\zeta$ becomes
timelike and where the limit $\zeta\rightarrow \infty$
is at infinite (timelike) distance. We will interpret region I, $\zeta_s < \zeta <
(2B)^{-1}$ as the inside region, and region II, $(2B)^{-1} < \zeta < \infty$ as 
the outside region, because it has an asymptotic boundary at infinite distance.

We now give the details and make further comments on the properties of the three-charge solution. Beginning by applying the coordinate change \eq{cchange} to the three-charge solution, the scalars
$q_0,q_1,q_2$ are found to be:
\begin{equation}
q_0 = \mp \frac{\Ham_0}{W^{1/2}}, \quad q_1 = \pm \frac{\Ham_1}{W^{1/2}}, \quad q_2 = \pm \frac{\Ham_2}{W^{1/2}},
\end{equation}
where $\Ham_0$ and $\Ham_1$ are defined exactly as before and we have further defined
\begin{equation*}
\Ham_2(\rho) := P^2 \left[\frac{1}{B} \sinh \bigg(\frac{B h^2}{P^2}\bigg) + \frac{e^{-\frac{Bh^2}{P^2}}}{\rho}\right].
\end{equation*}
The remaining scalars are given by
\begin{equation*}
q_A = \pm \frac{C}{g_A} \frac{1}{W^{1/2}}.
\end{equation*}
The metric degrees of freedom are
\begin{equation*}
  e^{-4\psi} = W(\rho)^{-2},
\end{equation*}
and
\begin{equation*}
\begin{aligned}
    e^\phi &= \half \left( \frac{\Ham_0 \Ham_1 \Ham_2 }{W^2} C f_3\left(\frac{1}{g_A} \right)\right)^{-\half}
      = \frac{W}{\Ham},
\end{aligned}
\end{equation*}
where we have defined the new function
\begin{equation*}
  \Ham(\rho) := 2 \sqrt{C f_3 \Ham_0 \Ham_1 \Ham_2} \; .
\end{equation*}
This produces the line element
\begin{equation}
\label{eq:3charge}
ds^2 = -\frac{W}{\Ham} dt^2 +\frac{\Ham}{W} \frac{d\rho^2}{\rho^4} + \Ham (dx^2 + dy^2).
\end{equation}

The Lagrangian (energy functional) for transverse geodesics in the metric \eq{3charge} is
\begin{equation*}
    \La = -\frac{W}{\Ham} \dot{t}^2 +\frac{\Ham}{W \rho^4} \dot{\rho}^2,
\end{equation*}
where the dot represents differentiation with respect to an affine parameter $\lambda$.
Null geodesics satisfy $\La=0$.
The corresponding constant of motion
\begin{equation*}
    E = \frac{W \dot{t}}{\Ham},
\end{equation*}
is rearranged to give
\begin{equation}
    \dot{\rho} = \pm \sqrt{ \rho^4 E^2}, \qquad \lambda = \pm \int \frac{d\rho}{E \rho^2} .
\end{equation}
This shows that light signals sent from $\rho > 2 B$ reach $\rho = \infty$ at finite affine parameter,
whereas $\rho\rightarrow 0$ is at infinite affine parameter. Therefore, $\rho\rightarrow 0$ should
be interpreted as being at infinite distance and $\rho<2B$ as the exterior region, while $\rho>2B$ is
the inside region.

Given this observation, we introduce the new transverse coordinate $\zeta=\rho^{-1}$ so that
infinity is now at $\zeta\rightarrow \infty$.
It is important to note that in order to reach the limit of $\zeta \rightarrow \infty$ we must cross the Killing horizon located at $\zeta = (2B)^{-1} =: \alpha^{-1}$ into a new, `exterior' region. In the exterior
$\zeta$ is a timelike coordinate, and since the line element depends explicitly on $\zeta$, the exterior is non-stationary and as such the solution is interpreted as cosmological.

We also note that for causal information coming from an asymptotic distance the Killing horizon is a cosmological horizon which is located at a point in time $\zeta = \alpha^{-1}$ and so will be necessarily crossed for all causal geodesics.\footnote{This mandatory crossing can be understood in the same way as all causal geodesics reaching the singularity for the Schwarzschild solution once the horizon has been crossed.} Once the horizon has been crossed, the timelike singularity can be avoided and geodesics may leave the static patch into a second dynamic patch of spacetime. This statement is justified with calculations later in the paper.

Following our coordinate transformation the metric can be written in the following form
\begin{equation}
\label{eq:3czmet}
ds^2 = -\frac{W(\zeta)}{\Ham(\zeta)} d\eta^2 +\frac{\Ham(\zeta)}{W(\zeta)} d \zeta^2 + \Ham(\zeta) (dx^2 + dy^2).
\end{equation}
Note that we have relabelled the coordinate $t$, which is interpreted as time in the interior, as $\eta$. This is because $t$ becomes a spacelike coordinate in the outside patch of spacetime. In the following we will use the neutral
notation $\eta, \zeta$ instead of $t, \rho$.
The metric functions are $W = 1 - \alpha \zeta$ and $\Ham(\zeta)$ and
\begin{equation*}
\begin{aligned}
    \Ham_0(\zeta) &= Q_0 \left[\frac{2}{\alpha} \sinh \bigg(\frac{\alpha h_0}{2 Q_0}\bigg) + e^{-\frac{\alpha h_0}{2 Q_0}} \zeta \right], \\
    \Ham_1(\zeta) &= P^1 \left[\frac{2}{\alpha} \sinh \bigg(\frac{\alpha h^1}{2 P^1}\bigg) + e^{-\frac{\alpha h^1}{2 P^1}} \zeta \right], \\
    \Ham_2(\zeta) &= P^2 \left[\frac{2}{\alpha} \sinh \bigg(\frac{\alpha h^2}{2 P^2}\bigg) + e^{-\frac{\alpha h^2}{2 P^2}} \zeta \right], \\
\end{aligned}
\end{equation*}
We will write this in a condensed format by redefining our integration constants
such that
\begin{equation*}
     \Ham_a = \left(\beta_a + \gamma_a \zeta \right),
\end{equation*}
for $a = 0,1,2$.

To demonstrate that the metric can be analytically continued to
$\zeta > \alpha^{-1}$ despite the coordinate singularity at $\zeta = \alpha^{-1}$ we
make an intermediate coordinate transformation to advanced Eddington-Finkelstein coordinates
\begin{equation*}
  v = \eta + \zeta^*, \qquad d\zeta^* = \frac{\Ham}{W}d\zeta,
\end{equation*}
where we have introduced the tortoise coordinate $\zeta^*$ such that the metric can be written in the form
\begin{equation*}
 ds^2 =  - \frac{W}{\Ham} dv^2 + 2 d\zeta dv + \Ham (dx^2 + dy^2),
\end{equation*}
This shows that the metric has no singularity at $\zeta = \alpha^{-1}$ and so we can analytically continue the coordinate $\zeta$ to $\zeta > \alpha^{-1}$, and then reverse the coordinate transformation
to obtain the metric for the dynamic patch of the spacetime
\begin{equation*}
    ds^2 = - \frac{{W(\zeta)}}{\Ham(\zeta)} d\eta^2 + \frac{\Ham(\zeta)}{{W(\zeta)}} d \zeta^2 + \Ham(\zeta) (dx^2 + dy^2)
\end{equation*}
for $\zeta > \alpha^{-1}$.
Note that $W(\zeta)$ is an everywhere negative function within the domain of the dynamic patch of the spacetime. To have a clearer picture of our spacetime we define a new, always positive function within this domain $\mathcal{W}(\zeta) := \alpha \zeta - 1$.

Using this, we can write down the metric for $\zeta > \alpha^{-1}$ where it is immediately obvious that the coordinate $\zeta$ is timelike. 

The exterior region ($I$) is the cosmological region where $\zeta$ is timelike and the metric is time-dependent. The inside region ($II$) where $\zeta<\alpha^{-1}$, $\eta$ is timelike, and as we will see
below spacetime ends at a timelike singularity located at $\zeta_s$, where $\zeta_s$ is the
first zero of $\Ham(\zeta)$. Their respective line elements are given by
\begin{equation}
\label{eq:3chargenew}
\begin{aligned}
    ds^2_{I} = - \frac{\Ham(\zeta)}{{\mathcal{W}(\zeta)}} d \zeta^2 + \frac{{\mathcal{W}(\zeta)}}{\Ham(\zeta)} d\eta^2 + \Ham(\zeta) (dx^2 + dy^2), \\
    ds^2_{II} = - \frac{{W(\zeta)}}{\Ham(\zeta)} d\eta^2 + \frac{\Ham(\zeta)}{{W(\zeta)}} d \zeta^2 + \Ham(\zeta) (dx^2 + dy^2). \\
\end{aligned}
\end{equation}
We postpone a discussion of Kruskal-like coordinates and of 
the Penrose-Carter diagram for after the discussion of the four-charge solution, as these two solutions lend themselves naturally to being discussed simultaneously.

Having found coordinates suitable for describing both regions of our solution, we can now
start analysing its properties.
We begin with the physical scalars
\begin{equation*}
  z_1 = -i \left( \frac{\Ham_0 \Ham_1 W^{1/2}}{\Ham_2 f_3(g)} \right)^{\half}, \quad z_2 = -i \left( \frac{\Ham_0 \Ham_2 W^{1/2}}{\Ham_1 f_3(g)} \right)^{\half}, \quad z_A = -i \left( \frac{\Ham_0 W^{-1/2}}{\Ham_1 \Ham_2 f_3(g)} \right)^{\half} \;.
\end{equation*}
The asymptotic behaviour in
the limit $\zeta \rightarrow \infty$ depends on whether the integration constants $h_0$, $h_1$, $h_2$ are zero or non-zero, and is summarised in the table below.

\renewcommand{\arraystretch}{1.2}
\begin{table}[h!]
\centering
\begin{tabular}{|c| c|c|c|}
\hline
        & $z_1$     & $z_2$     & $z_A$     \\ \hline \hline
All Finite   & $\zeta^{3/4}$ & $\zeta^{3/4}$ & $\zeta^{3/4}$    \\ \hline
$h_0 = 0$   & $\zeta^{1/4}$ & $\zeta^{1/4}$ & $\zeta^{5/4}$ \\ \hline
$h_1 = 0$   & $\zeta^{1/4}$ & $\zeta^{5/4}$ & $\zeta^{1/4}$ \\ \hline
$h_2 = 0$   & $\zeta^{5/4}$ & $\zeta^{1/4}$ & $\zeta^{1/4}$ \\ \hline
$h_0, h_1 = 0$ & $\zeta^{1/4}$ & $\zeta^{3/4}$ & $\zeta^{3/4}$ \\ \hline
$h_1, h_2 = 0$ & $\zeta^{3/4}$ & $\zeta^{3/4}$ & $\zeta^{1/4}$ \\ \hline
$h_0, h_2 = 0$ & $\zeta^{3/4}$ & $\zeta^{1/4}$ & $\zeta^{3/4}$ \\ \hline
All Zero    & $\zeta^{1/4}$ & $\zeta^{1/4}$ & $\zeta^{1/4}$ \\ \hline
\end{tabular}
\end{table}

Following standard calculations, we are able to find the curvature scalars corresponding to the metric \eq{3czmet}. The Ricci scalar is found to be
\begin{equation}
\label{eq:ricciscalar}
R = \frac{W \left(\Ham^{\prime 2}-2 \Ham \Ham''\right)}{2 \Ham^3},
\end{equation}
and the Kretschmann scalar, $K = R^{abcd}R_{abcd}$, is given by
\begin{equation}
\label{eq:krescalar}
\begin{aligned}
K &= \frac{3 W^2 \Ham''^2}{\Ham^4} - \frac{2 W \Ham' \Ham''
  \left(4W\Ham'+3 \alpha \Ham\right)}{\Ham^5}
\\ &+ \frac{\Ham'^2 (27
  W^2 \Ham'^2+ 44 \alpha W \Ham \Ham'+20 \alpha ^2
  \Ham^2)}{4\Ham^6},
\end{aligned}
\end{equation}
where the prime denotes a derivative with respect to $\zeta$. We find that both have singular behaviour for the limit of $\Ham(\zeta) \rightarrow 0$. As $\Ham(\zeta)$ is a polynomial of degree three which factorizes into three linear polynomials  it will in general have three 
distinct zeros at $\zeta = \gamma_a \beta_a^{-1}$. The boundary of the 
spacetime domain $\zeta < (2B)^{-1}$ is given by the largest of these zeros, 
or `first zero of ${\cal H}(\zeta)$', 
which we denote $\zeta_s$. We remark that $\zeta_s \leq 0$ for all values of the integration constants.

We can study the near horizon geometry of this solution employing the same method as for the
two-charge solution. The horizon values of the harmonic functions are:
\begin{gather*}
  \Ham_0(\alpha^{-1}) = \frac{Q_0}{\alpha} \exp\left(\frac{\alpha h_0}{2 Q_0} \right) , \quad
  \Ham_1(\alpha^{-1}) = \frac{P^1}{\alpha} \exp\left(\frac{\alpha h^1}{2 P^1} \right) , \\
  \Ham_2(\alpha^{-1}) = \frac{P^2}{\alpha} \exp\left(\frac{\alpha h^2}{2 P^2} \right).
\end{gather*}
Making a coordinate transformation
\begin{equation}
\label{eq:nhcc}
    \chi^2 = \zeta - \alpha^{-1}, \qquad d\zeta^2 = 4 \chi^2 d\chi^2 ,
\end{equation}
we find
\begin{equation}
W \simeq \alpha \chi^2, \quad \Ham (\alpha^{-1}) = \frac{Z\E}{\alpha^{3/2}} ,
\end{equation}
where we have defined the constants
\begin{equation}
Z := 2 \Lambda \sqrt{Q_0P^1P^2}, \quad \E := \exp\left(\frac{\alpha}{4} \left( \frac{h_0}{Q_0} +\frac{h^1}{P^1} + \frac{h^2}{P^2} \right)\right), \quad \Lambda = \sqrt{C f_3}.
\end{equation}
Substituting this in, we write down the near horizon line element
\begin{equation}
ds^2 = -\frac{\alpha^{5/2} \chi^2}{Z \E} d\eta^2 + \frac{4 Z\E}{\alpha^{5/2}} d\chi^2 + \frac{Z\E}{\alpha^{3/2}} (dx^2 + dy^2).
\end{equation}
To find the temperature of the Killing horizon we set
\begin{equation*}
  dR^2 = \left(\frac{4 Z \E}{\alpha^{5/2}} \right) d\chi^2,
\end{equation*}
and Wick rotate $\eta \rightarrow -i \eta_E$ to obtain the Hawking temperature
\begin{equation}
2\pi T_H = \frac{\alpha^{5/2}}{2 Z\E} \;.
\end{equation}
This shows that $\alpha$ should still be interpreted as the non-extremality parameter, with extremal limit
$\alpha\rightarrow 0$.
We can read off the entropy density of the solution as
\begin{equation}
s = \frac{ Z\E}{\alpha^{3/2}},
\end{equation}
and note that it diverges in the limit $\alpha \rightarrow 0$.
Equating the above two equations, we can solve for the integration constant\footnote{Here we replace $\alpha = 2B$ to allow for better comparison to our previous results.}
\begin{equation*}
  B = 2\pi s T_H,
\end{equation*}
which is the same relationship that we saw in the Nernst solution \cite{Dempster:2015} and the two-charge solution.

In region $I$ we can consider the asymptotic limit $\zeta \rightarrow \infty$,
which corresponds to future timelike infinity. Taking the limit of
the functions
\begin{equation*}
    \lim_{\zeta \rightarrow \infty}\frac{\mathcal{W}}{\Ham} = \frac{\alpha}{\Lambda \sqrt{\gamma_0 \gamma_1 \gamma_2}} \frac{1}{\sqrt{\zeta}}, \qquad \lim_{\zeta \rightarrow \infty} \Ham = \Lambda \sqrt{\gamma_0 \gamma_1 \gamma_2} \zeta^{\frac{3}{2}},
    \end{equation*}
we find the asymptotic form of the metric
\begin{equation}
    ds^2 = - \frac{\Lambda \sqrt{\gamma_0 \gamma_1 \gamma_2}}{\alpha} \sqrt{\zeta} d\zeta^{2} + \frac{\alpha}{\Lambda \sqrt{\gamma_0 \gamma_1 \gamma_2}} \frac{1}{\sqrt{\zeta}} d\eta^2 + \Lambda \sqrt{\gamma_0 \gamma_1 \gamma_2} \zeta^{\frac{3}{2}} (dx^2 + dy^2).
\end{equation}
Taking the coordinate change $\zeta = t^2$ and absorbing constant factors, we can write this metric as
\begin{equation}
    ds^2 = t^3 (-dt^2 + dx^2 + dy^2 ) + \frac{1}{t} d\eta^2,
\end{equation}
or taking
\begin{equation*}
    \zeta = \tau^{\frac{4}{5}} \qquad d\tau = \sqrt{\zeta} d\zeta^2,
\end{equation*}
to obtain a metric in the form
\begin{equation}
    ds^2 = -d\tau^{2} + \tau^{-\frac{2}{5}} d\eta^{2} + \tau^{\frac{6}{5}} (dx^2 + dy^2).
\end{equation}

%
%
\subsection{Four-Charge Solution}
It will turn out that the qualitative behaviour of the four-charge solution is the
same as that of the three-charge solution, and we proceed accordingly. 
We introduce the transverse coordinate $\zeta$ by 
\begin{equation*}
  e^{-2B\tau} = 1 - \alpha \zeta := W(\zeta),
\end{equation*}
and the horizon is located at $\zeta = \alpha^{-1} = (2 B)^{-1}$.  
Applying the same coordinate change to our scalars \eq{gensca} we obtain
\begin{equation}
q_a = \pm \frac{\Ham_a}{W^{1/2}},
\end{equation}
where we have defined the harmonic functions
\begin{equation*}
\Ham_a(\zeta) := |K_a| \left[\frac{2}{\alpha} \sinh \bigg(\frac{\alpha h_a}{2 |K_a|}\bigg) + e^{-\frac{\alpha h_a}{2 |K_a|}} \zeta \right], \quad a=0,\ldots, 3.
\end{equation*}
Using \eq{gensca} we obtain the following expressions for the physical scalars:
\begin{equation*}
  z^1 = -i \left( \frac{\Ham_0 \Ham_1 }{\Ham_2 \Ham_3} \right)^{\half}, \quad z^2 = -i \left( \frac{\Ham_0 \Ham_2 }{\Ham_1 \Ham_3} \right)^{\half}, \quad z^3 = -i \left( \frac{\Ham_0 \Ham_3 }{\Ham_1 \Ham_2} \right)^{\half}.
\end{equation*}
Taking the limit $\zeta \rightarrow \infty$ we find that
 \begin{equation*}
   \lim_{\zeta \rightarrow \infty} \Ham_a = K_a e^{-\frac{\alpha h_a}{2 |K_a|}} \zeta,
 \end{equation*}
and that the scalars all tend to a constant value as all $\Ham_a$ depend on $\zeta$ in the same manner.

We now wish to re-express the metric ansatz \eq{ansatz} with our new coordinates. We find that
\begin{equation}
\label{eq:4cmdf1}
e^{-4\psi} = e^{4B\tau} = \frac{1}{W^2},
\end{equation}
and
\begin{equation}
\label{eq:4cmdf2}
    e^\phi = \half (-q_0 q_1 q_2 q_3)^{-\half} = \frac{W}{\Ham},
\end{equation}
where we have defined:
\begin{equation}
\Ham (\zeta) := 2 \sqrt{\Ham_0 \Ham_1 \Ham_2 \Ham_3}    .
\end{equation}
Simply substituting in these into our ansatz \eq{ansatz} we obtain the line element for the interior region of the solution as
\begin{equation*}
ds^2_{II} = - \frac{W(\zeta)}{\Ham(\zeta)} d\eta^2 + \frac{\Ham(\zeta)}{W(\zeta)} d\zeta^2 + \Ham(\zeta) (dx^2 + dy^2).
\end{equation*}
We notice here the same functional form of the four-charge solution as the three-charge solution \eq{3czmet}, with the only difference coming from the polynomial order of $\Ham(\zeta)$. Analytically continuing to $\zeta > \alpha^{-1}$ we find that $W(\zeta)$ changes sign. This suggests we redfine $-W(\zeta) =: \mathcal{{W}}(\zeta) = \alpha \zeta - 1$ and now the metric between the asymptotic limit and the horizon is described by
\begin{equation}
\label{eq:4cmetricthing}
ds^2_{I} = - \frac{\Ham(\zeta)}{\mathcal{{W}}(\zeta)} d\zeta^2 + \frac{\mathcal{{W}}(\zeta)}{\Ham(\zeta)} d\eta^2 + \Ham(\zeta) (dx^2 + dy^2),
\end{equation}
where we notice that our metric depends only on the timelike coordinate $\zeta$. As the metric is the same in form as \eq{3czmet} the curvature scalars will be given by (\ref{eq:ricciscalar}, \ref{eq:krescalar}) and we see there is a curvature singularity when $\Ham=0$, which happens whenever $\Ham_a(\zeta) = 0$. 
Without loss of generality, we assume that the first zero of $\Ham$ will be for $\Ham_0=0$,
so that the singularity will occur at:
\begin{equation}
\label{eq:4csing}
    \zeta_s = \frac{1 - e^{\frac{\alpha h_0}{Q_0}}}{\alpha} = -\frac{\beta_0}{\gamma_0} \;.
\end{equation}

From the relations \eq{4dgauge} we can apply our coordinate transformation to write down the gauge fields in terms of the new $\zeta$ coordinate
\begin{equation}
\label{eq:new4dg}
    \begin{aligned}
        F^0_{\zeta \eta} = -\frac{Q_0}{2(\beta_0 + \gamma_0 \zeta)^2} \; ,\qquad \tilde{F}_{A |\zeta \eta} = \frac{P^A}{2(\beta_A + \gamma_A \zeta)^2} \; .
    \end{aligned}
\end{equation}

We can again study the near-horizon geometry with the coordinate change \eq{nhcc} and probing for when $\zeta \simeq \alpha^{-1}$
\begin{equation*}
  d\zeta^2 = 4\chi^2 dr^2, \quad \cW = \alpha \chi^{2}, \quad \Ham = \frac{2Z \E}{\alpha^2},
\end{equation*}
where we have defined
\begin{equation*}
  Z := \sqrt{Q_0P^1P^2P^3}, \quad \E := \exp \bigg(\frac{\alpha}{4} \bigg(\frac{h_0}{Q_0} + \frac{h^1}{P^1} + \frac{h^2}{P^2} + \frac{h^3}{P^3} \bigg) \bigg).
\end{equation*}
We now substitute these expressions into our metric to obtain the near-horizon line element
\begin{equation}
\label{eq:nearhorizon}
ds^2 = -\frac{\alpha^3 \chi^2}{2 Z \E} d\eta^2 + \frac{8 Z \E}{\alpha^3} d\chi^2 + \frac{2 Z \E}{\alpha^2} (dx^2 + dy^2).
\end{equation}
Following the method as before, we Wick rotate after the coordinate change to find the Hawking temperature associated to the brane
\begin{equation*}
    2\pi T_H = \frac{\alpha^3}{4 Z \E},
 \end{equation*}
and read off the entropy density
\begin{equation*}
    s = \frac{2 Z \E}{\alpha^2} .
\end{equation*}
We notice that as for the three-charge solution, the entropy density diverges in the extremal limit of $\alpha \rightarrow 0$. We also find that the equation of state is again given by
\begin{equation*}
    B = 2\pi s T_H,
\end{equation*}
which thereby is established as
a standard relation for the full set of solutions ranging from Nernst branes to the
four-charge solution. 

In the asymptotic limit, we take $\zeta \rightarrow \infty$ and we find that
\begin{equation*}
  \lim_{\zeta \rightarrow \infty} \Ham_a(\zeta) \simeq K_a e^{\frac{\alpha h_a}{2 K_a}}\zeta, \qquad \lim_{\zeta \rightarrow \infty} \Ham(\zeta) \simeq 2 Z \E \zeta^2, \qquad \lim_{\zeta \rightarrow \infty} \cW(\zeta) \simeq \alpha \zeta.
\end{equation*}
We use this to write down the asymptotic metric
\begin{equation*}
  ds^2 = - \frac{2 Z \E \zeta}{\alpha} d\zeta^2 + \frac{\alpha}{2 Z \E \zeta} d\eta^2 + 2 Z \E \zeta^2 (d\bar{x}^2 + d\bar{y}^2) ,
\end{equation*}
and with a simple change of coordinates to absorb all of the constants we find the asymptotic metric is in the form

\begin{equation}
\label{eq:asymet}
  ds^2 = - \bar{\zeta} d\bar{\zeta}^2 + \frac{1}{\bar{\zeta}} d\bar{\eta}^2 + \bar{\zeta}^2 (d{x}^2 + d{y}^2).
\end{equation}

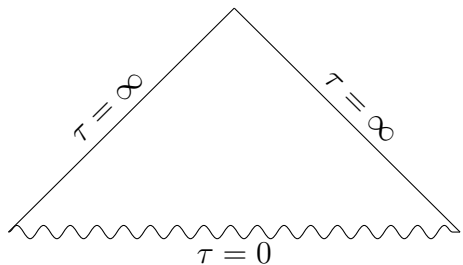
\begin{figure}[h]
\centering
\begin{tikzpicture}[scale = 0.7]
\node (I)  at (0,0)  {};

\path 
 (I) +(45:-6) coordinate (left)
   +(-45:6) coordinate (right)
   +(0:0) coordinate (top)
   ;
\draw (left) --node[midway, above, sloped] {$\tau=\infty$} (top);
\draw (right) -- node[midway, above, sloped] {$\tau=\infty$} (top);
\draw[decorate,decoration=snake, draw=black] (left) -- node[midway, below] {$\tau=0$}(right);
\end{tikzpicture}
\caption{Penrose diagram for type D Kasner solution}
\label{fig:kasnerpc}
\end{figure}

This can be identified with the planar Schwarzschild solution (AIII metric) \cite{Griffiths:2009dfa} with the mass $M=\frac{1}{2}$. Through the coordinate transformation
\begin{equation*}
  \bar{\eta} = \left(\frac{3}{2}\right)^{\frac{1}{3}} z, \qquad   \bar{\zeta} = \left(\frac{9}{4}\right)^{\frac{1}{3}} \tau^{\frac{2}{3}}, \qquad   (x,y) = \left(\frac{4}{9}\right)^{\frac{1}{3}} (x,y), \qquad
\end{equation*}
we can rewrite the asymptotic metric in the form
\begin{equation*}
  ds^2 = -d\tau^2 + \tau^{2/3} dz^2 + \tau^{4/3} (dx^2 + dy^2),
\end{equation*}
which is the type D vacuum Kasner solution \cite{Kasner:1921zz}. The Kasner solution is given generally by
\begin{equation*}
  ds^2 = -dt^2 + t^{2p_1} dx^2 + t^{2p_2} dy^2 + t^{2p_3} dz^2,
\end{equation*}
where the constants $(p_1,p_2, p_3)$ must satisfy the planar and spherical Kasner constraints
\begin{equation*}
  p_1 + p_2 + p_3 = 1, \qquad p_1^2 + p_2^2 + p_3^2 = 1,
\end{equation*}
and we see that our asymptotic solution is the case for $(p_1,p_2, p_3) = (\tfrac{2}{3},\tfrac{2}{3},-\tfrac{1}{3})$. The Penrose diagram for the vacuum Kasner type D solution is given by figure (\ref{fig:kasnerpc}).

%
%
\section{Causal Structure of Cosmological Solutions}
Of the three solutions that we have constructed, the higher charged solutions are more surprising. We have found that their static part is the interior of a solution that
on the outside is time-dependent and thus can be interpreted as a cosmological
solution. In this section, we further analyse the resulting spacetimes by studying the behaviour of causal geodesics and massive particles. We begin by outlining the coordinate transformations that lead to a Kruskal-type metric for the three- and four-charge solutions. This allows the construction of the Penrose-Carter diagram for these solutions and gives us an intuitive diagrammatic overview of the causal structure.
This representation also allows us to realise that the solutions we study for the three- and four-charge systems have an intersection with a class of cosmological solutions studied in \cite{Burgess:2002vu, Burgess:2003mk} for the case of generalised Einstein-Maxwell-Dilaton and the orientifold constructions in \cite{Cornalba:2003kd}.
We then study the static patch, and hence the singularity, in more detail by probing them with causal geodesics and the worldlines of stationary massive particles. We find that all timelike geodesics are repelled by the singularity and that stationary massive particles experience negative acceleration with respect to the singularity.

\subsection{Kruskal Coordinates}
While searching for an asymptotic region, we already 
found the Eddington-Finkelstein coordinates:
\begin{equation*}
 ds^2 \; = \; - \frac{W(\zeta)}{\Ham(\zeta)} dv^2 + 2 d\zeta dv + \Ham(\zeta) (dx^2 + dy^2),
\end{equation*}
where for the three- and four-charge solutions
\begin{equation*}
    \Ham^{(3)}(\zeta) = 2 \sqrt{C f_3 \Ham_0 \Ham_1 \Ham_2}, \qquad \Ham^{(4)}(\zeta) = 2\sqrt{\Ham_0 \Ham_1 \Ham_2 \Ham_3}.
\end{equation*}

To find Kruskal coordinates we first make a second coordinate change into light-cone coordinates
\begin{equation*}
    u = \eta - \zeta^*, \qquad ds^2 = -\frac{W}{\Ham} dv d\zeta + \Ham(dx^2 + dy^2).
\end{equation*}
The key to finding Kruskal coordinates is integrating $\zeta^*$
\begin{equation*}
    \zeta^* = \int d\zeta
 \frac{\Ham(\zeta)}{W(\zeta)},
    \end{equation*}
and picking a suitable $\lambda(\zeta)$ for the transformation to Kruskal-like coordinates
\begin{equation*}
 U = - e^{-\lambda u}, \quad V = e^{\lambda v},
\end{equation*}
where $U \leq 0$ and $V \geq 0$, and $\lambda$ is picked to remove the factor of $W$, and hence all zeros in the metric. Because of the form of the functions $W$ and $\Ham$ this procedure can become algebraically involved. However finding the 
coordinate transformation explicitly is not needed to infer its existence and
to draw the Penrose-Carter diagram. 

A simple example, where finding the explicit form of the transformation 
is not too involved, is the case where all functions  $\Ham_a$ are taken to be equal. This simplifies the calculation of $\zeta^*$ and hence allows for easy identification of $\lambda$. For those interested, the explicit expressions for this case are included in the appendix.

The final form to which the four-charge solution can be brought
is
\begin{equation}
 ds^2 = - \frac{1}{\lambda^2} \frac{e^{\xi(\zeta(U,V))}}{2(\beta + \gamma \zeta(U,V))^2} dU dV + 2(\beta + \gamma \zeta(U,V))^2 (dx^2 + dy^2),
\end{equation}
where $\xi$ is a function defined from the integrand of $\zeta^*$ and packages together the results of the coordinate change into an everywhere non-zero, monotonically increasing function for the global domain of $\zeta$, allowing to express $\zeta$ in terms of $U,V$. The constants $\beta,\gamma$ are the constants coming from the simplification $\gamma = \gamma_0 = \gamma_1 = \gamma_2 = \gamma_3$ and $\beta = \beta_0 = \beta_1 = \beta_2 = \beta_3$ ensuring all $\Ham_a$ are equal. 

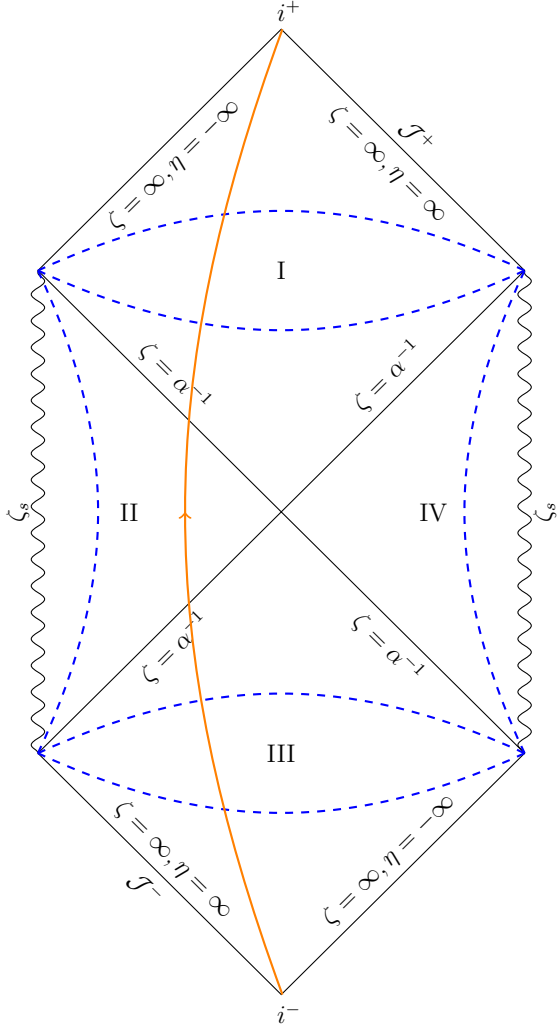
\begin{figure}[h!]
\centering
\begin{tikzpicture}[scale=0.8, every node/.style={scale=0.8}]
\node[above] (0) at (0.13,12) {$i^+$};
\node[below] (1) at (0.13,-4) {$i^-$};
\node (I) at (0,8) {I};
\node (II) at (0,0) {III};
\node (III) at (-2.5,4) {II};
\node (IV) at (2.5,4) {IV};
\path 
 (I) +(90:4) coordinate[label=90:] (IItop)
 +(-90:4) coordinate(IIbot)
 +(0:4) coordinate[label=360:] (IIright)
 +(180:4) coordinate[label=180:] (IIleft)
 ;
\draw (IIleft) --
 node[midway, below, sloped] {$\zeta=\infty, \eta = - \infty$}
 (IItop) --
 node[midway, below, sloped] {$\zeta=\infty, \eta = \infty$}
 (IIright)
     node[midway, above, sloped] {$\mathcal{J}^+$}
           --
 node[midway, above, sloped] {$\zeta = \alpha^{-1}$}
 (IIbot) --
 node[midway, above, sloped] {$\zeta = \alpha^{-1}$}
 (IIleft) -- cycle;

\path 
 (II) +(90:4) coordinate (Itop)
 +(-90:4) coordinate (Ibot)
 +(180:4) coordinate (Ileft)
 +(0:4) coordinate (Iright)
 ;
\draw (Ileft) --
 node[midway, below, sloped] {$\zeta = \alpha^{-1}$}
 (Itop) --
 node[midway, below, sloped] {$\zeta = \alpha^{-1}$}
 (Iright) --
 node[midway, above, sloped] {$\zeta=\infty, \eta = - \infty$}
 (Ibot) --
 node[midway, above, sloped] {$\zeta=\infty, \eta = \infty$}
 node[midway, below, sloped] {$\mathcal{J}^-$}
 (Ileft) -- cycle;

\draw[decorate,decoration=snake,draw=black] (Ileft) -- (IIleft)
 node[midway, above, sloped] {$\zeta_s$};

\draw[decorate,decoration=snake, draw=black] (Iright) -- (IIright)
 node[midway, below, sloped] {$\zeta_s$};

\draw[blue,dashed,bend right=25, thick] (Ileft) to (Iright);
\draw[blue,dashed,bend right=-25, thick] (Ileft) to (Iright);
\draw[blue,dashed,bend right=25, thick] (IIleft) to (IIright);
\draw[blue,dashed,bend right=-25, thick] (IIleft) to (IIright);
\draw[blue,dashed,bend right=25, thick] (Ileft) to (IIleft);
\draw[blue,dashed,bend right=-25, thick] (Iright) to (IIright);
\draw[->-,orange,bend right=-20, thick] (1) to (0);

\end{tikzpicture}
\caption{Penrose diagram for the planar cosmological solutions. Starting at $\zeta = \infty$ we have a cosmological spacetime (I) with a horizon located at a finite point in time; any observer must necessarily fall through the horizon. Passing through the horizon, the spacetime is static (II) with an avoidable (repulsive) naked singularity located at a point in space. Massive particles at rest experience negative acceleration and will leave the static region into a second dynamic spacetime. An example of a complete timelike geodesic is given in orange, spacelike hypersurfaces of constant time are given in blue.}
\label{fig:PC}
\end{figure}

If the global metric is known explicitly, it can
then be used to construct the Penrose-Carter diagram. If no explicit expression
is available, the causal diagram has to be constructed piece-wise, by patching
together regions separated by regular horizons. The diagram for the general 
three- and four-charge solution is 
given in figure (\ref{fig:PC}).  It looks like the Penrose-Carter diagram of the maximally extended
Schwarzschild spacetime, rotated by 90 degrees. Solutions with the same causal
structure have been found previously in \cite{Burgess:2002vu}. 
The regions
I and III are time-dependent and asymptotic to Kasner solutions at late and
early times, respectively. This cosmological solution is disturbed by 
the presence of two timelike singularities, which can be interpreted as brane-like
sources, which create the static regions II and IV and are separated from 
the cosmological regions I and III by a bifurcate Killing horizon. 

\subsection{Extremal Limit}
\label{sec:4Dext}
For our solutions, the extremal limit is defined by the vanishing of the temperature $T_H$, which is equivalent to taking the limit of $\alpha \rightarrow 0$ and is represented in the metric functions and integration constants by:
\begin{equation}
\label{eq:extlim}
    \alpha \rightarrow 0 \quad \Rightarrow \quad \cW(\zeta) \rightarrow -1, \qquad \beta_a \rightarrow h_a \zeta, \qquad     \gamma_a \rightarrow K_a .
\end{equation}
The resulting line element is given by
\begin{equation*}
ds^2 =  - \Ham^{-1}(\zeta) d\eta^2 + \Ham(\zeta) d\zeta^2  + \Ham(\zeta) (dx^2 + dy^2),
\end{equation*}
where $\eta, \zeta$ are now everywhere timelike and spacelike respectively. 

The extremal limit for the three- and four-charged
solutions has a dramatic effect on the causal structure 
of the spacetime. As the function $\cW$ becomes constant  we find that the location of the horizon is set by $\Ham^{-1} \rightarrow 0$, which occurs when $\zeta \rightarrow \infty$. The horizon location is pushed from $\alpha^{-1} \rightarrow \infty$ and the resulting spacetime is everywhere static with a naked singularity. This change in the causal structure is a general feature of the planar symmetric solutions we consider, and the simplest example is found for the planar symmetric Riessner-Nordstr\"om solution. Further discussion of the relationship between the causal structure and the extremal limit is left for section \ref{sect:conclusions} and the shifting of horizons under this limit is depicted in Figure \ref{fig:planarextremal}.

\subsection{Probing the Static Patch}

In this section, we study the static regions II and IV.
Obtaining a mass-like parameter for our solution is obstructed in two ways: firstly, the asymptotic region of the spacetime is time-dependent and so any conserved quantity associated to our Killing vector is going to be more physically related to momentum than mass.\footnote{Here we are interpreting our conserved charges as Noether charges. In static spacetimes, we associate time translation invariance with energy. In the dynamic region where our Killing vector is spacelike, invariance under spacelike translations is associated with conserved momentum.} Secondly, the solution is not asymptotically flat, and as such, there will be a complication as to how to properly normalise the norm of the Killing vector field and hence the conserved quantity associated to it.

To ameliorate these issues, we present two different methods of calculating a mass-like parameter. We first follow the work of \cite{Burgess:2002vu} and perform a space-dependent calculation for the Komar mass. Secondly, we employ the Brown-York formalism \cite{York1986} and calculate a quasi-local mass. Both techniques consistently 
give a mass-like parameter, which is negative and suggests that the 
singularity is repulsive to neutral massive particles. 
Following up on this observation, we show that there always exists a classical turning point for massive particles following geodesics. Further, we find that all massive particles at rest undergo negative acceleration. We conclude that the spacetime is 
`timelike geodesically complete,'  that is timelike geodesics can be extended to
infinite proper time and that the singularity is repulsive.

For the following section, both the three- and four-charge solutions are analysed simultaneously; when differences between the solution arise, this is expressed explicitly for the metric functions. The difference for our integration constants is always left implicit.

\subsubsection{Mass}

\subsection*{Komar Mass}
We begin our investigation of a local mass parameter within the static region by using the standard Komar integral
\begin{equation*}
    M_K = -\frac{1}{8 \pi} \int_{\Real^2} \star d k,
\end{equation*}
for a timelike Killing vector $k$. In more conventional solutions this integral is evaluated while taking the asymptotic limit where the Komar mass matches with that of the ADM mass for solutions with the appropriate asymptotic fall off. For our solution, the domain of the static region is finite, and hence we leave $k^2$ unnormalised to obtain a mass-like parameter dependent on the spacelike coordinate $\zeta$.

For a Killing vector: $k^\mu = (1,0,0,0)$ and taking the Hodge dual, with orientation set by $\epsilon_{\eta \zeta x y} = 1$ we find
\begin{equation*}
    (\star dk)_{\eta \zeta} = -\Ham \partial_\zeta \left( \frac{W}{\Ham} \right) .
\end{equation*}
The Komar integral is evaluated to
\begin{equation*}
        M_K = -\frac{1}{8\pi} \int_{\Real^2} \left(\alpha + \frac{W}{\Ham} \partial_{\zeta} \Ham \right) .
\end{equation*}
Due to the planar symmetry of our solution, this value is divergent when integrating over the plane, so we instead work with the mass density. The resulting position dependent mass density is
\begin{equation}
    m_K = - \left( \frac{\alpha}{8 \pi} + \frac{1}{8 \pi} \frac{W g(\zeta)}{\Ham^2} \right),
\end{equation}
where the function $g(\zeta)$ is related to the derivative of $\Ham$
\begin{equation*}
\partial_\mu \Ham = \frac{g(\zeta)}{\Ham}.
\end{equation*}
For the three-charge solution
\begin{equation*}
    g(\zeta) = 2 C f_3 (\gamma_0 \Ham_1 \Ham_2 + \gamma_1 \Ham_0 \Ham_2 + \gamma_2 \Ham_0 \Ham_1),
\end{equation*}
and for the four-charge solution
\begin{equation*}
    g(\zeta) = 2 (\gamma_0 \Ham_1 \Ham_2 \Ham_3 + \gamma_1 \Ham_0 \Ham_2 \Ham_3 + \gamma_2 \Ham_0 \Ham_1 \Ham_3 + \gamma_3 \Ham_0 \Ham_1 \Ham_2).
\end{equation*}
Within the domain of the static region $\Ham, g, W > 0$. As $\alpha$ is always positive, the Komar mass will be everywhere negative within the static patch of the spacetime, regardless of the overall normalisation of $k^2$.

\subsection*{Brown-York Mass}

Alternatively, we can calculate the Quasi-Local mass of the spacetime by using the Brown-York formalism \cite{Brown:1992br}. The Brown-York quasi-local energy is found from\footnote{We note here the inclusion of the lapse function $N$. For asymptotically flat spacetimes $\lim_{\zeta \rightarrow \infty} N = 1$ and so $N$ is absent from many papers in the literature. The inclusion of $N$ is talked about in more detail when considering non-asymptotically flat spacetimes, asymptotically AdS spaces being the most common example of this currently \cite{Lu:2013ura}.}

\begin{equation}
\label{eq:BYmass}
E = -\frac{1}{8 \pi} \int_{\Real^2} \sqrt{\sigma} N (\mathtt{k} - \mathtt{k}_0),
\end{equation}

In this formalism, we consider a physical spacetime $M$, which is topologically a hypersurface $\Sigma$, foliated over a real line interval. The boundary of $\Sigma$ is $B$. Taking the product of $B$ with the timelike worldlines orthogonal to $\Sigma$ produces the codimension-1 hypersurface ${}^3B$, a component of the 3-boundary of $M$. The full boundary of $M$ includes the end points of timelike worldlines.

The spacetime $M$ is equipped with the metric $g_{\mu\nu}$ and Levi-Civita connection $\nabla_\mu$. To calculate \eq{BYmass}, we will need the geometric data of $B$ in terms of the known data of $(M,g)$.
We take a future pointing unit vector $u^\mu$, normal to the foliation $\Sigma$. A tensor $T$ is said to be spatial when $T\cdot u=0$. The metric $g_{\mu \nu}$ induces 
a metric on $\Sigma$, which, when regarded as a tensor $h_{\mu \nu}$ on $M$, is 
a spatial tensor. The induced covariant derivative $\D_\mu$ for spatial tensors is found through projection $\D_\mu = h^\nu_\mu \nabla_\nu$. The extrinsic curvature of $\Sigma$ 
as an embedded submanifold of $M$ is denoted $K_{\mu \nu}$. We use
the notation $h_{ij}, \; K_{ij}$, where $i,j$ run from one to the dimension of $\Sigma$, when regarding the metric and extrinsic curvature
of $\Sigma$ as tensors on $\Sigma$.

The ADM decomposition of the metric is given by
\begin{equation}
\label{eq:ADM}
    ds^2 = -N^2 dt^2 + h_{ij} (dx^i + V^i dt)(dx^j + V^j dt),
\end{equation}
for a lapse function $N$ and shift vector $V^i$.

We proceed in the same way with the 3-boundary ${}^3B$ by considering the outward pointing unit vector $n^\mu$, normal to ${}^3B$. The metric induced by $g_{\mu \nu}$ is denoted $\gamma_{mn}$ when
regarded as a tensor on ${}^3B$ and $\gamma_{\mu \nu}$ when regarded 
as a horizontal tensor on $M$, \textit{i.e.} as a tensor $T$ on $M$ satisfying $n\cdot T =0$.

The boundary $B$, which is the intersection of $\Sigma$ and ${}^3B$ has a metric $\sigma_{\mu\nu}$ which can be induced from either of the codimension-1 manifolds or the spacetime itself. The extrinsic curvature $\mathtt{k}_{\mu\nu}$ of $B$ – the vital part needed to calculate \eq{BYmass} – is computed using the embedding of $B$ in $\Sigma$:
\begin{equation}
\begin{aligned}
\label{eq:extexplicit}
\mathtt{k}_{\mu \nu} &= \sigma^\alpha_\mu \D_\alpha n_\nu \\
&= \gamma^\alpha_\mu h^\beta_\nu h^\rho_\alpha \nabla_\rho n_\beta    
\end{aligned}
\end{equation}
We will also need the trace $\mathtt{k} = \sigma^{\mu\nu} \mathtt{k}_{\mu \nu}$ in our later calculations.

The quasi-local energy is evaluated by studying the Hamiltonian that generates a unit time translation orthogonal to $\Sigma$ at the boundary $B$. This is related to the action together with a normalisation coming from a chosen background metric denoted with a subscript $0$. For more details on the full derivation of \eq{BYmass} see \cite{Brown:1992br}.

Comparing our metric with the ADM decomposition \eq{ADM} we identify

\begin{equation*}
N^2 = \frac{W}{\Ham}, \quad V^i = 0, \quad \sigma_{xx} = \sigma_{yy} = 2\Ham.
\end{equation*}

The quasi-local energy is then simply calculated using these quantities together with the trace of the extrinsic curvature \eq{extexplicit}

\begin{equation}
\mathtt{k} = \frac{1}{\Ham} \sqrt{\frac{W}{\Ham}}\partial_{\zeta} \Ham.
\end{equation}

As there is no divergent contribution from the calculation there is no natural normalisation choice for $\mathtt{k}_0$, and so, for now, we set it to zero; simplifying our calculations

\begin{equation}
\begin{aligned}
E_{BY} &= -\frac{1}{8 \pi} \int d^2 x \sqrt{\sigma} N \mathtt{k}, \\
        &= - \frac{1}{4 \pi} \int d^2x \frac{W g(\zeta)}{\Ham^2}.
\end{aligned}
\end{equation}

When the spacetime is static, the Brown-York energy is equivalent to the mass of the spacetime. Removing the divergent factor related to the planar symmetry and again looking at the mass density
we find
\begin{equation}
m_{BY} = - \frac{1}{4 \pi} \frac{W g}{\Ham^2}.
\end{equation}

This is negative definite in the static domain due to identical reasoning as for the Komar calculation.

We note here that despite both being everywhere negative $m_K \neq m_{BY}$. We could have handpicked $\mathtt{k}_0$ to match our result to the Komar calculation but without an asymptotic limit to properly normalise the Killing vector $k^\mu$ there is no reason pick either of our results as `correct'. This undecidability of overall normalisation is not important for our current discussion as we focus on that the calculated mass-like parameter is everywhere negative throughout the static region rather than the precise value.

\subsubsection{Geodesic Motion}
We now turn our attention to studying the motion of causal geodesics within the static region of our spacetime. Using the metric, we can write down the Lagrangian of our system
\begin{equation}
    s = \La = - \frac{W(\zeta)}{\Ham(\zeta)} \dot{\eta}^2 + \frac{\Ham(\zeta)}{W(\zeta)} \dot{\zeta}^2 + \Ham (\dot{x}^2 + \dot{y}^2),
\end{equation}
where $s=0,-1$ for null and timelike geodesics respectively. We calculate the constants of motion as
\begin{equation*}
    E = \frac{W}{\Ham} \dot{\eta}, \qquad a = \Ham \dot{x}, \qquad b = \Ham \dot{y} ,
\end{equation*}
allowing us to rewrite the Lagrangian as
\begin{equation}
    s\frac{W}{\Ham} = -E^2 + \dot{\zeta}^2 + (a^2 + b^2)\frac{W}{\Ham^2}.
\end{equation}
This can be rearranged into the familiar form
\begin{equation}
    \dot{\zeta}^2 = E^2 - V(\zeta), \qquad V(\zeta) = \frac{W}{\Ham} \left(-s + \frac{(a^2 + b^2)}{\Ham} \right),
\end{equation}
which can be interpreted as the equation of motion for a particle with mass $m=2$. We 
rearrange this equation and package together the function $V(\zeta)$ to explicitly highlight that this piece can be interpreted as an effective `potential' of the system. The domain of validity for the equation of motion is restricted by the inequality
\begin{equation*}
    V(\zeta) \leq E^2.
\end{equation*}
The point at which $V(\zeta_0)=E$ is interpreted as the classical truning point of the particle's trajectory. The domain is further restricted by the presence of the singularity such that $\zeta > \zeta_s$ and so the domain of $\zeta$ in region II is given by
\begin{equation}
 \alpha^{-1} > \zeta > -\frac{\beta_0}{\gamma_0}.
\end{equation}

Now, by studying the potential $V(\zeta)$ of our spacetime for the correct domain of $\zeta$, we can look at the motion of causal information along geodesics.

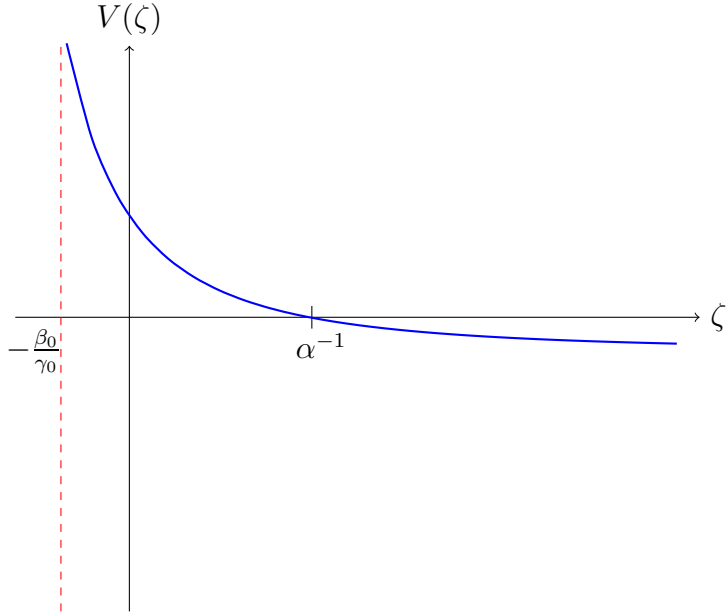
\begin{figure}[h!]
\centering
\begin{tikzpicture}[scale=1.5]
 \draw[->] (-1,0.6) -- (5,0.6) node[right] {$\zeta$};
 \draw[->] (0,-2) -- (0,3) node[above] {$V(\zeta)$};
 \draw[dashed,red] (-.6,-2) -- (-.6,3);
 \draw[black] (1.6,0.5) -- (1.6,0.7);
     \draw[domain=-.55:4.8,smooth,variable=\x,blue,thick] plot({\x},{(30 -\x)/((3*\x+5)*(3*\x+5)) + 0.3});
     \node[left] (one) at (-0.5,0.3) {$-\frac{\beta_0}{\gamma_0}$};
     \node[left] (one) at (2,0.35) {$\alpha^{-1}$};
\end{tikzpicture}
\caption{Behaviour of the effective `potential' as a function of $\zeta$ for the set of casual geodesics excluding null transverse geodesics, for which $V=0$.}
\label{fig:3pot}
\end{figure}
In figure \ref{fig:3pot} we plot $V(\zeta)$ and see that for region $II$ of our spacetime the potential is everywhere positive and therefore repulsive. When decreasing $\zeta$ from $\alpha^{-1}$ to $\zeta_s$ we see that the potential monotonically increases until it diverges in the limit of the singularity. As such we are guaranteed a unique solution for $
    V(\zeta_0) = E^2$, and hence the existence of a classical turning point.

There is one exception to this, the case when $s = a = b = 0$, specific to the motion along transverse null geodesics where the potential is everywhere zero. We see that our spacetime is not geodesically complete as transverse null rays can reach the singularity in a finite proper time.

We conclude that for non-zero potentials a particle will arrive from $\mathcal{J}^-$ and necessarily fall through the horizon at $\zeta = \alpha^{-1}$. The particle will then continue towards the singularity to a minimum distance $\zeta_0$. At this point, it will then be reflected and continue off through the Killing horizon into a second dynamic spacetime towards $\mathcal{J}^+$. The only causal geodesics which do not follow these trajectories are those for which $V(\zeta)=0$. These are precisely the transverse null geodesics which fall through the horizon from $\mathcal{J}^-$ and straight into the singularity.

\subsubsection{Proper Acceleration}
Seeing that all timelike geodesics are repelled by the singularity, it is interesting to also study the acceleration of massive particles at rest within the static patch of the spacetime.

Particles at rest follow orbits of the stationary Killing vector field $k^\mu$ with a proper velocity defined by
\begin{equation*}
    u^\mu = \frac{k^\mu}{\sqrt{-k^2}},
\end{equation*}
where the normalisation has been chosen such that $u^2 = -1$. From this, the proper acceleration can be found
\begin{equation}
\label{eq:propacc}
    A^\mu = u^\nu \nabla_\nu u^\mu = \frac{1}{2} \partial^\mu \log(-k^2).
\end{equation}
The metric for the three- and four-charge solution is in the standard static form
\begin{equation}
    ds^2 = - f(\zeta) d\eta^2 + f(\zeta)^{-1} d\zeta^2 + g(\zeta) (dx^2 + dy^2).
\end{equation}
The Killing vector is given by
\begin{equation*}
    k^\mu = (1,0,0,0) \qquad \Rightarrow \qquad k^2 = -f(\zeta),
\end{equation*}
allowing us to calculate the proper acceleration of a massive particle at rest
\begin{equation*}
    A^\mu = \half g^{\mu \nu} \partial_\nu \log \left( f(\zeta) \right).
\end{equation*}
For these symmetric, static solutions the metric depends only on one coordinate, $\zeta$, and the only non zero component of the proper acceleration is
\begin{equation*}
        A^\zeta = \half f(\zeta)
 \partial_\zeta \log \left( f(\zeta) \right)
 = \half \partial_\zeta f(\zeta) .
\end{equation*}
For the case of the three- and four-charge solutions we have
\begin{equation*}
    f(\zeta) = \frac{W(\zeta)}{\Ham(\zeta)},
\end{equation*}
and so the proper acceleration is found to be
\begin{equation}
A^\zeta = - \frac{\alpha \Ham + W \partial_\zeta \Ham}{\Ham^2}.
\end{equation}
As the functions $W$, $\Ham$ and $\partial_\zeta \Ham$ are everywhere positive in the static region of spacetime we see that a particle at rest always experiences a force repelling it from the singularity. We remark that the qualitative behaviour of geodesics and
Killing orbits is the same as for the interior region of the non-extremal
Reissner-Nordstr\"om solution (usually called region $III$). Moreover in both 
cases, this static interior region is related by 
horizons both in the past and in the future to regions where the Killing vector
field becomes spacelike. The main difference is that the Reissner-Nordstrom solution
has a third type of region (usually called region $I$) which is static and asymptotically 
flat. We will come back to this comparison in Section \ref{sect:conclusions}.

%
%
\section{Dimensional Lifting of the Cosmological STU Solution}

To better understand the physical origin of the four-charge solution, we turn our attention to finding consistent higher dimensional embeddings. This is motivated by the success of \cite{Dempster:2016} which offered a new understanding of the Nernst solution \cite{Dempster:2015} by using a five-dimensional embedding. We are further motivated by the work of \cite{Cornalba:2003kd} and comments made by \cite{Burgess:2002vu} which link cosmological solutions to higher dimensional theories reduced on orientifolds.

We remark that 
the cosmological solution of the STU model found in  \cite{Fre:2008zd} can be shown
to describe part of the space-time of our four-charge solution, using different
coordinates.\footnote{The explicit relation between their solution and ours is complicated, and we will not
need it in the following.} Their solution covers part of our dynamic patch and does not
include the Killing horizon. The higher-dimensional interpretation of their
solution was through a lift from $4D$ to $10D$  on the orientifold 
$K3 \times T^2 / \mathbb{Z}_2$. We will instead consider simpler, 
toroidal lifts, which will allow to relate our cosmological solution to 
black hole solutions of the STU model and to make contact with six-dimensional
BPS solutions. We will follow
 the oxidation prescription of \cite{Chow:2013gba} to write down consistent truncated string/M-theory Lagrangians and their corresponding metric and gauge field content.

This section is organised as follows: first, we rewrite our Lagrangian \eq{4dlag} in a form that allows a direct
comparison with \cite{Chow:2013gba}. We then uplift our non-extremal planar solutions of the STU model
from $D=4$ to $D=5,6,10,11$, expressing our solutions as embedded into truncations of string-/M-theory.
Upon taking the extremal limit of the $4D$ solution, we make contact with
well-known brane configurations in string/M-theory models. Additional fine-tuning of the $4D$
electric charges is shown to make its $6D$ uplift supersymmetric in addition to being extremal.
This is an interesting result as we did not utilise Killing spinor equations and therefore the existence of supersymmetric limit was
not guaranteed.

\subsection{Rewriting the Lagrangian for Uplift}
\label{sec:4Dlag}
Our starting point is the Lagrangian \eq{4dlag}, repeated here for reference
\begin{equation*}
 e_4^{-1} \La = -\frac{1}{2}R - g_{A\bar{B}} \partial_\mu z^A \partial^\mu \bar{z}^B + \frac{1}{4} \I_{IJ} F^I_{\mu \nu} F^{J|\mu \nu} + \frac{1}{4} \cR_{IJ} F^I_{\mu \nu} \tilde{F}^{J|\mu \nu}.
\end{equation*}
Explicit expressions for the gauge couplings are obtained from the prepotential $F(X)$ using
standard special geometry formulae. We use the same conventions as \cite{Cortes:2009cs}.
When imposing the `purely imaginary' conditions on the scalars the gauge couplings take the form:
\begin{equation}
\label{eq:stucoup1}
 \cR_{IJ} = 0, \qquad  \I_{IJ} = \text{diag}\left( - s t u, -\frac{ t u}{s}, -\frac{s u}{t}, -\frac{ s t}{u} \right),
 \end{equation}
 \begin{equation}
 \label{eq:stucoup2}
 g_{A\bar{B}} = \text{diag}\left( \dfrac{1}{4s^2}, \dfrac{1}{4t^2}, \dfrac{1}{4u^2}\right),
\end{equation}
where
\begin{equation*}
 s = -\text{Im}(z^1) , \quad t = -\text{Im}(z^2), \quad u = -\text{Im}(z^3) .
\end{equation*}
When evaluating the couplings $\I_{IJ}$ on our solution, we find, using the $\zeta$-coordinate system \eq{4cmetricthing}
\begin{equation*}
    \I_{00}^2 = \frac{\Ham_0^3}{\Ham_1 \Ham_2 \Ham_3}, \qquad \I_{11}^2 = \frac{\Ham_0 \Ham_2 \Ham_3}{\Ham_1^3}, \qquad \I_{22}^2 = \frac{\Ham_0 \Ham_1 \Ham_3}{\Ham_2^3}, \qquad \I_{33}^2 = \frac{\Ham_0 \Ham_1 \Ham_2}{\Ham_3^3}.
\end{equation*}
After redefining our scalars
\begin{equation*}
 s = e^{-\phi_1}, \qquad t = e^{-\phi_2}, \qquad u = e^{-\phi_3},
\end{equation*}
the Lagrangian takes the following form
\begin{equation}
\label{eq:stulag}
 \begin{aligned}
 e_4^{-1} \La = -\frac{1}{2}R - \frac{1}{4} \partial_\mu \phi_i \partial^\mu \phi_i
 - \frac{1}{4} e^{-\phi_1 - \phi_2 - \phi_3} \left[ (F^0)^2 + e^{2 \phi_A } (F^A)^2 \right] ,
 \end{aligned}
\end{equation}
\noindent where we sum over $i=1,2,3$. Using the STU couplings (\ref{eq:stucoup1}) we can evaluate the scalars $\varphi_i$
on our solution and thus express them
as functions of $\zeta$
\begin{equation*}
    e^{2\varphi_1} = \frac{\I_{11}}{\I_{00}} = \frac{\Ham_2 \Ham_3}{\Ham_0 \Ham_1}
    , \quad e^{2\varphi_2} = \frac{\I_{22}}{\I_{00}} = \frac{\Ham_1 \Ham_3}{\Ham_0 \Ham_2}
    , \quad e^{2\varphi_3} = \frac{\I_{33}}{\I_{00}} =\frac{\Ham_1 \Ham_2}{\Ham_0 \Ham_3} \;.
\end{equation*}

To embed our solution into higher dimensions, we will use
various ans\"atze given in \cite{Chow:2013gba} as we obtain the results for 10D and 11D solutions via 6D and 5D solutions respectively. The
relevant truncation of the $4D$ STU Lagrangian given in \cite{Chow:2013gba} is
\begin{equation}
\label{eq:4dchow}
     e_4^{-1} \La_4 = -R - \frac{1}{2} \partial_\mu \varphi_i \partial^\mu \varphi_i
 - \frac{1}{4} e^{-\varphi_1 - \varphi_2 - \varphi_3} \left[ (\bF^4)^2 + e^{2 \varphi_i } (\tilde{\bF}_i)^2 \right] .
\end{equation}
This is related to our Lagrangian \eq{stulag} by an overall factor of 2, together with the following rescaling of the gauge fields and the scalars
\begin{equation*}
    F^0 = \frac{1}{\sqrt{2}} \bF^4, \qquad F^A = \frac{1}{\sqrt{2}} \tilde{\bF}_i, \qquad \phi_i = \varphi_i .
\end{equation*}
We will need to keep track of these factors while oxidising and insert the exact values for the gauge fields into the ansatz of \cite{Chow:2014cca}.

\subsection{Oxidation to Five Dimensions}
The STU model can be consistently embedded into five dimensions with the Lagrangian
\begin{equation}
\label{eq:5dchow}
    \La_5 = - R \star 1 - \half h_i^{-2} \left(\star dh_i \wedge dh_i + \star \tilde{\bF}_i \wedge \tilde{\bF}_i \right) + \tilde{\bF}_1 \wedge \tilde{\bF}_2 \wedge \tilde{\mathbb{A}}_3,
\end{equation}
where the five-dimensional scalars $h^i$ satisfy the constraint $h_1 h_2 h_3 = 1$. Using the Kaluza-Klein reduction ansatz
\begin{equation}
\label{eq:5dan}
    ds_5 = f^{-1} ds_4^2 + f^2 (dz_5 - \mathbb{A}^4)^2 \;,\qquad \qquad \tilde{\mathbb{A}}_{(5D)i} = \tilde{\mathbb{A}}_i \;,
\end{equation}
we obtain the four-dimensional Lagrangian \eq{4dchow} when we make the choice $f h_i= e^{-\varphi_i}$. The vector field $\mathbb{A}^4$ is the KK vector field, while the vector fields
$\mathbb{A}_i$ descend from the $5D$ vector fields.

Introducing new linear combinations $\sigma, \varphi, \lambda$ for the three independent real
four-dimensional scalars by
\begin{equation*}
     \varphi_1 = -\frac{2}{\sqrt{6}} \sigma + \frac{1}{\sqrt{3}} \lambda, \quad \varphi_2 = -\frac{1}{\sqrt{2}} \phi + \frac{1}{\sqrt{6}} \sigma + \frac{1}{\sqrt{3}} \lambda, \quad \varphi_3 = \frac{1}{\sqrt{2}} \phi + \frac{1}{\sqrt{6}} \sigma + \frac{1}{\sqrt{3}} \lambda,
\end{equation*}
the five-dimensional constrained scalars $h^i$ can be expressed in terms of 2 independent fields
\begin{equation*}
 h_1 = e^{2 \sigma/\sqrt{6}}, \qquad h_2 = e^{\phi/\sqrt{2} - \sigma/\sqrt{6}}, \qquad h_3 = e^{-\phi/\sqrt{2} - \sigma/\sqrt{6}} \; .
\end{equation*}
Combining these two relations we obtain
\begin{equation*}
    \begin{aligned}
        h_1 &= \exp \left(-\frac{2\varphi_1}{3} + \frac{\varphi_2}{3} + \frac{\varphi_3}{3} \right) = \left( \frac{\I_{22} \I_{33} }{\I^2_{11}} \right)^{\frac{1}{6}}, \\
        h_2 &= \exp \left(\frac{\varphi_1}{3} - \frac{2\varphi_2}{3} + \frac{\varphi_3}{3} \right) = \left( \frac{\I_{11} \I_{33} }{\I^2_{22}} \right)^{\frac{1}{6}}, \\
        h_3 &= \exp \left(\frac{\varphi_1}{3} + \frac{\varphi_2}{3} - \frac{2 \varphi_3}{3} \right) = \left( \frac{\I_{11} \I_{22} }{\I^2_{33}} \right)^{\frac{1}{6}} ,
    \end{aligned}
\end{equation*}
and expressing the gauge couplings in terms of the harmonic functions $\Ham_i$,
\begin{equation}
\label{hi_Harmonic}
        h_i = \frac{\Ham_i}{(\Ham_1 \Ham_2 \Ham_3)^{\frac{1}{3}}} \; ,
\end{equation}
allows us to write down the Kaluza-Klein scalar $f$ in terms of $\zeta$
\begin{equation}
\label{eq:5dkk}
    f = e^{-\varphi_i} h_i^{-1} = \left(\frac{\I_{00}^3}{\I_{11} \I_{22} \I_{33}} \right)^{\frac{1}{6}}
    = \left( \frac{ \Ham_0^3}{\Ham_1 \Ham_2 \Ham_3} \right)^{1/6} \;.
\end{equation}
\subsection*{Five-Dimensional Metric}
Using \eq{5dan} together with \eq{5dkk} and $\mathbb{A}^4 = \sqrt{2} A^0$, as well as
collecting common factors, we obtain the following five-dimensional metric for the uplift
of our four-charge solution:
\begin{equation}
\begin{aligned}
        ds^2_5 = (\Ham_1 \Ham_2 \Ham_3)^{-\frac{1}{3}} &\bigg[\Ham_0 dz_5^2 + \frac{\cW}{2\Ham_0}\left(\cW \frac{\gamma_0^2}{Q_0^2} + 1 \right) d\eta^2 - \frac{2 \cW \gamma_0}{\sqrt{2} Q_0} d\eta dz_5 \\ &+ 2\Ham_1 \Ham_2 \Ham_3 \left(- \frac{ d\zeta^2}{\cW} + dx^2 + dy^2 \right) \bigg]
\end{aligned}
\end{equation}

\subsection*{Five-Dimensional Gauge Potential}

To obtain expressions for the five-dimensional gauge potentials,
it is necessary to express all the gauge fields in our solution in terms of electric components. This
requires replacing the dual vector potentials $\tilde{A}_A$ by the `standard'
 vector potentials $A^A$. The associated field strength $\tilde{F}_A$ and $F^A$ are
 related by Hodge duality together with multiplication by inverse gauge coupling matrix
 \begin{equation*}
    F^A = -\I^{AB} \star \tilde{F}_B.
\end{equation*}
Using the form of the gauge potential found in \eq{new4dg}, standard calculations give their form
\begin{equation}
\label{eq:5dfield}
    F^A = -\I^{AB} \star \tilde{F}_B = - \frac{P^A}{2\sqrt{\gamma_0 \gamma_1 \gamma_2 \gamma_3}} dx \wedge dy.
\end{equation}
Integrating and relating to the gauge fields in the ansatz we obtain the form of the three $5D$ vector potentials
\begin{equation}
\label{eq:5dvec}
    \tilde{\mathbb{A}}_i = \sqrt{2} A^A = \mathfrak{p}_a \left(y dx - x dy \right), \qquad \mathfrak{p}_a = \frac{ P^a}{2 \sqrt{2 \gamma_0 \gamma_1 \gamma_2 \gamma_3}} .
\end{equation}
We will return to these gauge fields before uplifting the solution to six dimensions; when it will be necessary to Hodge dualise in 5D to obtain a two-form potential.

\subsection*{Extremal Limit}
We now investigate the effect of the $4D$ extremal limit defined in section \ref{sec:4Dext} for the following higher-dimensional lifts.\footnote{We note here that we use the same symbols $h_a$ for the integration 
constants in \eq{extlim} and the constrained five-dimensional scalars, as this allows us to match notation
with the literature on five-dimensional solutions. We trust that the reader will infer
from context which quantity is meant in a particular expression.} Just as in the $4D$ case, the horizon for the $5D$ solution is pushed out to $\zeta \rightarrow \infty$ and the static region takes up the entirety of our spacetime; in other words, the extremal limit results in a solution containing a naked singularity. Simplifying the metric functions using the expressions from \eq{extlim} we can write down the $5D$ line element in the form
\begin{equation*}
    ds^2_5 = (\Ham_1\Ham_2\Ham_3)^{-\frac{1}{3}} \left[d\eta dz_5 + \Ham_0 dz_5^2 + \Ham_1\Ham_2\Ham_3 (d\zeta^2 + dx^2 + dy^2) \right].
\end{equation*}
We will consider the extremal limit for each of the following uplifts in term as we further oxidise the STU model.

\subsection{Oxidation to Eleven Dimensions}
To uplift our solution to eleven dimensions, we start with the bosonic part of the $11D$ supergravity Lagrangian
\begin{equation*}
    \La_{11} = -R \star 1 - \half \star \mathcal{F} \wedge \mathcal{F} - \frac{1}{6} \mathcal{F} \wedge \mathcal{F} \wedge \mathcal{A},
\end{equation*}
where $\mathcal{A}$ is the three-form such that $\mathcal{F} =d\mathcal{A}$ is the four-form field strength. We can directly embed the $5D$ STU model into this theory through a Kaluza-Klein reduction on $T^6$ with the ansatz
\begin{equation*}
    ds_{11}^2 = ds_5^2 + h_1 (dy_1^2 + dy_2^2) + h_2 (dy_3^2 + dy_4^2) + h_3 (dy_5^2 + dy_6^2),
\end{equation*}
\begin{equation*}
    \mathcal{A} = \tilde{\mathbb{A}}_1 \wedge dy^1 \wedge dy^2 + \tilde{\mathbb{A}}_2 \wedge dy^3 \wedge dy^4 + \tilde{\mathbb{A}}_3 \wedge dy^5 \wedge dy^6.
\end{equation*}
In a consistent truncation to five-dimensional minimal supergravity, the volume of the torus corresponds
to a scalar in a hypermultiplet, while its shape is encoded by scalars in vector multiplets.
This factorization imposes the condition $h_1h_2h_3=1$ on the scalars $h_i$. By restricting
to backgrounds where hypermultiplets are trivial, we can consistently truncate out the
hypermultiplets and remain with the five-dimensional STU model with two vector multiplets.

We now combine this 5D/11D lift with the previous 4D/5D lift.
In our four-dimensional solution we can express the $h_i$ as functions of $\zeta$ through the
harmonic functions $\Ham_i$, see (\ref{hi_Harmonic}).
The three-form gauge potential is found directly from the components of \eq{5dvec}. Thus the full line element for the $11D$ lift of the non-extremal planar solution to the $4D$ STU model is
\begin{equation}
\label{eq:11Dmet}
\begin{aligned}
        ds^2_{11} = (\Ham_1 \Ham_2 \Ham_3)^{-\frac{1}{3}} &\bigg[\Ham_0 dz_5^2 + \frac{\cW}{2\Ham_0}\left(\cW \frac{\gamma_0^2}{Q_0^2} + 1 \right) d\eta^2 - \frac{\cW \gamma_0}{\sqrt{2} Q_0} d\eta dz_5 \\ &+ 2\Ham_1 \Ham_2 \Ham_3 \left(- \frac{ d\zeta^2}{\cW} + dx^2 + dy^2 \right) \\
        &+ \Ham_1 (dy_1^2 + dy_2^2) + \Ham_2 (dy_3^2 + dy_4^2) + \Ham_3 (dy_5^2 + dy_6^2)
        \bigg] \;.
\end{aligned}
\end{equation}

\subsection*{Eleven-Dimensional Extremal Limit}
By again substituting in the limit \eq{extlim} we can write down \eq{11Dmet} in the extremal limit to find
\begin{equation}
\begin{aligned}
    ds_{11}^2 = &(\Ham_1\Ham_2\Ham_3)^{-\frac{1}{3}} \big[ d\eta dz_5 + \Ham_0 dz_5^2 + \Ham_1\Ham_2\Ham_3 (d\zeta^2 + dx^2 + dy^2) \\
    &+ \Ham_1(dy_1^2 + dy_2^2) + \Ham_2(dy_3^2 + dy_4^2) + \Ham_3(dy_5^2 + dy_6^2) \big]
\end{aligned}
\end{equation}
This looks like a standard BPS solution of eleven-dimensional supergravity,
a configuration of three stacks of M5-branes, encoded by $\Ham_1, \Ham_2,\Ham_3$
which triple intersect over a string, and with a gravitational wave, encoded by $\Ham_0$,
superimposed along the string \cite{Tseytlin:1996bh}. Compactification on $T^6\times S^1$ gives
rise to four-charged BPS black holes when the branes are delocalised along $y^1, \ldots, y^6$
but localised in the remaining three spacelike directions \cite{Behrndt:1996jn}. In our solutions
the M5-branes have in addition been delocalised in two of the non-compact directions, giving
rise to planar rather than spherical symmetry.

\subsection{Oxidation to Six Dimensions}

Lifting the four-dimensional Lagrangian to six dimensions by extending the 4D/5D lift
requires a tweak of the five-dimensional Lagrangian, namely
to Hodge-dualise one of the three vector potentials into a two-form $B$. The reason is that the six-dimensional supergravity is chiral and both the supergravity multiplet and tensor multiplets contain self-dual or anti-self-dual tensor fields which do not admit a standard Lagrangian
description. However, in supergravity coupled to one tensor multiplet (plus vector and hypermultiplets),
one self-dual and one anti-self-dual tensor combine into an unconstrained tensor, which allows
a standard Lagrangian description. String compactifications to six dimensions are of this type,
with the tensor field descending from the ten-dimensional Kalb-Ramond field.

Matching our conventions with the work of \cite{Chow:2013gba} we define the three-form from the dualisation of the two-form field strength in five dimensions
\begin{equation}
\label{eq:hodtra}
    \tilde{\bF}_3 = d\tilde{\mathbb{A}}_3 = -h_1^{-2} h_2^{-2} \star_5 \mathbb{H} \;.
\end{equation}
Making this transformation and substituting into the Lagrangian results in
\begin{equation}
\begin{aligned}
        \label{eq:5dchow2}
    \La_5 = &- R \star 1 - \half h_i^{-2} \star dh_i \wedge dh_i +\half h_1^{-2} \star \tilde{\bF}_1 \wedge \tilde{\bF}_1 \\
    &+ \half h_2^{-2} \star \tilde{\bF}_2 \wedge \tilde{\bF}_2 - \half h_1^{-2} h_2^{-2} \star \mathbb{H} \wedge \mathbb{H} \;.
\end{aligned}
\end{equation}
We can now use the results of \cite{Chow:2014cca}, and first work with the new three-form field strength in five dimensions in terms of $\zeta$.

\subsection*{Dualization of the Five-Dimensional Gauge Field}
Taking the Hodge dual of \eq{hodtra} again we find the three-form
\begin{equation*}
        \star_5 \mathbb{H} = - h^2_1 h^2_2 \tilde{\bF}_3 \;,\qquad
        \star_5 \star_5 \mathbb{H} = - \star_5(h^2_1 h^2_2 \tilde{\bF}_3) \;, \qquad
        \mathbb{H} = \star_5(h^2_1 h^2_2 \tilde{\bF}_3) \;,
\end{equation*}
where we have used that for a $k$-form $\omega$ in $n$ dimensions in the Lorentzian signature $\star \star \omega = (-1)^{k(n-k)+1} \omega$. Substituting in \eq{5dfield} together with:
\begin{equation*}
    \sqrt{-g_5} = 2 (\Ham_1 \Ham_2 \Ham_3)^{\frac{2}{3}}
, \qquad \epsilon_{\eta \zeta x y z_5} = 1 \;,
\end{equation*}
\begin{equation*}
    h_1^2 h_2^2 = h_3^{-2} = (\Ham_1 \Ham_2)^{\frac{2}{3}} \Ham_3^{-\frac{4}{3}} \;, \qquad g^{xx} = g^{yy} = \half(\Ham_1 \Ham_2 \Ham_3)^{-\frac{2}{3}} \;,
\end{equation*}
we find that the three-form is
\begin{equation*}
        \mathbb{H} = -\left( \frac{ \mathfrak{p}_3}{\Ham_3^2} \right) d\eta \wedge d\zeta \wedge dz_5 \;.
\end{equation*}

\subsection*{Lift to Six Dimensions}

The six-dimensional Lagrangian is
\begin{equation*}
    \La_6 = - R\star 1 - \half \star d\phi \wedge d\phi - \half e^{-\sqrt{2}\phi} \star H \wedge H,
\end{equation*}
where $H = dB$ is a three-form field strength. The reduction ansatz \cite{Chow:2013gba} which
reproduces our five-dimensional Lagrangian (\ref{eq:5dchow2}) is
\begin{equation*}
    ds^2_6 = e^{\sigma/\sqrt{6}} ds_5^2 + e^{-3\sigma/\sqrt{6}} (dz_6 + \tilde{\mathbb{A}}_1)^2 \;,\qquad B_{(6D)} = B + \tilde{\mathbb{A}}_2 \wedge (dz_6 + \tilde{\mathbb{A}}_1),
\end{equation*}
with the field strengths decomposed as
\begin{equation*}
H_{(6D)} = \mathbb{H} + \tilde{\bF}_2 \wedge (dz_6 + \tilde{\mathbb{A}}_1) \;,    \qquad \mathbb{H} = dB - \tilde{\mathbb{A}}_2 \wedge \tilde{\bF}_1 ,\qquad \tilde{\bF}_i = d \tilde{\mathbb{A}}_i \;.
\end{equation*}
We see that from our parameterisation of the $h_i$ we can write the $6D$ Kaluza-Klein scalar as
\begin{equation*}
    e^{\sigma/\sqrt{6}} = \sqrt{h_1} = \left( \frac{\Ham_1^3}{\Ham_1 \Ham_2 \Ham_3} \right)^{\frac{1}{6}} \;.
\end{equation*}

We are now in the position to combine these results to write down the $6D$ metric for our embedded solution:
\begin{equation*}
\begin{aligned}
        ds^2_6 = (\Ham_2 \Ham_3)^{-\frac{1}{2}} &\bigg[\Ham_0 dz_5^2 + \frac{\cW}{2\Ham_0}\left(\cW \frac{\gamma_0^2}{Q_0^2} + 1 \right) d\eta^2 - \frac{2\cW \gamma_0}{\sqrt{2} Q_0} d\eta dz_5 \\ &+ 2\Ham_1 \Ham_2 \Ham_3 \left(- \frac{ d\zeta^2}{\cW} + dx^2 + dy^2 \right) \bigg] + \frac{(\Ham_2 \Ham_3)^\half}{\Ham_1} (dz_6 + \tilde{\mathbb{A}}_1)^2 \;,
\end{aligned}
\end{equation*}
where the determinant of the metric is
\begin{equation*}
    \sqrt{-g_6} = 2 \Ham_1 \sqrt{\Ham_2 \Ham_3} \;.
\end{equation*}
The piece containing the gauge field $\mathbb{A}_1$ can be expanded
\begin{equation*}
\begin{aligned}
        (dz_6 + \tilde{\mathbb{A}}_1)^2 &= (dz_6 + (\tilde{\mathbb{A}}_1)_x dx + (\tilde{\mathbb{A}}_1)_y dy)^2 \\
        &= (dz_6 + \mathfrak{p}_1(ydx - xdy))^2 \;.
\end{aligned}
\end{equation*}

\subsection*{Six-Dimensional Gauge Fields}
We now take the gauge fields and express them as a function of the $6D$ coordinates. We see that for the two remaining one-form potentials nothing has been changed compared to the lower dimensional solutions
\begin{equation*}
    \tilde{\mathbb{A}}_1 = \sqrt{2} A^1, \qquad \tilde{\mathbb{A}}_2 = \sqrt{2} A^2 \;.
\end{equation*}
The three-form $H$ is found from two pieces
\begin{equation*}
    H_{(6D)} = \mathbb{H} + \tilde{\mathbb{F}}_2 \wedge ( dz_6 + \tilde{\mathbb{A}}_1) \;.
\end{equation*}
This is simplified as the term
\begin{equation*}
    \tilde{\mathbb{F}}_2 \wedge \tilde{\mathbb{A}}_1 = 2 \mathfrak{p}_2 dx \wedge dy \wedge \mathfrak{p}_1 \left(y dx - xdy \right) = 0,
\end{equation*}
is zero due to anti-symmetry. Using the work from the $5D$ calculations the $6D$ three-form field strength is therefore given by:
\begin{equation*}
\begin{aligned}
        H_{(6D)} = -\left( \frac{ \mathfrak{p}_3}{\Ham_3^2} \right) d\eta \wedge d\zeta \wedge dz_5
 -( 2\mathfrak{p}_2) dx \wedge dy \wedge dz_6 \;.
\end{aligned}
\end{equation*}

\subsection*{Six-Dimensional Extremal Limit}
Taking the limit \eq{extlim} the six-dimensional line element is given by
\begin{equation}
\label{eq:6dmet}
    ds_6^2 = \sqrt{\frac{\Ham_2}{\Ham_3}} \left[\Ham_2^{-1} \left(\Ham_0 dz_5^2 + dz_5d\eta \right) + \Ham_3 \Ham_1 (d\zeta^2 + dx^2 + dy^2) + \Ham_3 \Ham_1^{-1}(dz_6 +\tilde{\mathbb{A}}^1)^2 \right].
\end{equation}
The three-form in this limit is given by
\begin{equation*}
        H_{(6D)} = -\left( \frac{ \mathfrak{p}_3}{\Ham_3^2} \right) d\eta \wedge d\zeta \wedge dz_5
 -( 2\mathfrak{p}_2) dx \wedge dy \wedge dz_6, \qquad \mathfrak{p}_a = \frac{ P^a}{2 \sqrt{2 Q_0 P^1 P^2 P^3}}.
\end{equation*}

\subsection{Oxidation to Ten Dimensions}
\label{sec:to10d}
The six-dimensional STU model is a consistent truncation of the reduction of IIB supergravity
on $T^4$.
To lift our solution, we only need to include the overall volume of the $T^4$ as a modulus
\begin{equation*}
    ds_{10}^2 = ds_6^2 + e^{\phi / \sqrt{2} }(dy_1^2 + dy_2^2 + dy_3^2 + dy_4^2), \qquad \Phi = \frac{\phi}{\sqrt{2}}, \qquad C \equiv B.
\end{equation*}
The expression for the six-dimensional dilaton $\phi$ in terms of $\zeta$ is
\begin{equation*}
    e^{\sqrt{2} \phi} = \frac{h_2}{h_3} = \left(\frac{\I_{33} }{\I_{22}}\right)^{\frac{1}{2}} \quad \Rightarrow \quad e^{\phi / \sqrt{2} } = \left(\frac{\I_{33} }{\I_{22}}\right)^{\frac{1}{4}} = \sqrt{\frac{\Ham_2}{\Ham_3}} \;.
\end{equation*}
All other data follow straight from the six-dimensional solutions.
The ten-dimensional dilaton is given by
\begin{equation*}
    \Phi = \half \log \left(\frac{\Ham_2}{\Ham_3} \right).
\end{equation*}
The ten-dimensional line element is given by
\begin{equation*}
\begin{aligned}
        ds^2_{10} = \sqrt{\frac{\Ham_2}{\Ham_3}} &\bigg[\Ham_0 \Ham_2^{-1} dz_5^2 + \frac{\cW}{2\Ham_0 \Ham_2}\left(\cW \frac{\gamma_0^2}{Q_0^2} + 1 \right) d\eta^2 - \frac{\cW \gamma_0}{\sqrt{2} Q_0 \Ham_2} d\eta dz_5 \\ &+ 2\Ham_1 \Ham_3 \left(- \frac{ d\zeta^2}{\cW} + dx^2 + dy^2 \right) + \frac{\Ham_3}{\Ham_1} (dz_6 + \tilde{\mathbb{A}}_1)^2 \\
        &+ dy_1^2 + dy_2^2 + dy_3^2 + dy_4^2 \bigg].
\end{aligned}
\end{equation*}

\subsection*{Ten-Dimensional Extremal Limit}
Uplifting the extremal 6D solution using the same methods as (\ref{sec:to10d}) we find that the line element is
\begin{equation}
\begin{aligned}
        ds^2_{10} = &\sqrt{\frac{\Ham_2}{\Ham_3}} \bigg[\Ham_2^{-1} \left(\Ham_0 dz_5^2 + dz_5d\eta \right) + \Ham_3 \Ham_1 (d\zeta^2 + dx^2 + dy^2) \\ &+ \Ham_3 \Ham_1^{-1}(dz_6^2 + \tilde{\mathbb{A}}^1)^2 + dy_1^2 + dy_2^2 + dy_3^2 + dy_4^2\bigg].
\end{aligned}
\end{equation}
which is  the intersection of a D1 and D5 brane with momentum along the common direction and a Taub-NUT space.

%
%
\section{Supersymmetry in Six Dimensions}

Supersymmetric solutions of six-dimensional supergravity have been classified in detail. The first classification of supersymmetric solutions in the minimal ungauged six-dimensional theory, with a self-dual three-form, was constructed in \cite{Gutowski:2003rg}.
Following on from this, the supersymmetric solutions of six-dimensional $U(1)$, and $SU(2)$ gauged supergravity were classified in \cite{Cariglia:2004kk}.
This analysis was done using the spinor bilinears method.
Supersymmetric solutions of more general theories coupled to arbitrary vector and tensor multiplets were classified using spinorial geometry methods in \cite{Akyol:2010iz}; see also \cite{Akyol:2012cq, Akyol:2013ana, Cano:2018wnq, Lam:2018jln}.
These classifications have been used to find many new examples of solutions, and we shall show that in a certain limit, the 6D uplift solution we have constructed satisfies the necessary and sufficient conditions for supersymmetry.

\subsection{Conditions Required for Supersymmetry}

We now turn our attention to the $6D$ uplift of our solution and test to see whether there is a configuration of integration constants such that the solution is supersymmetric.
In this particular case, the theory of interest is the $U(1)$ gauged supergravity
whose supersymmetric solutions were classified in \cite{Cariglia:2004kk}, in the special case for which the $U(1)$ gauge parameter
is set to zero.
The bosonic content of this theory is the metric $g$, a real three-form $G$, and a dilaton $\phi$. The geometry of these solutions was also considered in \cite{Bena:2011dd}. Before considering the $6D$ uplift in detail, we first summarise the necessary and sufficient conditions on the bosonic fields in order for a generic solution of this theory to be supersymmetric.

The metric for the supersymmetic solutions is given by
\begin{equation}
\label{eq:benamet}
    ds^2_6 = -2H^{-1} (dv + \beta) (du + \omega + \half \mathcal{F}(dv+\beta)) + H ds_4^2.
\end{equation}
The metric for the four-dimensional base space $\mathcal{B}$ is written as
\begin{equation}
    ds^2_4 = h_{mn}dx^m dx^n,
\end{equation}
with $\beta=\beta_m dx^m$ and $\omega = \omega_m dx^m$ regarded as one-forms on $\mathcal{B}$. The vector ${\partial \over \partial u}$ corresponds to a Killing spinor bilinear
and the Killing spinor equations imply that this vector is an isometry, and moreover that
the Lie derivative of the three-form $G$ and the dilaton $\Phi$ with respect to ${\partial \over \partial u}$ vanish. However, in general, the metric, the three-form and the dilaton
may depend on the $v$ and the $x^m$ coordinates.

Analysis of the algebraic properties of the spinor bilinears by considering the Fierz identities implies that there
are 3 anti-self-dual two-forms on the base ${\cal{B}}$: $J^{(A)}$, $A=1,2,3$, which
satisfy the algebra of the imaginary unit quaternions; ${\cal{B}}$ therefore admits an almost hyper-K\"ahler structure. In addition, the
gravitino Killing spinor equations imply that
\begin{eqnarray}
\label{almosthk}
{\tilde{d}} J^{(A)} = \partial_v \bigg( \beta \wedge J^{(A)} \bigg),
\end{eqnarray}
where ${\tilde{d}}$ denotes the exterior derivative restricted to surfaces of constant $u$ and $v$;
and $\partial_v$ denotes the Lie derivative with respect to ${\partial \over \partial v}$.
It is also useful to define the differential operator $D$ by
\begin{eqnarray}
D \chi = {\tilde{d}} \chi - \beta \wedge \partial_v \chi,
\end{eqnarray}
where $\chi$ is a $u$-independent differential form on ${\cal{B}}$.
Then supersymmetry implies that
\begin{eqnarray}
D \beta = \star_4 D \beta,
\end{eqnarray}
where $\star_4$ denotes the Hodge dual on ${\cal{B}}$. This exhausts the conditions on the
geometry obtained from the gravitino Killing spinor equation. It remains to consider
the conditions on the fluxes.

The Killing spinor equations determine the components of the three-form $G$ as
\begin{equation}
\begin{aligned}
    e^{\sqrt{2} \Phi } G &= \tfrac{1}{2} \star_4 \left(DH + H \partial_v \beta - \sqrt{2} H D \Phi \right) \\
    &- \tfrac{1}{2} e^+ \wedge e^- \wedge \left(H^{-1} DH + \partial_v \beta + \sqrt{2} D \Phi \right) \\
    &- e^+ \wedge (-H \psi + \tfrac{1}{2} (D\omega)^- -K) + \tfrac{1}{2} H^{-1} e^-\wedge D \beta,
\end{aligned}
\label{susythree}
\end{equation}
where $K$ is a self-dual form on the base ${\cal{B}}$, $\psi$ is expressed as
\begin{eqnarray}
\psi = {1 \over 16} H \epsilon^{ABC} J^{(A) mn} (\partial_v J^{(B)})_{mn} J^{(C)},
\label{psidef}
\end{eqnarray}
and we have adopted the null basis
\begin{eqnarray}
e^+ = H^{-1} \big(dv+\beta), \qquad e^-= du+\omega +{1 \over 2} {\cal{F}} H e^+, \qquad
e^a= H^{1 \over 2} {\tilde{e}}^a,
\end{eqnarray}
in which the metric is
\begin{eqnarray}
ds^2_6 =-2 e^+ e^- + \delta_{ab} e^a e^b,
\end{eqnarray}
and the basis ${\tilde{e}}^a= {\tilde{e}}^a{}_m dx^m$ is a basis for the base ${\cal{B}}$.

On imposing the Bianchi identity $dG=0$, the following conditions are obtained
\begin{eqnarray}
\label{bian61}
D \bigg(H^{-1} e^{\sqrt{2} \Phi} \big(K-H {\cal{G}}-H \psi \big) \bigg)
+{1 \over 2}\partial_v \star_4 \bigg(D \big(H e^{\sqrt{2} \Phi}\big)+H e^{\sqrt{2}\Phi} \partial_v \beta \bigg)
\nonumber \\
-H^{-1} e^{\sqrt{2} \Phi} (\partial_v \beta) \wedge \big(K-H {\cal{G}}-H \psi \big)=0,
\end{eqnarray}
and
\begin{eqnarray}
\label{bian62}
-D \bigg(H^{-1} e^{-\sqrt{2} \Phi} \big(K+H {\cal{G}}+H \psi \big) \bigg)
+{1 \over 2}\partial_v \star_4 \bigg(D \big(H e^{-\sqrt{2} \Phi}\big)+H e^{-\sqrt{2}\Phi} \partial_v \beta \bigg)
\nonumber \\
+H^{-1} e^{-\sqrt{2} \Phi} (\partial_v \beta) \wedge \big(K+H {\cal{G}}+H \psi \big)=0,
\nonumber \\
\end{eqnarray}
where
\begin{eqnarray}
{\cal{G}}= {1 \over 2H} \bigg( \big(D \omega\big)^+ +{1 \over 2}{\cal{F}} D \beta \bigg),
\end{eqnarray}
and $(D \omega)^\pm$ denote the self-dual and anti-self dual parts of $D \omega$.

The gauge field equations, $d (e^{2 \sqrt{2} \Phi} \star_6 G)=0$ also imply the following conditions
\begin{eqnarray}
\label{gauge61}
D \star_4 \bigg(D \big(H e^{\sqrt{2} \Phi} \big) + H e^{\sqrt{2} \Phi} \partial_v \beta \bigg)
= 2 H^{-1} e^{\sqrt{2} \Phi} \big(K-H {\cal{G}} \big) \wedge D \beta,
\end{eqnarray}
and
\begin{eqnarray}
\label{gauge62}
D \star_4 \bigg(D \big(H e^{-\sqrt{2} \Phi} \big) + H e^{-\sqrt{2} \Phi} \partial_v \beta \bigg)
= -2 H^{-1} e^{-\sqrt{2} \Phi} \big(K+H {\cal{G}} \big) \wedge D \beta,
\end{eqnarray}

As noted in \cite{Cariglia:2004kk}, imposing these conditions implies that the dilaton field equation is automatically satisfied, and also all but one component of the Einstein field equations also hold. The remaining $++$ component of the Einstein equations must be imposed
as an additional condition. On defining
\begin{eqnarray}
L=\partial_v \omega +{1 \over 2} {\cal{F}} \partial_v \beta -{1 \over 2}D {\cal{F}},
\end{eqnarray}
this component of the Einstein equation is given by
\begin{eqnarray}
\star_4 D \star_4 L &=&{1 \over 2} h^{mn} \partial_v^2 \big(H h_{mn} \big)
+{1 \over 2} \partial_v \big(H h^{mn} \big) \partial_v \big(H h_{mn} \big)\nonumber \\
&-& {1 \over 2} H^{-2} \big(D \omega +{1 \over 2} {\cal{F}} D \beta \big)^2 -2 L^m ({\partial_v} \beta)_m+2 H^2 (\partial_v \Phi)^2
\nonumber \\
&+& 2 H^{-2} \big(K-H \psi +{1 \over 2} (D \omega)^- \big)^2.
\label{nulleinst}
\end{eqnarray}
where we adopt the convention that if $X$ is a two-form on ${\cal{B}}$ then $X^2={1 \over 2} X_{mn} X^{mn}$.

\subsection{Matching the Solutions}
We begin by taking the $\alpha \rightarrow 0$ limit; in four dimensions we can think of this limit as taking the blackening factor to zero and thus, being associated with extremality.

In this limit the resulting metric was found to be \eq{6dmet} and for convenience, we repeat the full expression for the six-dimensional three-form
\begin{equation}
\begin{aligned}
        H_{(6D)} = -\frac{P^3}{2\Ham_3^2 \sqrt{2 Q_0 P^1 P^2 P^3}} d\eta \wedge d\zeta \wedge dz_5 - \frac{P^2}{\sqrt{2 Q_0 P^1 P^2 P^3}} dx \wedge dy \wedge dz_6.
\end{aligned}
\label{threeform6}
\end{equation}
Comparing our metric with the metric \eq{benamet} we extract a four-dimensional base space:
\begin{equation}
    ds^2_6 = (\Ham_2 \Ham_3)^{-\half} dz_5 (d\eta + \Ham_0 dz_5) + (\Ham_2 \Ham_3)^{\half} ds_4^2 ,
\end{equation}
in the form
\begin{equation}
\label{eq:4dbase}
    ds^2_4 = \Ham_1 ( d\zeta^2 + dx^2 + dy^2) + \Ham_1^{-1} (dz_6 + \mathbb{A}^1)^2 .
\end{equation}
Direct comparison to \eq{benamet} shows that we should make the following identifications:
\begin{equation*}
 \beta = \omega = 0, \qquad H = \sqrt{\Ham_2 \Ham_3}, \qquad \mathcal{F} = \Ham_0, \qquad dv = dz_5, \qquad 2du = d\eta    ,
\end{equation*}
with all components of the metric and three-form independent of the $v$ coordinate. The basis vectors are given as:
\begin{equation*}
    \begin{aligned}
        e^+ = (\Ham_2 \Ham_3)^{-\half} dz_5, \qquad e^- = \half d\eta + \half \Ham_0 dz_5, \qquad e^a = (\Ham_1 \Ham_2)^{\frac{1}{4}}\tilde{e}^a_m dx^m .
    \end{aligned}
\end{equation*}

We begin by looking more closely at the base space
\eq{4dbase}
\begin{equation*}
    ds^2_4 = \Ham_1 ( d\zeta^2 + dx^2 + dy^2) + \Ham_1^{-1} (dz_6 + \mathbb{A}^1)^2 ,
\end{equation*}
which has a set of basis vectors:
\begin{equation*}
    \begin{aligned}
        e^1 &= \Ham_1^{\half} d\zeta, \qquad e^2 &&= \Ham_1^{\half} dx ,\\
        e^3 &= \Ham_1^{\half} dy, \qquad e^4 &&= \Ham_1^{-\half} (dz_6 + \mathbb{A}^1) ,\\
    \end{aligned}
\end{equation*}
with
\begin{equation}
\Ham_1 = h_1 + P^1 \zeta, \qquad \mathbb{A}^1={P^1 \over 2 \sqrt{2 Q_0 P^1 P^2 P^3}} \big(y dx-x dy \big).
\end{equation}
As the solution is independent of the $v$ coordinate, the condition ({\ref{almosthk}}) implies that the base is hyper-K\"ahler. In particular, we require that the Ricci scalar of the base must vanish, which imposes the following condition
\begin{equation}
2 Q_0 P^1 P^2 P^3 =1,
\end{equation}
which we can interpret as a condition for the integration constant
\begin{equation}
    Q_0 = \frac{1}{2 P^1 P^2 P^3} ,
\end{equation}
and so we see that the supersymmetric limit occurs by fine-tuning of the $4D$ electric charge or alternatively the KK momentum in $5/6D$.

Given this fine tuning condition, the base metric is then given by
\begin{equation}
ds^2_4 = \big(h_1 + P^1 \zeta\big)\bigg( d\zeta^2 + dx^2 + dy^2 \bigg)
+\big(h_1 + P^1 \zeta\big)^{-1} \bigg(dz^6 +{1 \over 2} P^1 \big(ydx-x dy\big) \bigg)^2.
\end{equation}
This metric is in the form of the Gibbons-Hawking instanton solution
\cite{Gibbons:1979zt, Gibbons:1987sp}
\begin{equation*}
ds^2_{GH} = U^{-1} (d\tau + \omega)^2 + U d\vec{x} \cdot d\vec{x}    ,
\end{equation*}
where $\tau=z^6$ is the direction corresponding to the tri-holomorphic isometry
${\partial \over \partial \tau}$ of the hyper-K\"ahler structure, and
$U=h_1+P^1 \zeta$ is a linear harmonic function of the Cartesian coordinates $\{\zeta, x, y \}$ on $\mathbb{R}^3$, and the one-form $\omega = dz^6 +{1 \over 2} P^1 \big(ydx-x dy\big) $ is a $U(1)$ connection on $\mathbb{R}^3$ which satisfies
\begin{equation}
    dU = \star_3 d \omega.
\end{equation}
This base space corresponds to a constant density planar distribution of Taub-NUT instantons. Moreover, the conditions imposed on the three-form given in
({\ref{susythree}}) are consistent with the three-form obtained from
the uplift in ({\ref{threeform6}}), on setting $K=0$ in ({\ref{susythree}}),
and also identifying
\begin{equation}
\Phi=-{1 \over 2 \sqrt{2}} \log \bigg({\Ham_2 \over \Ham_3} \bigg).
\end{equation}

We remark that the dilaton which appears in the classification of
\cite{Cariglia:2004kk}, which we have denoted by $\Phi$, differs from the
dilaton $\phi$ appearing in previous sections by a scaling
\begin{equation*}
    \Phi = -\half \phi = -\frac{1}{2\sqrt{2}} \log \left(\frac{\Ham_2}{\Ham_3} \right) \quad \Rightarrow \quad e^{\sqrt{2} \Phi} = \left( \frac{\Ham_2}{\Ham_3}\right)^{-\half}.
\end{equation*}

With these identifications, it is straightforward to match ({\ref{susythree}})
with ({\ref{threeform6}}), on making use of the identities
\begin{equation}
\begin{aligned}
        d\zeta &= \Ham_1^{-\half} e^1, \qquad
        \star_4 d\zeta &= - \Ham_1^{-\half} e^2 \wedge e^3 \wedge e^4
        = - dx \wedge dy \wedge dz_6.
\end{aligned}
\label{hzeta}
\end{equation}
and
\begin{equation}
e^{-\sqrt{2} \Phi} \star_4 \big(dH-\sqrt{2} H d \Phi \big)=- P^2 dx \wedge dy \wedge dz_6,
\end{equation}
and
\begin{equation}
e^{-\sqrt{2} \Phi} e^+ \wedge e^- \wedge \big(H^{-1} dH +\sqrt{2} d \Phi \big)
={P^3 \over 2 \Ham_3^2} dz_5 \wedge d \eta \wedge d \zeta.
\end{equation}

In addition, the Bianchi identities ({\ref{bian61}}) and ({\ref{bian62}}) hold with no further conditions imposed, as all terms are independent of $v$, and also $K=0$,
${\cal{G}}=0$ and $\psi=0$. The condition $\psi=0$ follows from ({\ref{psidef}}), on using the fact that the hyper-complex structures are independent of $v$.

It remains to consider the gauge field equations ({\ref{gauge61}}) and ({\ref{gauge62}}). The RHS of these equations vanishes identically, as a consequence of the fact that $h=0$. The remaining content of the gauge field equations is that $He^{\pm \sqrt{2} \Phi}$ be harmonic on the base space. This holds automatically as a consequence of the previously obtained conditions, because
$He^{\sqrt{2} \Phi}= \Ham_3$ and $He^{-\sqrt{2} \Phi}= \Ham_2$, and $\zeta$ is harmonic on the base space as a consequence of ({\ref{hzeta}}). Similarly, the Einstein equation ({\ref{nulleinst}}) holds automatically, because all terms on the RHS vanish individually, and also $L=-{1 \over 2} Q_0 d \zeta$ which is co-closed on the base, again as a consequence of ({\ref{hzeta}}).

\subsection{Analysis of the Spacetime}
Now we have shown that by fine-tuning our integration constants we can obtain a supersymmetric solution, it is interesting to look at the geometric properties of this spacetime.

Our analysis is focused on the simplified metric
\begin{equation*}
        ds^2_6 = (\Ham_2 \Ham_3)^{-\half} dz_5 (d\eta + \Ham_0 dz_5) + (\Ham_2 \Ham_3)^{\half} \left[\Ham_1 ( d\zeta^2 + dx^2 + dy^2) + \Ham_1^{-1} (dz_6 + \mathbb{A}^1)^2 \right].
\end{equation*}

In the limit of $\zeta \rightarrow \infty$ we find that the Riemann tensor falls off as $\zeta^{-n}$ for $n \geq 1$. The Ricci curvature of the spacetime is
\begin{equation*}
    R_{(6D)} = \frac{(h_3 P^2 - h_2 P^3)^2}{4\Ham_1(\Ham_2 \Ham_3)^{\frac{5}{2}}},
\end{equation*}
and we notice here that we have the option to pick either a charge $P^{2/3}$ or $h_{2/3}$ value such that the spacetime is Ricci flat
\begin{equation*}
    h_3 = \frac{h_2 P^3}{P^2} \quad \Leftrightarrow \quad R_{(6D)} = 0.
\end{equation*}
We can understand this condition by looking back at the harmonic functions
\begin{equation*}
\begin{aligned}
        \Ham_2 \Ham_3 &= (h_2 + P^2 \zeta)(h_3 + P^3 \zeta) = \left(h_2 + P^2 \zeta \right) \left(\frac{h_2 P^3}{P^2} + P^3 \zeta \right) \\
        &= P^2 P^3 \left(\zeta + \frac{h_2}{P^2}\right)^2.
\end{aligned}
\end{equation*}
and we see that picking the right integration constants we allow the zeros of both $\Ham_2$ and $\Ham_3$ to simultaneously occur.

%
%
\section{Conclusions and Outlook \label{sect:conclusions}}

We have seen that, surprisingly, a method designed to produce static solutions has
provided us with a class of cosmological solutions. In the four-charge case, we were
able to lift these solutions to five and six, and then to ten and eleven dimensions; allowing an embedding into string theory and M-theory. In the extremal limit, we recover two of the best known intersecting brane solutions, which give rise
to four-dimensional BPS black hole solutions upon dimensional reduction together
with the usual delocalisation of the branes along the compact directions. The extremal
limits of our planar solution are related to these configurations by the additional delocalisation
along two of the non-compact spatial directions, which changes the symmetry from spherical
to planar. The dimensional lift and embedding into string theory does not provide by itself
any insight into why we obtain cosmological rather than black brane solutions since the
additional dimensions are just spectators. Instead, we learn an interesting lesson about
the importance of being able to make brane configurations non-extremal. The compactified
BPS brane solutions used to obtain four-dimensional BPS black holes have the same
causal structure as the extremal Reissner-Nordstr\"om solution, which is embedded as a
`double-extreme' limit \cite{Ferrara:1997tw}, where all four-dimensional scalars are constant. The essential features of our cosmological solutions can be understood using the
charged electro-vac solutions of Einstein-Maxwell theory.

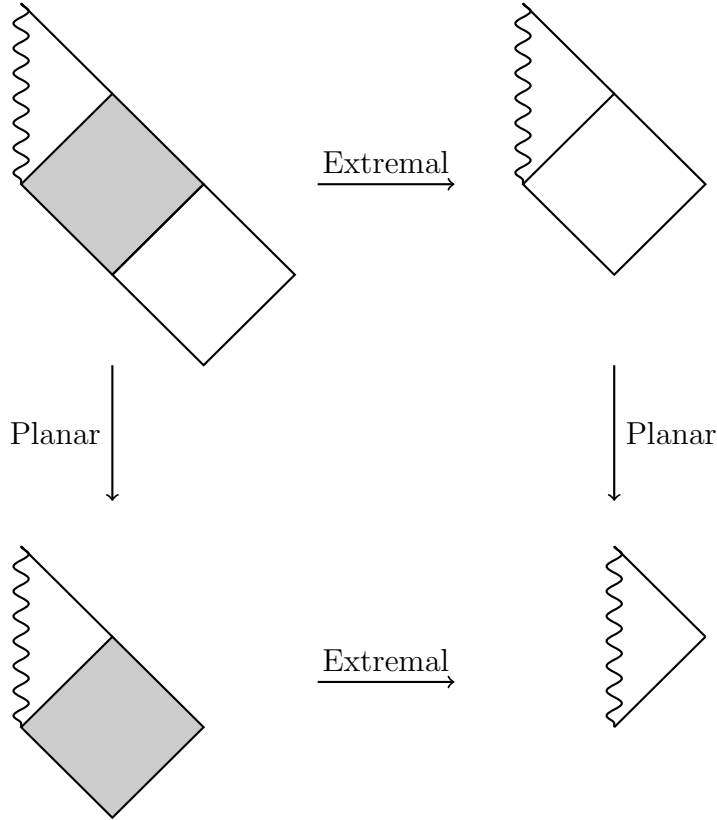
\begin{figure}[h]
\centering
\begin{tikzpicture}[scale = 0.3]

\node (topLeft2) at (0,4) {};
\node (topLeft1) at (-4,8) {};
\node (topLeft3) at (-8,12) {};

\node (topRight1) at (18,8) {};
\node (topRight3) at (14,12) {};

\node (botLeft1) at (-4,-16) {};
\node (botLeft3) at (-8,-12) {};

\node (botRight) at (18,-12) {};

\draw[thick, ->] (5,8)-- node[midway, above] {Extremal} (11,8);
\draw[thick, ->] (5,-14)--node[midway, above] {Extremal} (11,-14);

\draw[thick, ->] (-4,0)--node[midway, left] {Planar} (-4,-6);
\draw[thick, ->] (18,0)--node[midway, right] {Planar}(18,-6);

\path
 (topLeft1) +(90:4) coordinate[label=90:] (topLeft1top)
 +(-90:4) coordinate(topLeft1bot)
 +(0:4) coordinate[label=360:] (topLeft1right)
 +(180:4) coordinate[label=180:] (topLeft1left)
 ;
\draw[thick, fill=gray!40!white]  (topLeft1left)
        -- node[midway, below, sloped] {}
         (topLeft1top)
         -- node[midway, below, sloped] {}
         (topLeft1right)
             node[midway, above, sloped] {}
           -- node[midway, above, sloped] {}
         (topLeft1bot)
         -- node[midway, above, sloped] {}
         (topLeft1left)
         -- cycle;

\path
 (topLeft2) +(90:4) coordinate[label=90:] (topLeft2top)
 +(-90:4) coordinate(topLeft2bot)
 +(0:4) coordinate[label=360:] (topLeft2right)
 +(180:4) coordinate[label=180:] (topLeft2left)
 ;
\draw[thick]  (topLeft2left)
        -- node[midway, below, sloped] {}
         (topLeft2top)
         -- node[midway, below, sloped] {}
         (topLeft2right)
             node[midway, above, sloped] {}
           -- node[midway, above, sloped] {}
         (topLeft2bot)
         -- node[midway, above, sloped] {}
         (topLeft2left)
         -- cycle;

 \path
 (topLeft3) +(90:4) coordinate[label=90:] (topLeft3top)
 +(-90:4) coordinate(topLeft3bot)
 +(0:4) coordinate[label=360:] (topLeft3right)
 ;
\draw[thick] (topLeft3top) -- (topLeft3right);
\draw[decorate,decoration={snake, segment length=3.3mm, amplitude=1mm},thick]  (topLeft3top) -- (topLeft3bot);

\path
 (topRight1) +(90:4) coordinate[label=90:] (topRight1top)
 +(-90:4) coordinate(topRight1bot)
 +(0:4) coordinate[label=360:] (topRight1right)
 +(180:4) coordinate[label=180:] (topRight1left)
 ;
\draw[thick]  (topRight1left)
        -- node[midway, below, sloped] {}
         (topRight1top)
         -- node[midway, below, sloped] {}
         (topRight1right)
             node[midway, above, sloped] {}
           -- node[midway, above, sloped] {}
         (topRight1bot)
         -- node[midway, above, sloped] {}
         (topRight1left)
         -- cycle;

 \path
 (topRight3) +(90:4) coordinate[label=90:] (topRight3top)
 +(-90:4) coordinate(topRight3bot)
 +(0:4) coordinate[label=360:] (topRight3right)
 ;
\draw[thick] (topRight3top) -- (topRight3right);
\draw[decorate,decoration={snake, segment length=3.3mm, amplitude=1mm},thick]  (topRight3top) -- (topRight3bot);

\path
 (botLeft1) +(90:4) coordinate[label=90:] (botLeft1top)
 +(-90:4) coordinate(botLeft1bot)
 +(0:4) coordinate[label=360:] (botLeft1right)
 +(180:4) coordinate[label=180:] (botLeft1left)
 ;
\draw[thick, fill=gray!40!white]  (botLeft1left)
        -- node[midway, below, sloped] {}
         (botLeft1top)
         -- node[midway, below, sloped] {}
         (botLeft1right)
             node[midway, above, sloped] {}
           -- node[midway, above, sloped] {}
         (botLeft1bot)
         -- node[midway, above, sloped] {}
         (botLeft1left)
         -- cycle;

 \path
 (botLeft3) +(90:4) coordinate[label=90:] (botLeft3top)
 +(-90:4) coordinate(botLeft3bot)
 +(0:4) coordinate[label=360:] (botLeft3right)
 ;
\draw[thick] (botLeft3top) -- (botLeft3right);
\draw[decorate,decoration={snake, segment length=3.3mm, amplitude=1mm},thick]  (botLeft3top) -- (botLeft3bot);

\path
 (botRight)
 +(+90:4) coordinate[label=360:] (botRighttop)
 +(-90:4) coordinate(botRightbot)
 +(0:4) coordinate[label=180:] (botRightright)
 ;

 \draw[decorate,decoration={snake, segment length=3.3mm, amplitude=1mm},thick] (botRighttop) -- (botRightbot);
 \draw[thick] (botRighttop) -- (botRightright);
  \draw[thick] (botRightbot) -- (botRightright);

\end{tikzpicture}
\caption{Comparison of the conformal diagrams for spherical and planar Reissner-Nordstr\"om-like spacetimes. We only display one copy of each type of region. Shaded regions are where the spacetime is dynamical (no timelike Killing vector).}
\label{fig:planarextremal}
\end{figure}

We start with the spherically symmetric extremal Reissner Nordstr\"om solution, whose causal
structure is shared by a large class of BPS solutions obtained by compactifying brane configurations.
Its maximal analytical extension is a sequence of two types of regions, both static: one containing
an asymptotically flat exterior, the other – the interior – containing a timelike singularity which is repulsive
to massive neutral particles. In other words, timelike geodesics are infinitely extendable.
For our purposes, we focus on just a single pair of such regions, see Figure (\ref{fig:planarextremal})
for illustration.
If the solution is made
non-extremal, a third type of region occurs, which is dynamical (non-stationary) and located
between the two types of static patches. Let us now consider the effect of replacing spherical
by planar symmetry, or, in brane language, of delocalisation of the constituent branes
along two non-compact spatial directions. In this case, the solution cannot be asymptotically
flat any more. For brane-type solutions, it is a well-known feature that asymptotic flatness
requires more than two transverse dimensions: `large branes' (those with two or less
transverse dimensions, like the D7-brane in type-IIB) cannot be asymptotically flat. In terms
of the causal structure, we loose the static, asymptotically flat patch and remain with a static
patch containing the singularity, and a dynamical patch. More precisely, by maximal analytic
extension, we end up with two patches of each type, resulting in a conformal diagram which
is the same as Schwarzschild rotated by 90 degrees, see Figure (\ref{fig:PC}).  If we now perform an extremal limit,
we also lose the dynamical patch and remain with a static patch containing a singularity.
Comparing the four types of conformal diagrams, we see that going from spherical to
planar symmetry removes the asymptotically flat region, while the existence of a dynamical
patch depends on non-extremality. Viewed from this perspective, the presence of a cosmological
patch in our solutions is completely natural, and results from physics already present in
Einstein-Maxwell theory. These features are robust under dimensional lifting and persist
for the three-charged solution, which is a solution of gauged supergravity and have the same
conformal diagram. However, the Nernst brane solutions \cite{Dempster:2015,Dempster:2016,Barisch:2011ui,Cardoso:2015wcf}
illustrate that these features do not
persist if we modify essential features. Like the three- and four-charge solutions, the single-charged Nernst branes are planar and not asymptotically flat, but they do not share the `inside-out' feature of a singularity at a finite distance inside the static patch. The essential difference is that
Nernst branes require a non-constant scalar fields, and therefore
there is no limit in which
they become solutions of four-dimensional Einstein-Maxwell theory. Instead, as shown
in \cite{Dempster:2016}, they lift to boosted AdS-Schwarzschild black brane solutions
of five-dimensional AdS gravity.

The close relation of our cosmological solutions to the planar Reissner-Nordstr\"om
solution also settles the question of whether we need to interpret it as being sourced by
negative tension branes. We have found that the local Komar mass is negative in the static patch, which
is consistent with the repulsive character of the singularity. However, this feature is also present
in the spherical Reissner-Nordstr\"om solution, the only difference being that with planar symmetry we
loose the asymptotically flat region, and hence the ability to define a `proper' mass by evaluating the
Komar expression at asymptotic infinity. This reflects the general insight, reviewed recently in \cite{Deser:2019acl}, that the definition of global
quantities through conservation laws a la Noether requires that general diffeomorphism invariance
is `broken naturally' by the presence of extra structure, such as boundary conditions.
That we do not have a static asymptotic region does not provide a good reason to assign negative tension to the sources, because locally the situation is not different from Reissner-Nordstr\"om. Moreover, for the cases
where we can lift to ten or eleven dimensions, the sources reveal themselves as conventional, positive tension branes.

The only caveat is that our four-dimensional solutions admit other embeddings into string theory,
which might change their higher-dimensional interpretation. In particular, it can be shown that the solution found in \cite{Fre:2008zd} describes a region of our solution, although in different
coordinates, where the existence of a Killing horizon is not obvious. The solution of \cite{Fre:2008zd}
admits an uplift over the orientifold $K3 \times \mathbb{T}^2 / \mathbb{Z}_2$. This alternative
embedding, which we have not analysed in detail in this paper, is interesting because it starts
with a compactification which has less than maximal supersymmetry. In contrast, in our uplift we have used
toroidal compactifications and start with maximally supersymmetric theories in ten and eleven
dimensions. Therefore reduction to a four-dimensional $\N=2$ theory requires one to truncate the field content
after compactification. In \cite{Fre:2008zd} the sources are orientifolds, rather than D-branes or M-branes. Some authors \cite{Cornalba:2003kd}
have argued that in string theory, orientifolds naturally give rise to cosmological solutions.
We leave the investigation of this alternative lift for future work.

Another aspect of our solutions which we have only noted in passing is the reduction of the
number of integration constants resulting from imposing the presence of a regular Killing horizon.
This is related to the question of whether there is an analogue or generalisation of the attractor
mechanism for non-extremal solutions. The attractor mechanism \cite{Ferrara:1995ih,Ferrara:1996dd,Strominger:1996kf,Ferrara:1996um}
forces scalar field to
attain unique values, determined by the charges, at the event horizon of static extremal 
BPS black holes.\footnote{For non-BPS extremal black holes the horizon values 
of some scalar fields may remain un-fixed, as long as the variation of these values
does not change the black hole entropy \cite{Sen:2005wa}.}
This mechanism reduces the number of integration constants in the second order scalar field equations by a factor of one half, since only the asymptotic values of the scalars at
infinity remain integration constants that can be chosen arbitrarily. When constructing
solutions using the Killing spinor equations, or more generally, by imposing that
the scalar field equations reduce to first order gradient flow equations, this reduction 
is automatic. When solving the second order field equations directly, the reduction 
in the number of integration constants enters when imposing that the scalars 
should take regular values at the horizon, rather than displaying run-away 
behaviour. Interestingly, this link between regularity and the reduction of the 
number of integration constants also exists for non-extremal solutions,
as we have seen in Section \ref{Sect:4dSolutions}. While naively we could have
expected to obtain solutions with two integration constants per scalar field,
there is only one, corresponding to the fields value at infinity, and one
additional constant, which corresponds to the non-extremality parameter.
While the values of the scalars at the horizon are not exclusively determined
by the charges, they are still fixed and completely 
determined by the other integration constants. 
Similar observations were made in \cite{Errington:2014bta,Mohaupt:2010fk} for 
non-extremal five- and four-dimensional black holes, and in \cite{Dempster:2015}
for Nernst branes. In \cite{Mohaupt:2010fk} this behaviour was dubbed 
`dressed attractor mechanism', since the horizon values of the scalars 
are given by the same expressions as in the extremal limit, with the charges
replaced by dressed charges which depend on the other integration constants.
These observations are consistent with the idea of `hot attractors', which 
was advocated in \cite{Goldstein:2014gta,Goldstein:2014qha,Goldstein:2018mwt},
and support the idea that the attractor mechanism is relevant, in a modified
form, for non-extremal solutions.

Another loose end is the question of whether the Killing horizons of our solutions admit any thermodynamic interpretation. In an upcoming companion paper, we will show that despite our solutions being cosmological and despite the lack of any standard spacelike asymptotics (such as asymptotically flat or asymptotically AdS), a version of the first law of horizon mechanics can be established.

In the second part of this paper, we have focussed on the four-charge solution, since it
admits various higher dimensional lifts and embeddings into string/M-theory. It would be
interesting to do something similar for the three-charge solution, which however is a
solution of a gauged supergravity, for which, to our knowledge, it is not known how to
obtain it as a consistent truncation of a higher-dimensional theory.

One aspect which we have not investigated in this paper is the question of whether our solutions are stable. For this we refer to the discussion in
\cite{Burgess:2002vu,Burgess:2003mk} which have addressed some aspects of
the stability of the horizon. They found that the situation for the first horizon is the same as for the inner horizon of non-extremal Reissner-Nordstrom solution, while
for the second horizon no indication for an instability was found.

While our analysis disfavours interpreting the sources of our solutions as negative tension
branes, it has been argued that negative
branes exist in string theory \cite{Dijkgraaf:2016lym}. In \cite{Hull:1998ym} it was shown that when admitting timelike T-duality, the web of string/M-theories contains exotic theories with twisted supersymmetry and negative kinetic energy for some of the fields (type-II$^*$).
Moreover, there exists at least one version of type-II string theory for any possible space-time signature.  According to \cite{Dijkgraaf:2016lym}, some of the branes of these exotic theories appear as `negative branes' when viewed from the point of view of a dual theory. This could allow the construction of new, genuinely stringy cosmological solutions, and
our formalism could easily be tweaked to study these solutions.

\section*{Acknowledgements}

JG is supported by the STFC Consolidated Grant ST/L000490/1.

%
%

\appendix
\newpage

%
\section{Kruskal Cordinates}
%
We detail here the simplest case of calculating global coordinates for the four-charge solution using a Kruskal like coordinate change. Here we assume that the harmonic functions $\Ham_a$ are all equal; physically this is understood as trivialising the scalar fields within the spacetime, but it is a decision made to simplify the integral of $\zeta^*$ to highlight how to choose $\lambda$ to obtain global coordinates.

As all $\Ham_a$ are equal, we rewrite the function
\begin{equation*}
    \Ham = 2(\beta + \gamma \zeta)^2,
\end{equation*}
and begin with the metric for $\zeta < \alpha^{-1}$
\begin{equation*}
 ds^2 = - \frac{1-\alpha \zeta}{2(\beta + \gamma \zeta)^2} d\eta^2 + \frac{2(\beta + \gamma \zeta)^2}{1 - \alpha \zeta} d\zeta^2 + 2(\beta + \gamma \zeta)^2 (dx^2 + dy^2).
\end{equation*}
We make the coordinate transformation into Eddington-Finkelstein coordinates by using the advanced coordinates
\begin{equation*}
 v = \eta + \zeta^*, \qquad d\zeta^* = \frac{2(\beta + \gamma z)^2}{1 - \alpha \zeta} d\zeta,
\end{equation*}
where we have introduced the tortoise coordinate $z^*$ such that the metric can be written in the form
\begin{equation*}
 ds^2 = - \frac{1 - \alpha \zeta}{2(\beta + \gamma \zeta)^2} dv^2 + 2 d\zeta dv + 2(\beta + \gamma \zeta)^2 (dx^2 + dy^2),
\end{equation*}
and we can integrate up to find
\begin{equation*}
 \zeta^* = -\frac{2(\alpha \beta +\gamma )^2}{\alpha ^3} \log (1-\alpha \zeta)-\frac{\gamma \zeta}{ \alpha ^2} (4 \alpha
 \beta +2 \gamma +\alpha \gamma \zeta).
\end{equation*}
We can also define the advanced coordinate $u = t - \zeta^*$ to write the metric in lightcone coordinates
\begin{equation*}
 ds^2 = - \frac{1 - \alpha \zeta}{2(\beta + \gamma \zeta)^2} dv du + 2(\beta + \gamma \zeta)^2 (dx^2 + dy^2).
\end{equation*}
We make the Kruskal-like change
\begin{equation*}
 U = - e^{-\lambda u}, \quad V = e^{\lambda v},
\end{equation*}
such that $U \leq 0$ and $V \geq 0$. Taking derivatives we find $ dUdV = \lambda^2 UV du dv$, where
\begin{equation*}
 UV = (1-\alpha \zeta)^{-\frac{4 \lambda}{\alpha^3}(\alpha \beta + \gamma)^2} \exp\left(-\frac{2\lambda \gamma \zeta}{\alpha^2} (4\alpha \beta + 2 \gamma + \alpha \gamma \zeta) \right).
\end{equation*}
To find the form of $\lambda$ we substitute this all into the metric and pick $\lambda$ to remove $(1-\alpha \zeta)$ from the metric to ensure that there globally are no zeros of the metric
\begin{equation*}
 \begin{aligned}
 ds^2 &= \frac{1}{\lambda^2 UV} \frac{1 - \alpha \zeta}{2(\beta + \gamma \zeta)^2} dU dV + 2(\beta + \gamma \zeta)^2 (dx^2 + dy^2) \\
 &= - \frac{(1 - \alpha \zeta)^{1+\frac{4 \lambda}{\alpha^3}(\alpha \beta + \gamma)^2}}{2 \lambda^2 (\beta + \gamma \zeta)^2} \exp \left[\frac{2 \lambda \gamma \zeta}{\alpha^2} (4\alpha \beta + 2 \gamma + \alpha \gamma \zeta) \right] dU dV + 2(\beta + \gamma \zeta)^2 (dx^2 + dy^2).
 \end{aligned}
\end{equation*}
Making the choice
\begin{equation*}
 \lambda = - \frac{\alpha^3}{4}(\alpha \beta + \gamma)^{-2},
\end{equation*}
we obtain the metric
\begin{equation}
 ds^2 = - \frac{1}{\lambda^2} \frac{e^{\xi(\zeta(U,V))}}{2(\beta + \gamma \zeta(U,V))^2} dU dV + 2(\beta + \gamma \zeta(U,V))^2 (dx^2 + dy^2),
\end{equation}
where the new function $\xi(U,V)$ is an everywhere non-zero function in the global domain of $\zeta$
\begin{equation*}
 \xi(\zeta(U,V)) = -\frac{\alpha \gamma \zeta (4 \alpha \beta +2 \gamma +\alpha \gamma \zeta)}{2 \alpha (\alpha \beta
 +\gamma )^2}.
\end{equation*}

\section{Planar Einstein-Maxwell Solution}

If we take the limit of setting the physical scalars of the theory to be constant, the geometry of the four-charge solution becomes that of vacuum solution to the Einstein-Maxwell equations with planar symmetry. This behaviour is expected as the Reissner-Nordstr\"om solution is the resulting geometry for the spherically symmetric solution to the STU model with constant physical scalars, also known as the double-extremal limit \cite{Ferrara:1997tw}.

The physical scalars are given by
\begin{equation*}
    z^A = -i \Ham_A \bigg(\frac{\Ham_0}{\Ham_1 \Ham_2 \Ham_3} \bigg)^{\frac{1}{2}},
\end{equation*}
and we see that they are everywhere constant under the restriction that $\Ham_0 = \Ham_1 = \Ham_2 = \Ham_3$. This means the integration constants must be fine-tuned such that $Q_0 = P^1 = P^2 = P^3 = K$ and $h_0 = h^1 = h^2 = h^3 = h$.

Trivialising the constants in this way allows us to easily see the recovery of the Einstein-Maxwell system through studying the $4D$ Lagrangian in section (\ref{sec:4Dlag}). The kinetic term for the scalars will vanish; the choice that all integration constants are the same reduces the number of charge gauge fields from one to four, and the term in front of the gauge field can be simply removed through the redefinition of the remaining gauge potential.

This transition from the STU model to the Einstein-Maxwell system is also mirrored in our geometry. When we take the above limit for our integration constants, we recover the line element for the Einstein-Maxwell solution with planar symmetry. The metric for the static patch of the spacetime in the main body of the paper, repeated here
\begin{equation}
\label{eq:genmet}
    ds^2 = - \frac{W(\zeta)}{\Ham(\zeta)} d\eta^2 + \frac{\Ham(\zeta)}{W(\zeta)} d\zeta^2  + \Ham(\zeta) (dx^2 + dy^2),
\end{equation}
changes at the level of these functions, which are now given by
\begin{equation*}
 W(\zeta) = 1 - \alpha \zeta, \qquad \Ham(\zeta) = (\beta + \gamma \zeta)^2,
\end{equation*}
with constants simplified as
\begin{equation*}
    \alpha = 2B, \quad \beta =\frac{2 K}{\alpha} \sinh\left(\frac{\alpha h}{2 K}\right), \quad \gamma = \exp\left(-\frac{\alpha h}{2 K}\right), \qquad \alpha,\beta,\gamma \in (0, \infty).
\end{equation*}
The metric written in terms of these new constants for $\zeta < \alpha^{-1}$ is given by
\begin{equation}
\label{eq:global}
  ds^2 = - \frac{1 - \alpha \zeta }{2(\beta + \gamma \zeta)^2} d\eta^2 + \frac{2(\beta + \gamma \zeta)^2}{(1 - \alpha \zeta )} d\zeta^2  + 2(\beta + \gamma \zeta)^2(dx^2 + dy^2).
\end{equation}

The solution to Einstein-Maxwell's equations with planar symmetry is generally given in the form \cite{Griffiths:2009dfa}
\begin{equation}
\label{eq:planarEM}
  ds^2 =  - f(r) dt^2 + \frac{dr^2}{f(r)}  + r^2(dx^2 + dy^2), \qquad     f(r) = -\frac{2M}{r} + \frac{e^2}{r^2}.
\end{equation}

We can show the equivalence of our solution \eq{genmet} and \eq{planarEM} by making the following coordinate transformations
\begin{equation*}
    2(\beta + \gamma \zeta)^2 = \tilde{r}^2 \; \Rightarrow \; \tilde{r} = \sqrt{2} (\beta + \gamma \zeta), \qquad \zeta = \frac{1}{\gamma} \left(\frac{\tilde{r}}{\sqrt{2}} -\beta \right), \qquad d\zeta = \frac{d\tilde{r}}{\sqrt{2} \gamma},
\end{equation*}
we can then rewrite parts of the line element as
\begin{equation*}
\begin{aligned}
         \frac{1 - \alpha \zeta }{2(\beta + \gamma \zeta)^2} d\eta^2 &= \left(-\frac{\alpha}{\sqrt{2} \gamma \tilde{r}} + \frac{\alpha \beta + \gamma}{\gamma \tilde{r}^2} \right) d\eta^2, \\
         \frac{2(\beta + \gamma \zeta)^2}{(\alpha \zeta - 1)} d\zeta^2 &=  \left(-\frac{\alpha}{\sqrt{2} \gamma \tilde{r}} + \frac{\alpha \beta + \gamma}{\gamma \tilde{r}^2} \right)^{-1} \frac{d\tilde{r}^2}{2 \gamma^2}.
\end{aligned}
\end{equation*}
To ensure that the functions preceding the $d\eta^2$ and $d\tilde{r}^2$ are each other's multiplicative inverse we rescale $\tilde{r}$ such that
\begin{equation*}
    r = \frac{\tilde{r}}{\sqrt{2} \gamma}, \qquad dr = \frac{d\tilde{r}}{\sqrt{2} \gamma}, \qquad \tilde{r} = \gamma \sqrt{2} r.
\end{equation*}
Allowing us to write down the metric in the form
\begin{equation*}
  ds^2 = - f(r) d\eta^2 + \frac{dr^2}{f(r)} + 2 \gamma^2 t^2(dx^2 + dy^2),
\end{equation*}
where we have defined the function
\begin{equation*}
    f(r) := - \frac{\alpha}{2 \gamma^2} \frac{1}{r} + \frac{\alpha \beta + \gamma}{2 \gamma^3} \frac{1}{r^2}.
\end{equation*}
Finally we rescale the $x$ and $y$ coordinates to re-absorb the $2 \gamma^2$ factor and rename $\eta$ to $t$ to arrive at the metric
\begin{equation}
  ds^2 = - f(r) dt^2 + \frac{dr^2}{f(r)} + r^2(dx^2 + dy^2).
\end{equation}
Looking at the function $f(r)$ and comparing this to \eq{planarEM} we can relate the integration constants from our solution to the ``mass"\footnote{$M$ is much more loosely related to the mass for planar solutions as in the spherically symmetric case. As the solution is not asymptotically flat, we cannot identify the integration constant $M$ with the Newtonian limit as is done for the Schwarzschild or Reissner-Nordstr\"om solutions. In contrast, the electric charge can still be set via Gauss' law and so is easier to pin down.} and electric charge of the solution.
\begin{equation*}
    M = \frac{\alpha}{4 \gamma^2}, \qquad e^2 = \frac{\alpha \beta + \gamma }{2\gamma^3}.
\end{equation*}

\section{Spherically Symmetric Solutions for STU Supergravity from the C-Map}

While working on the oxidation of the planar STU model we noticed that the $4D$ metric with planar symmetry was only superficially different from the solution found in \cite{Errington:2014bta} where spherical symmetry was imposed on a class of prepotentials which included the STU model.

Due to the simplicity of the generalisation of the uplift and the popularity of spherically symmetric solutions for supergravity theories we have chosen to include here the uplift of the non-extremal STU model with spherical symmetry in ten and eleven dimensions. We hope that the line elements and gauge field content for these theories could be interesting for those looking at non-extremal STU models in the future.

In this appendix, we first show how the spherically symmetric solution is related to the four-charge solution solved in the main body of the paper. We then write down the metric and gauge field content of both string/M-theory embeddings. Finally, we show that taking the 4D extremal limit the resulting metrics are now in the form of the intersecting M5 branes \cite{Behrndt:1996jn} in $11D$ and the $D1$-$D5$-$P$-$KK$ solutions in $10D$. Unlike the planar solution, the harmonic functions of the spherically symmetric solution diverge for $\rho=0$ and so we obtain an interpretation for the position of the intersecting branes for each higher dimensional theory.

It is also interesting to note here how the two solutions differ; that when changing the geometric ansatz from spherical to planar the asymptotic region of the spacetime changes from static to dynamic.

We begin this section referring to \cite{Dempster:2015} where the STU prepotential is picked out by setting $n=3$ in (5.22) with the resulting four-dimensional metric
\begin{equation}
    ds_4^2 = -\frac{W(\rho)dt^2}{\sqrt{-H_0 H^1 H^2 H^3}} + \sqrt{-H_0 H^1 H^2 H^3} \left(\frac{d\rho^2}{W(\rho)} + \rho^2 d\Omega_2^2 \right).
\end{equation}
The functions are:
\begin{equation*}
    W(\rho) = 1 - \frac{2c}{\rho},
\end{equation*}
\begin{equation*}
    H_0(\rho) = -\sqrt{2} Q_0 \left[\frac{1}{c} \sinh\left( \frac{ch_0}{Q_0} \right) + e^{- \frac{ch_o}{Q_0}} \rho^{-1} \right],
\end{equation*}
\begin{equation*}
    H^A(\rho) = \sqrt{2} P^A \left[\frac{1}{c} \sinh\left( \frac{ch^A}{P^A} \right) + e^{- \frac{ch^A}{P^A}} \rho^{-1} \right],
\end{equation*}
for $A=1,2,3$. These should remind the reader of the functions $W(\zeta)$ and $\Ham_a(\zeta)$ with $\zeta \rightarrow \rho^{-1}$ and $\alpha \leftrightarrow c$. Of course these two metrics are not related by $\zeta \rightarrow \rho^{-1}$ as this would affect $d\zeta^{2} \rightarrow \frac{d\rho}{\rho^4}$.

The physical scalars are given by (5.24) in \cite{Errington:2014bta}
\begin{equation*}
    z^A = -i H_A \sqrt{\frac{-H_0}{H^1 H^2 H^3}},
\end{equation*}
and the gauge field strengths
\begin{equation*}
    F^0 = \half \frac{Q_0}{q_0^2} dt\wedge d\tau, \qquad F^A = -\half P^a \sin\theta d\theta\wedge d\phi, \qquad q_0^2 = \frac{H_0^2}{2 W}.
\end{equation*}
The physical scalars for the spherically symmetric model are in an identical form but with $H_a \leftrightarrow \Ham_a$. In the asymptotic limit $H_a$ tend to constants, where as $\Ham_a$ diverge $\Op(\zeta)$ and although visually similar the physical behaviour of the scalars between the two solutions will be different. In particular, assuming that $h_a \neq 0$ all physical scalars are asymptotic to constant values.

Integrating up and applying boundary conditions gauge potential is found to be
\begin{equation*}
    A^0 = \frac{W c}{Q_0} \left(e^{\frac{2c h_0}{Q_0}} -W \right)^{-1} d\eta, \qquad A^A = \half P^A \cos \theta d\phi,
\end{equation*}
which can be manipulated into
\begin{equation*}
    A^0 = -\frac{W \gamma_0}{\sqrt{2} Q_0 H_0} d\eta, \qquad A^A = \half P^A \cos \theta d\phi, \qquad \gamma_0 = Q_0 e^{- \frac{ch^A}{P^A}}.
\end{equation*}
We see that $A^0$ has the same form as $A^0$ from the planar solution and the $A^A$ are now constants as before but over the two-sphere and not the two-plane.

As we are not required to take derivatives of the metric functions during the oxidation procedure, we find that the uplift of the spherically symmetric solution is unaffected by the difference in form of the harmonic functions $H_a$. This allows us to simply write down the higher dimensional uplifts of this solution straight from the work in the main body of the text.

\subsection*{Oxidation to Five Dimensions}
\begin{equation}
\begin{aligned}
        ds^2_5 = (H_1 H_2 H_3)^{-\frac{1}{3}} &\bigg[H_0 dz_5^2 + \frac{W}{2H_0}\left( W \frac{\gamma_0^2}{Q_0^2} - 1\right) dt^2 + \frac{2 W \gamma_0}{\sqrt{2} Q_0} dt dz_5 \\ &+ 2H_1 H_2 H_3 \left( \frac{ d\rho^2}{W} + \rho^2 d\Omega_2^2 \right) \bigg],
\end{aligned}
\end{equation}
where $d\Omega_2^2$ is the line element for the two sphere
\begin{equation}
    \tilde{\mathbb{A}}_i = \sqrt{2} A^A = \mathfrak{p}_a \cos \theta d\phi, \quad \mathfrak{p}_a = \frac{P^A}{\sqrt{2}}.
\end{equation}

\subsection*{Oxidation to Eleven Dimensions}
Using an identical procedure, the uplift to $11D$ is trivial and given by
\begin{equation*}
    ds_{11}^2 = ds_5^2 + h_1 (dy_1^2 + dy_2^2) + h_2 (dy_3^2 + dy_4^2) + h_3 (dy_5^2 + dy_6^2),
\end{equation*}
\begin{equation*}
    \mathcal{A} = \tilde{\mathbb{A}}_1 \wedge dy^1 \wedge dy^2 + \tilde{\mathbb{A}}_2 \wedge dy^3 \wedge dy^4 + \tilde{\mathbb{A}}_3 \wedge dy^5 \wedge dy^6.
\end{equation*}
This compactification is subject to the constraint that the torus has constant volume (which is equivalent to $h_1h_2h_3 =1$). The explicit $\zeta$ dependence of $h_i$ is
\begin{equation*}
        h_i = \frac{H_i}{(H_1 H_2 H_3)^{\frac{1}{3}}} \; ,
\end{equation*}
and the three-form gauge potential is found simply from the components of \eq{5dvec}. Thus the full line element for the non-extremal planar STU model is
\begin{equation*}
\begin{aligned}
        ds^2_5 = (H_1 H_2 H_3)^{-\frac{1}{3}} &\bigg[H_0 dz_5^2 + \frac{W}{2H_0}\left( W \frac{\gamma_0^2}{Q_0^2} -1 \right) dt^2 + \frac{2 W \gamma_0}{\sqrt{2} Q_0} dt dz_5 \\ &+ 2H_1 H_2 H_3 \left( \frac{ d\rho^2}{W} + \rho^2 d\Omega_2^2 \right) \\
        &+ H_1 (dy_1^2 + dy_2^2) + H_2 (dy_3^2 + dy_4^2) + H_3 (dy_5^2 + dy_6^2) \bigg].
\end{aligned}
\end{equation*}

\subsection*{Oxidation to Six Dimensions}
The appropriate reduction ansatz \cite{Chow:2013gba} to arrive back at the 5D Lagrangian is given by
\begin{equation*}
    ds^2_6 = e^{\sigma/\sqrt{6}} ds_5^2 + e^{-3\sigma/\sqrt{6}} (dz_6 + \tilde{\mathbb{A}}_1)^2, \qquad B_{(6D)} = B + \tilde{\mathbb{A}}_2 \wedge (dz_6 + \tilde{\mathbb{A}}_1).
\end{equation*}
with the field strengths decomposed as
\begin{equation*}
H_{(6D)} = \mathbb{H} + \tilde{\bF}_2 \wedge (dz_6 + \tilde{\mathbb{A}}_1)    ,\qquad \mathbb{H} = dB - \tilde{\mathbb{A}}_2 \wedge \tilde{\bF}_1 ,\qquad \tilde{\bF}_i = d \tilde{\mathbb{A}}_i.
\end{equation*}
The metric is given in the same form as the main body of the text
\begin{equation*}
\begin{aligned}
        ds^2_6 = (H_2 H_3)^{-\frac{1}{2}} &\bigg[H_0 dz_5^2 + \frac{W}{2H_0}\left(W \frac{\gamma_0^2}{Q_0^2} -1\right) d\eta^2 + \frac{2W \gamma_0}{\sqrt{2} Q_0} d\eta dz_5 \\ &+ 2H_1 H_2 H_3 \left( \frac{ d\rho^2}{W} + d\Omega^2_2 \right) \bigg] + \frac{(H_2 H_3)^\half}{H_1} (dz_6 + \tilde{\mathbb{A}}_1)^2.
\end{aligned}
\end{equation*}
The gauge content is very similar. The gauge fields
\begin{equation*}
    \tilde{\mathbb{A}}_1 = \sqrt{2} A^1, \qquad \tilde{\mathbb{A}}_2 = \sqrt{2} A^2,
\end{equation*}
are identical to the $4D$ solutions. Using the work from the $5D$ calculations the $6D$ three-form field strength is given by:
\begin{equation*}
\begin{aligned}
        H_{(6D)} = -\left( \frac{ \mathfrak{p}_3}{\Ham_3^2} \right) d\eta \wedge d\rho \wedge dz_5
 + ( 2\mathfrak{p}_2) \sin \theta d\theta \wedge d\phi \wedge dz_6.
\end{aligned}
\end{equation*}

\subsection*{Oxidation to Ten Dimensions}
 The reduction ansatz to uplift the solution to $10D$ is given by
\begin{equation*}
    ds_{10}^2 = ds_6^2 + e^{\phi / \sqrt{2} }(dy_1^2 + dy_2^2 + dy_3^2 + dy_4^2), \qquad \Phi = \frac{\phi}{\sqrt{2}}, \qquad C \equiv B,
\end{equation*}
where $\phi$ is the dilaton from the $6D$ theory and can be found in terms of $\rho$ by using
\begin{equation*}
    e^{\sqrt{2} \phi} = \frac{h_2}{h_3} = \left(\frac{\I_{33} }{\I_{22}}\right)^{\frac{1}{2}} \quad \Rightarrow \quad e^{\phi / \sqrt{2} } = \left(\frac{\I_{33} }{\I_{22}}\right)^{\frac{1}{4}} = \sqrt{\frac{H_2}{H_3}},
\end{equation*}
and everything else is found simply from the $6D$ analysis. The Dilaton is given by
\begin{equation*}
    \Phi = \half \log \left(\frac{H_2}{H_3} \right).
\end{equation*}
The line element is given by
\begin{equation*}
\begin{aligned}
        ds^2_{10} = \sqrt{\frac{H_2}{H_3}} &\bigg[H_0 H_2^{-1} dz_5^2 + \frac{W}{2H_0 H_2}\left( W \frac{\gamma_0^2}{Q_0^2} -1 \right) d\eta^2 + \frac{W \gamma_0}{\sqrt{2} Q_0 H_2} d\eta dz_5 \\ &+ 2H_1 H_3 \left(\frac{ d\rho^2}{W} + d\Omega^2_2 \right) + \frac{H_3}{H_1} (dz_6 + \tilde{\mathbb{A}}_1)^2
        + dy_1^2 + dy_2^2 + dy_3^2 + dy_4^2 \bigg].
\end{aligned}
\end{equation*}

\subsection*{Extremal Limit}
We can again take the limit of $c\rightarrow 0$ for the $4D$ solution to obtain the extremal limit of the spherically symmetric solution.

Uplifting the extremal solution to $11D$ results in the following line element
\begin{equation}
\begin{aligned}
    ds_{11}^2 = &(H_1H_2H_3)^{-\frac{1}{3}} \big[ d\eta dz_5 + H_0 dz_5^2 + H_1H_2H_3 (d\rho^2 + d\Omega^2_2) \\
    &+ H_1(dy_1^2 + dy_2^2) + H_2(dy_3^2 + dy_4^2) + H_3(dy_5^2 + dy_6^2) \big],
\end{aligned}
\end{equation}
which matches exactly with equation (4.1) in \cite{Behrndt:1996jn}. This allows us to identify this solution as the intersection of three $M5$ branes with momentum along the common intersection.

Uplifting the extremal solution to $10D$ we find that the line element is
\begin{equation}
\begin{aligned}
        ds^2_{10} = &\sqrt{\frac{H_2}{H_3}} \bigg[H_2^{-1} \left(H_0 dz_5^2 + dz_5d\eta \right) + H_3 H_1 (d\rho^2 + d\Omega^2_2) \\ &+ H_3 H_1^{-1}(dz_6^2 + \tilde{\mathbb{A}}^1)^2 + dy_1^2 + dy_2^2 + dy_3^2 + dy_4^2\bigg],
\end{aligned}
\end{equation}
which is the intersection of a D1 and D5 brane with momentum along the common direction and a Taub-NUT space.

\pagebreak

\begin{thebibliography}{10}

\bibitem{Mohaupt:2011aa}
T.~Mohaupt and O.~Vaughan, {\it {The Hesse potential, the c-map and black hole
  solutions}},  {\em JHEP} {\bf 1207} (2012) 163,
  [\href{http://arxiv.org/abs/1112.2876}{{\tt arXiv:1112.2876}}].

\bibitem{Errington:2014bta}
D.~Errington, T.~Mohaupt, and O.~Vaughan, {\it {Non-extremal black hole
  solutions from the c-map}},  {\em JHEP} {\bf 05} (2015) 052,
  [\href{http://arxiv.org/abs/1408.0923}{{\tt arXiv:1408.0923}}].

\bibitem{Dempster:2015}
P.~Dempster, D.~Errington, and T.~Mohaupt, {\it {Nernst branes from special
  geometry}},  {\em JHEP} {\bf 05} (2015) 079,
  [\href{http://arxiv.org/abs/1501.0786}{{\tt arXiv:1501.0786}}].

\bibitem{Dempster:2016}
P.~Dempster, D.~Errington, J.~Gutowski, and T.~Mohaupt, {\it {Five-dimensional
  Nernst branes from special geometry}},  {\em JHEP} {\bf 11} (2016) 114,
  [\href{http://arxiv.org/abs/1609.0506}{{\tt arXiv:1609.0506}}].

\bibitem{Barisch:2011ui}
S.~Barisch, G.~Lopes~Cardoso, M.~Haack, S.~Nampuri, and N.~A. Obers, {\it
  {Nernst branes in gauged supergravity}},  {\em JHEP} {\bf 1111} (2011) 090,
  [\href{http://arxiv.org/abs/1108.0296}{{\tt arXiv:1108.0296}}].

\bibitem{Cardoso:2015wcf}
G.~L. Cardoso, M.~Haack, and S.~Nampuri, {\it {Nernst branes with Lifshitz
  asymptotics in N=2 gauged supergravity}},
  \href{http://arxiv.org/abs/1511.0767}{{\tt arXiv:1511.0767}}.

\bibitem{Burgess:2002vu}
C.~P. Burgess, F.~Quevedo, S.~J. Rey, G.~Tasinato, and I.~Zavala, {\it
  {Cosmological space-times from negative tension brane backgrounds}},  {\em
  JHEP} {\bf 10} (2002) 028,
  [\href{http://arxiv.org/abs/hep-th/0207104}{{\tt hep-th/0207104}}].

\bibitem{Burgess:2003mk}
C.~P. Burgess, C.~Nunez, F.~Quevedo, G.~Tasinato, and I.~Zavala, {\it {General
  brane geometries from scalar potentials: Gauged supergravities and
  accelerating universes}},  {\em JHEP} {\bf 08} (2003) 056,
  [\href{http://arxiv.org/abs/hep-th/0305211}{{\tt hep-th/0305211}}].

\bibitem{Cornalba:2003kd}
L.~Cornalba and M.~S. Costa, {\it {Time dependent orbifolds and string
  cosmology}},  {\em Fortsch. Phys.} {\bf 52} (2004) 145--199,
  [\href{http://arxiv.org/abs/hep-th/0310099}{{\tt hep-th/0310099}}].

\bibitem{Akyol:2010iz}
M.~Akyol and G.~Papadopoulos, {\it {Spinorial geometry and Killing spinor
  equations of 6-D supergravity}},  {\em Class. Quant. Grav.} {\bf 28} (2011)
  105001, [\href{http://arxiv.org/abs/1010.2632}{{\tt arXiv:1010.2632}}].

\bibitem{Akyol:2012cq}
  M.~Akyol and G.~Papadopoulos,
  {\it (1,0) superconformal theories in six dimensions and Killing spinor equations,}
  {\em JHEP} {\bf 07} (2012) 070, [\href{https://arxiv.org/abs/1204.2167}{{\tt arXiv:1204.2167}}].

\bibitem{Akyol:2013ana}
  M.~Akyol and G.~Papadopoulos,
  {\it Brane solitons of (1, 0) superconformal theories in six dimensions with hyper-multiplets,}
  {\em Class.\ Quant.\ Grav.}  {\bf 31} (2014) 065012,
  [\href{https://arxiv.org/abs/1307.1041}{{\tt arXiv:1307.1041}}].

\bibitem{Gillard:2004xq}
  J.~Gillard, U.~Gran and G.~Papadopoulos,
  {\it The Spinorial geometry of supersymmetric backgrounds,}
  {\em Class.\ Quant.\ Grav.}  {\bf 22} (2005) 1033,
  [\href{https://arxiv.org/abs/hep-th/0410155}{{\tt hep-th/0410155}}].

\bibitem{Cano:2018wnq}
  P.~A.~Cano and T.~Ort\'in,
  {\it All the supersymmetric solutions of ungauged $\mathcal{N} = (1,0),d=6$ supergravity,}
  [\href{https://arxiv.org/abs/1804.04945}{{\tt arXiv:1804.04945}}].

\bibitem{Lam:2018jln}
H.~Het~Lam and S.~Vandoren, {\it {BPS solutions of six-dimensional (1, 0)
  supergravity coupled to tensor multiplets}},  {\em JHEP} {\bf 06} (2018) 021,
  [\href{http://arxiv.org/abs/1804.0468}{{\tt arXiv:1804.0468}}].


\bibitem{Cariglia:2004kk}
M.~Cariglia and O.~A.~P. Mac~Conamhna, {\it {The General form of supersymmetric
  solutions of N=(1,0) U(1) and SU(2) gauged supergravities in
  six-dimensions}},  {\em Class. Quant. Grav.} {\bf 21} (2004) 3171--3196,
  [\href{http://arxiv.org/abs/hep-th/0402055}{{\tt hep-th/0402055}}].

\bibitem{Horowitz:2017fyg}
G.~T. Horowitz, H.~K. Kunduri, and J.~Lucietti, {\it {Comments on Black Holes
  in Bubbling Spacetimes}},  {\em JHEP} {\bf 06} (2017) 048,
  [\href{http://arxiv.org/abs/1704.0407}{{\tt arXiv:1704.0407}}].

\bibitem{Breunholder:2018roc}
V.~Breunhlder and J.~Lucietti, {\it {Supersymmetric black hole non-uniqueness
  in five dimensions}},  {\em JHEP} {\bf 03} (2019) 105,
  [\href{http://arxiv.org/abs/1812.0732}{{\tt arXiv:1812.0732}}].

\bibitem{Gauntlett:2004wh}
J.~P. Gauntlett and J.~B. Gutowski, {\it {Concentric black rings}},  {\em Phys.
  Rev.} {\bf D71} (2005) 025013,
  [\href{http://arxiv.org/abs/hep-th/0408010}{{\tt hep-th/0408010}}].

\bibitem{Gauntlett:2004qy}
J.~P. Gauntlett and J.~B. Gutowski, {\it {General concentric black rings}},
  {\em Phys. Rev.} {\bf D71} (2005) 045002,
  [\href{http://arxiv.org/abs/hep-th/0408122}{{\tt hep-th/0408122}}].

\bibitem{Bena:2007kg}
I.~Bena and N.~P. Warner, {\it {Black holes, black rings and their
  microstates}},  {\em Lect. Notes Phys.} {\bf 755} (2008) 1--92,
  [\href{http://arxiv.org/abs/hep-th/0701216}{{\tt hep-th/0701216}}].

\bibitem{Bena:2004de}
I.~Bena and N.~P. Warner, {\it {One ring to rule them all ... and in the
  darkness bind them?}},  {\em Adv. Theor. Math. Phys.} {\bf 9} (2005), no.~5
  667--701, [\href{http://arxiv.org/abs/hep-th/0408106}{{\tt
  hep-th/0408106}}].

\bibitem{Cardoso:2012nh}
G.~L. Cardoso, B.~de~Wit, and S.~Mahapatra, {\it {Non-holomorphic deformations
  of special geometry and their applications}},  {\em Springer Proc.Phys.} {\bf
  144} (2013) 1--58, [\href{http://arxiv.org/abs/1206.0577}{{\tt
  arXiv:1206.0577}}].

\bibitem{Griffiths:2009dfa}
J.~B. Griffiths and J.~Podolsky, {\em {Exact Space-Times in Einstein's General
  Relativity}}.
\newblock Cambridge University Press, Cambridge, 2009.

\bibitem{Kasner:1921zz}
E.~Kasner, {\it {Geometrical theorems on Einstein's cosmological equations}},
  {\em Am. J. Math.} {\bf 43} (1921) 217--221.

\bibitem{York1986}
J.~W. York, {\it Boundary terms in the action principles of general
  relativity},  {\em Foundations of Physics} {\bf 16} (Mar, 1986) 249--257.

\bibitem{Brown:1992br}
J.~D. Brown and J.~W. York, Jr., {\it {Quasilocal energy and conserved charges
  derived from the gravitational action}},  {\em Phys. Rev.} {\bf D47} (1993)
  1407--1419, [\href{http://arxiv.org/abs/gr-qc/9209012}{{\tt
  gr-qc/9209012}}].

\bibitem{Lu:2013ura}
  H.~L\"u, Y.~Pang and C.~N.~Pope,
  {\it AdS Dyonic Black Hole and its Thermodynamics,}
  {\em JHEP} {\bf 11} (2013) 033,
  [\href{https://arxiv.org/abs/1307.6243}{{\tt arXiv:1307.6243}}].

\bibitem{Fre:2008zd}
P.~Fre and J.~Rosseel, {\it {On full-fledged supergravity cosmologies and their
  Weyl group asymptotics}},  [\href{http://arxiv.org/abs/0805.4339}{{\tt
  arXiv:0805.4339}}].

\bibitem{Chow:2013gba}
D.~D.~K. Chow and G.~Comp{\`e}re, {\it {Dyonic AdS black holes in maximal
  gauged supergravity}},  {\em Phys. Rev.} {\bf D89} (2014), no.~6 065003,
  [\href{http://arxiv.org/abs/1311.1204}{{\tt arXiv:1311.1204}}].

\bibitem{Cortes:2009cs}
V.~Cortes and T.~Mohaupt, {\it {Special Geometry of Euclidean Supersymmetry
  III: The Local r-map, instantons and black holes}},  {\em JHEP} {\bf 07}
  (2009) 066, [\href{http://arxiv.org/abs/0905.2844}{{\tt
  arXiv:0905.2844}}].

\bibitem{Chow:2014cca}
D.~D.~K. Chow and G.~Comp{\`e}re, {\it {Black holes in N=8 supergravity from
  SO(4,4) hidden symmetries}},  {\em Phys. Rev.} {\bf D90} (2014), no.~2
  025029, [\href{http://arxiv.org/abs/1404.2602}{{\tt arXiv:1404.2602}}].

\bibitem{Tseytlin:1996bh}
A.~A. Tseytlin, {\it {Harmonic superpositions of M-branes}},  {\em Nucl. Phys.}
  {\bf B475} (1996) 149--163,
  [\href{http://arxiv.org/abs/hep-th/9604035}{{\tt hep-th/9604035}}].
  [,286(1996)].

\bibitem{Behrndt:1996jn}
K.~Behrndt, G.~Lopes~Cardoso, B.~de~Wit, R.~Kallosh, D.~Lust, and T.~Mohaupt,
  {\it {Classical and quantum N=2 supersymmetric black holes}},  {\em Nucl.
  Phys.} {\bf B488} (1997) 236--260,
  [\href{http://arxiv.org/abs/hep-th/9610105}{{\tt hep-th/9610105}}].

\bibitem{Cvetic:1995bj}
M.~Cvetic and A.~A. Tseytlin, {\it {Solitonic strings and BPS saturated dyonic
  black holes}},  {\em Phys. Rev.} {\bf D53} (1996) 5619--5633,
  [\href{http://arxiv.org/abs/hep-th/9512031}{{\tt hep-th/9512031}}].
  [Erratum: Phys. Rev.D55,3907(1997)].

\bibitem{Gutowski:2003rg}
J.~B. Gutowski, D.~Martelli, and H.~S. Reall, {\it {All Supersymmetric
  solutions of minimal supergravity in six- dimensions}},  {\em Class. Quant.
  Grav.} {\bf 20} (2003) 5049--5078,
  [\href{http://arxiv.org/abs/hep-th/0306235}{{\tt hep-th/0306235}}].



\bibitem{Bena:2011dd}
I.~Bena, S.~Giusto, M.~Shigemori, and N.~P. Warner, {\it {Supersymmetric
  Solutions in Six Dimensions: A Linear Structure}},  {\em JHEP} {\bf 03}
  (2012) 084, [\href{http://arxiv.org/abs/1110.2781}{{\tt
  arXiv:1110.2781}}].

\bibitem{Gibbons:1979zt}
G.~W. Gibbons and S.~W. Hawking, {\it {Gravitational Multi - Instantons}},
  {\em Phys. Lett.} {\bf 78B} (1978) 430.

\bibitem{Gibbons:1987sp}
G.~W. Gibbons and P.~J. Ruback, {\it {The Hidden Symmetries of Multicenter
  Metrics}},  {\em Commun. Math. Phys.} {\bf 115} (1988) 267.

\bibitem{Ferrara:1997tw}
S.~Ferrara, G.~W. Gibbons, and R.~Kallosh, {\it Black holes and critical points
  in moduli space},  {\em Nucl. Phys.} {\bf B500} (1997) 75--93,
  [\href{http://arxiv.org/abs/hep-th/9702103}{{\tt hep-th/9702103}}].

\bibitem{Deser:2019acl}
S.~Deser, {\it {Energy in Gravitation and Noether's Theorems}},
  \href{http://arxiv.org/abs/1902.0510}{{\tt arXiv:1902.0510}}.

\bibitem{Ferrara:1995ih}
S.~Ferrara, R.~Kallosh, and A.~Strominger, {\it N=2 extremal black holes},
  {\em Phys. Rev.} {\bf D52} (1995) 5412--5416,
  [\href{http://arxiv.org/abs/hep-th/9508072}{{\tt hep-th/9508072}}].

\bibitem{Ferrara:1996dd}
S.~Ferrara and R.~Kallosh, {\it {Supersymmetry and Attractors}},  {\em Phys.
  Rev.} {\bf D54} (1996) 1514--1524,
  [\href{http://arxiv.org/abs/hep-th/9602136}{{\tt hep-th/9602136}}].

\bibitem{Strominger:1996kf}
A.~Strominger, {\it {Macroscopic Entropy of $N=2$ Extremal Black Holes}},  {\em
  Phys. Lett.} {\bf B383} (1996) 39--43,
  [\href{http://arxiv.org/abs/hep-th/9602111}{{\tt hep-th/9602111}}].

\bibitem{Ferrara:1996um}
S.~Ferrara and R.~Kallosh, {\it {Universality of Supersymmetric Attractors}},
  {\em Phys. Rev.} {\bf D54} (1996) 1525--1534,
  [\href{http://arxiv.org/abs/hep-th/9603090}{{\tt hep-th/9603090}}].

\bibitem{Sen:2005wa}
A.~Sen, {\it {Black hole entropy function and the attractor mechanism in higher
  derivative gravity}},  {\em JHEP} {\bf 0509} (2005) 038,
  [\href{http://arxiv.org/abs/hep-th/0506177}{{\tt hep-th/0506177}}].

\bibitem{Mohaupt:2010fk}
T.~Mohaupt and O.~Vaughan, {\it {Non-extremal Black Holes, Harmonic Functions,
  and Attractor Equations}},  {\em Class. Quant. Grav.} {\bf 27} (2010) 235008,
  [\href{http://arxiv.org/abs/1006.3439}{{\tt arXiv:1006.3439}}].

\bibitem{Goldstein:2014gta}
K.~Goldstein, V.~Jejjala, and S.~Nampuri, {\it {Hot Attractors}},  {\em JHEP}
  {\bf 01} (2015) 075, [\href{http://arxiv.org/abs/1410.3478}{{\tt
  arXiv:1410.3478}}].

\bibitem{Goldstein:2014qha}
K.~Goldstein, S.~Nampuri, and {\'A}.~V{\'e}liz-Osorio, {\it {Heating up branes
  in gauged supergravity}},  [\href{http://arxiv.org/abs/1406.2937}{{\tt
  arXiv:1406.2937}}].

\bibitem{Goldstein:2018mwt}
K.~Goldstein, V.~Jejjala, J.~Mashiyane, James, and S.~Nampuri, {\it
  {Generalized Hot Attractors}},  {\em JHEP} {\bf 03} (2019) 188,
  [\href{http://arxiv.org/abs/1811.0496}{{\tt arXiv:1811.0496}}].

\bibitem{Dijkgraaf:2016lym}
R.~Dijkgraaf, B.~Heidenreich, P.~Jefferson, and C.~Vafa, {\it {Negative Branes,
  Supergroups and the Signature of Spacetime}},  {\em JHEP} {\bf 02} (2018)
  050, [\href{http://arxiv.org/abs/1603.0566}{{\tt arXiv:1603.0566}}].

\bibitem{Hull:1998ym}
C.~Hull, {\it {Duality and the signature of space-time}},  {\em JHEP} {\bf
  9811} (1998) 017, [\href{http://arxiv.org/abs/hep-th/9807127}{{\tt
  hep-th/9807127}}].

\end{thebibliography}
\providecommand{\href}[2]{#2}\begingroup\raggedright\endgroup

\end{document}